\documentclass[reprint,
nofootinbib,
 amsmath,amssymb,
 aps,
]{revtex4-2}

\usepackage{graphicx}% Include figure files
\usepackage{dcolumn}% Align table columns on decimal point
\usepackage{bm}% bold math
\usepackage{ulem}
\usepackage{multirow}
\usepackage[dvipsnames]{xcolor}
\usepackage{braket}
\begin{document}
\newcommand{\vP}{\mathbf{P}}
\newcommand{\vers}{\mathbf{r''}}
\newcommand{\vRs}{\mathbf{R''}}
\newcommand{\verp}{\mathbf{r'}}
\newcommand{\vRp}{\mathbf{R'}}
\newcommand{\Wcm}{\;\mathrm{W/cm}^2}
\newcommand{\eV}{\;\mathrm{eV}}
\newcommand{\Rep}{\mathrm{Re}\,}
\newcommand{\Imp}{\mathrm{Im}\,}
\newcommand{\vk}{{\mathbf{k}}}
\newcommand{\vkst}{\mathbf{k}_\textrm{st}}
\newcommand{\vqst}{\mathbf{q}_\textrm{st}}
\newcommand{\vi}{\hat{\mathbf{i}}}
\newcommand{\vj}{\hat{\mathbf{j}}}
\newcommand{\vS}{{\mathbf S}}
\newcommand{\vH}{\mathbf{H}}
\newcommand{\vv}{\mathbf{v}}
\newcommand{\ve}{\hat{\mathbf{e}}}
\newcommand{\0}{\mathbf{0}}
\newcommand{\vE}{\mathbf{E}}
\newcommand{\vA}{\mathbf{A}}
\newcommand{\ver}{\mathbf{r}}
\newcommand{\vd}{\mathbf{d}}
\newcommand{\va}{\mathbf{a}}
\newcommand{\vD}{\mathbf{D}}
\newcommand{\vp}{\mathbf{p}}
\newcommand{\vR}{\mathbf{R}}
\newcommand{\vq}{\mathbf{q}}
\newcommand{\vrho}{\mbox{\boldmath{$\rho$}}}
\newcommand{\del}{\mbox{\boldmath{$\nabla$}}}
\newcommand{\valpha}{\mbox{\boldmath{$\alpha$}}}
\newcommand{\vRR}{\{\vR\}}
\newcommand{\vRRp}{\{\vRp\}}
\newcommand{\Ip}{I_\mathrm{p}}
\newcommand{\Up}{U_\mathrm{p}}
\newcommand{\calT}{\mathbf{\cal T}}
\newcommand{\calF}{\mathbf{\cal F}}
\newcommand{\calU}{\mathbf{\cal U}}
\newcommand{\et}{\tilde{e}}
\newcommand{\cm}{\mathrm{c.m.}}
\newcommand{\vro}{\mathbf{\rho}}
\newcommand{\tos}{t_{0s}}
\newcommand{\ts}{t_{s}}
\newcommand{\Ep}{E_{\mathbf{p}}}

\preprint{APS/123-QED}

\title{Below threshold nonsequential double ionization with linearly polarized two-color fields II: Quantum interference }% Force line breaks with \\
\author{S. Hashim$^1$, D. Habibovi\'{c}$^{2}$, and C. Figueira de Morisson Faria$^{1}$}\thanks{Corresponding author: c.faria@ucl.ac.uk}
\affiliation{$^1$Department of Physics and Astronomy, University College London, Gower Street, London, WC1E 6BT, UK\\$^2$
University of Sarajevo, Faculty of Science, Zmaja od Bosne 35, 71000 Sarajevo, Bosnia and Herzegovina}

\date{\today}% It is always \today, today,

\begin{abstract}
We perform a systematic analysis of intra-channel quantum interference in laser-induced nonsequential double ionization with linearly polarized bichromatic fields, focusing on the recollision-excitation with subsequent ionization (RESI) mechanism, and employing the strong-field approximation. We generalize and elaborate several analytic interference conditions for RESI in arbitrary driving fields, with a focus on the interference arising from the specific symmetries of bichromatic fields. For example, for waves of comparable strengths, multiple events per half cycle for the direct electron must be considered. Furthermore, interference breaks some of the symmetries arising from the field.  We detangle the superimposed interference fringes originating from phase differences related to symmetrization due to electron exchange, temporal shifts and a combination of exchange and event interference. We show that the hierarchy between exchange-only and exchange-temporal interference is fluid and can be manipulated by an appropriate choice of driving-field parameters. This is enabled by different types of interference occupying specific regions of the plane spanned by the electron momentum components parallel to the driving-field polarization. 
\end{abstract}

\maketitle

%\tableofcontents

\section{\label{sec:intro}Introduction}

Quantum interference is central to strong-field and attosecond physics. Examples include the high-order harmonic phase, used in attosecond pulse generation \cite{Gaarde2008}, above-threshold ionization (ATI) peaks from inter-cycle interference \cite{Lewenstein1995,Becker2002Review}, and structural molecular signatures in photoelectron and harmonic spectra \cite{Lein2007,Augstein2012}. It also underpins ultrafast photoelectron holography \cite{Faria2020} and measurement techniques such as frequency-resolved optical gating (FROG) \cite{Gagnon2008,Orfanos2019}, reconstruction of attosecond bursts by interference of two-photon transitions (RABBIT) \cite{Haessler2010}, the attosecond streak camera \cite{Itatani2002}, spectral phase interferometry for direct electric field reconstruction (SPIDER) \cite{Cormier2005}, and phase-of-the-phase spectroscopy using collinear \cite{Skruszewicz2015,Almajid2017,Wuerzler2020} or circularly polarized \cite{Tulsky2018,Tulsky2020} two-color fields. All these methods exploit phase differences between quantum pathways leading to the same final state.

The above-mentioned pathways can be steered with tailored fields  \cite{Habibovic2024}. For instance, orthogonally polarized fields leave their imprint in high-order harmonic generation (HHG) \cite{Das2013,Das2015} and above-threshold ionization (ATI) \cite{Xie2015,Gong2017,Han2017,Han2018,Habibovic2021,Kim2022}, and bicircular fields create interference vortices \cite{NgokoDjiokap2015,NgokoDjiokap2016,yuan2016,NgokoDjiokap2018,Pengel2017,Pengel2017a,Li2018, Kerbstadt2019,Kerbstadt2019a, Armstrong2019b,Kang2021,Maxwell2021}. 
This control is enabled by the physical mechanisms behind strong-field phenomena, namely the laser-induced recombination or rescattering of an electron with its parent ion \cite{Corkum1993}. Recombination leads to the emission of high-harmonic radiation \cite{Lewenstein1994}, while rescattering leads to high-energy photoelectrons \cite{Paulus1994,Becker2018,Becker2002Review,MilosReviewATI}.  If an electron is freed in the continuum without further interaction with the core, direct ATI occurs, while elastic rescattering gives rise to high-order ATI. Finally, the electron may rescatter inelastically with the parent ion, so that other electrons are released. This causes nonsequential double and multiple ionization (NSDI, NSMI) \cite{Faria2011,Becker2012}.

NSDI in tailored fields has been extensively studied in the literature. However, the vast majority of such investigations have been performed using classical-ensemble computations, and have focused on the shapes of the electron-momentum distributions. These studies include few-cycle pulses \cite{Liu2004,Bergues2012,Huang2016,Kubel2016,Chen2017}, circularly polarized fields \cite{Fu2012,Huang2018}, polarization gated fields \cite{Quan2009}, OTC fields \cite{Zhang2014,Mancuso2016,Song2018}, or few-cycle counter-rotating two-color circularly or elliptically polarized (TCCP, TCEP) laser fields \cite{Pang2020,Ge2023,Liu2024}. This brings with it the  question of whether NSDI is essentially classical or quantum. 

At high enough driving-field intensities, the returning electron's energy exceeds the second ionization potential, and electron-impact (EI) ionization dominates NSDI. In EI, the second electron leaves without delay. Furthermore, integrating over transverse momenta suppresses quantum interference, as shown by the excellent agreement between the full solution of the time-dependent Schr\"odinger equation (TDSE)  \cite{Parker2006} and classical models \cite{Ye2008}. Thus, TDSE studies have focused on how the shapes of the electron momentum distributions are affected by the type of electron-electron interaction \cite{Lein2000,Parker2006} or the field \cite{Baier2006,Baier2007}. The same holds for early work using the strong-field approximation (SFA) \cite{Faria2004,Faria2004b,Faria2005,Faria2008}, or studies employing the quantitative rescattering theory (QRS) \cite{Chen2010,Chen2019,Chen2021,Chen2022}. 
However, for lower, below-threshold driving-field intensities, the second electron is freed by recollision-excitation with subsequent ionization (RESI). In RESI, the first electron, upon recollision, excites a second electron, which is freed with a time delay, and the target's excited states behave as intermediate pathways. In the parameter range where RESI prevails,  quantum-mechanical SFA studies have found evidence that quantum interference is more robust than anticipated. 

For instance, in \cite{Hao2014}, it was shown that quantum interference between excitation channels can significantly reshape RESI distributions, breaking the fourfold symmetry seen in $p_{1\parallel} p_{2\parallel}$ parallel momentum plane for linearly polarized monochromatic fields. This symmetry arises from three field symmetries: half-cycle symmetry (translation plus time reflection), and reflection symmetry about both field maxima and zero crossings \cite{Rook2022}. By controlling quantum interference, correlated or anti-correlated RESI distributions were also achieved.

These findings were extended in our previous work, in which we identified and classified different types of quantum interference in RESI. The patterns encountered are associated with inter- and intrachannel interference, and survive focal averaging and integration over the transverse momentum components \cite{Maxwell2016}. 
Even if only a single excitation channel is considered, interference from electron exchange, time-delayed events, and both physical origins will give rise to a wide range of shapes. Examples are bright fringes at both diagonals $p_{1\parallel}=\pm p_{2\parallel}$ in the $p_{1\parallel}p_{2\parallel}$ plane, and hyperbolic structures. 
Analytic expressions for these interference patterns have been derived in \cite{Maxwell2015} for monochromatic fields and, in \cite{Hashim2024},  generalized for arbitrary driving fields, and investigated for few-cycle pulses.  In \cite{Maxwell2015}, we have verified that inter-channel interference leads to hyperbolic fringes whose width is inversely proportional to the energy difference of the two interfering channels, although an analytical expression remains challenging. Our results also back those in \cite{Hao2014}, as, once quantum interference is incorporated, the RESI distributions obtained with the SFA for a linearly polarized monochromatic field are no longer fourfold symmetric. This invites the question of how quantum interference would affect the symmetries in RESI distributions for linearly polarized bichromatic fields, for which some of the three symmetries mentioned above may be broken. 

In \cite{Hashim2024b}, using the symmetries analyzed in \cite{Rook2022}, we have steered the dominant events and altered the shapes of the RESI distributions for linearly polarized bichromatic fields composed of a wave of frequency $\omega$ and its second or third harmonics, commonly known as ($\omega$,$2\omega$) and $(\omega$,$3\omega$) fields. By an appropriate choice of driving-field parameters such as the phase difference between the two driving waves, the RESI distributions may be confined to specific quadrants of the $p_{1\parallel}p_{2\parallel}$ plane. This is enabled by specific features of two-color fields, which are absent for monochromatic waves and few-cycle pulses. First, for bichromatic fields, there may be more than one relevant ionization event per half a cycle for the second electron. Second, the probability densities for the second electron may not be centered at vanishing parallel momentum. 
However, the transition amplitudes associated with each event and those arising from the electron-momentum symmetrization, required due to their indistinguishability, have been summed incoherently. 
In particular, if a bichromatic field allows one to confine the relevant RESI distributions to a specific momentum region, the contributions from different events will strongly overlap and are expected to interfere substantially. 

Here, we revisit RESI in the $(\omega, 2\omega ) $ and $(\omega, 3 \omega)$ fields used in \cite{Hashim2024b}, but focus on intra-channel quantum interference.  Thus, the present paper and \cite{Hashim2024b} complement each other.  We perform coherent sums of events and symmetrization, investigate how the different interference conditions derived in \cite{Hashim2024} manifest in a two-color scenario, and whether the patterns encountered can be controlled with the specific properties of these fields. We employ the RESI transition amplitude within the SFA, which, albeit reliant on several simplifications, such as the neglect of the residual binding potential in the continuum, allows us to focus on the specific process at hand. In contrast, the outcome of \textit{ab-initio} methods such as the TDSE works as a numerical experiment, in which the different physical mechanisms can be hard to disentangle. 
This article is organized as follows. In Sec.~\ref{sec:backgd}, we review the SFA transition amplitude and the saddle-point equations for RESI. In Sec.~\ref{sec:expinterf}, the interference conditions are adapted to arbitrary fields with more than one ionization event per half cycle for the second electron and applied to the $(\omega,2\omega)$ and $(\omega,3\omega)$ cases.  Subsequently, in Sec.~\ref{sec:coherent}, different types of single-channel quantum interference are investigated in RESI electron momentum distributions. Finally, in Sec.~\ref{sec:conclusions}, we state our conclusions.  Unless otherwise stated, we employ atomic units.

\section{\label{sec:backgd} Background}
\subsection{Transition amplitude}

Within the SFA framework, the RESI transition amplitude reads \cite{Shaaran2010,Shaaran2010a}
\begin{eqnarray}
&&M^{(\mathcal{C})}(\mathbf{p}_{1},\mathbf{p}_{2})=\hspace*{-0.2cm}\int_{-\infty }^{\infty
}dt\int_{-\infty }^{t}dt^{^{\prime }}\int_{-\infty }^{t^{\prime
}}dt^{^{\prime \prime }}\int d^{3}k  \notag \\
&&\times V^{(\mathcal{C})}_{\mathbf{p}_{2}e}V^{(\mathcal{C})}_{\mathbf{p}_{1}e,\mathbf{k}g}V^{(\mathcal{C})}_{\mathbf{k}%
	g}\exp [iS^{(\mathcal{C})}(\mathbf{p}_{1},\mathbf{p}_{2},\mathbf{k},t,t^{\prime },t^{\prime
	\prime })],  \label{eq:Mp}
\end{eqnarray}
where the index $(\mathcal{C})$ denotes the $\mathcal{C}^{th}$ excitation channel. In Eq.~\eqref{eq:Mp} the action is given by
\begin{eqnarray}
&&S^{(\mathcal{C})}(\mathbf{p}_{1},\mathbf{p}_{2},\mathbf{k},t,t^{\prime },t^{\prime \prime
})=  \notag \\
&&\quad E^{(\mathcal{C})}_{\mathrm{1g}}t^{\prime \prime }+E^{(\mathcal{C})}_{\mathrm{2g}}t^{\prime
}+E^{(\mathcal{C})}_{\mathrm{2e}}(t-t^{\prime })-\int_{t^{\prime \prime }}^{t^{\prime }}%
\hspace{-0.1cm}\frac{[\mathbf{k}+\mathbf{A}(\tau )]^{2}}{2}d\tau  \notag \\
&&\quad -\int_{t^{\prime }}^{\infty }\hspace{-0.1cm}\frac{[\mathbf{p}_{1}+%
	\mathbf{A}(\tau )]^{2}}{2}d\tau -\int_{t}^{\infty }\hspace{-0.1cm}\frac{[%
	\mathbf{p}_{2}+\mathbf{A}(\tau )]^{2}}{2}d\tau  .\label{eq:singlecS}
\end{eqnarray} 
Equations~\eqref{eq:Mp} and \eqref{eq:singlecS} have been constructed for the
process in which two electrons are initially bound. At a
time $t^{\prime\prime}$, one of the electrons tunnels from the ground state, whose energy is $-E^{(\mathcal{C})}_{1g}$, to a Volkov state with intermediate momentum $\mathbf{k}$. At a time   $t^{\prime}$, this electron recollides with the singly ionized target and excites a second electron from a state with energy  $-E^{(\mathcal{C})}_{2g}$ to a state with energy  $-E^{(\mathcal{C})}_{2e}$. Upon recollision, the first electron acquires a momentum $\mathbf{p}_1$, with which it will eventually reach the detector. The second electron tunnels at a later time $t$, and reaches a Volkov state with final momentum $\mathbf{p}_2$. 

The prefactors $V^{(\mathcal{C})}_{\mathbf{k}g}$, $V^{(\mathcal{C})}_{\bm{p}_1e,\mathbf{k}g}$ and $V^{(\mathcal{C})}_{\bm{p}_2e}$ contain all information about the geometry of the electronic bound states and the interactions involved in RESI \cite{Shaaran2010,Maxwell2015,Hashim2024}. Explicitly, 
 \begin{eqnarray}\label{eq:pre3}
V^{(\mathcal{C})}_{\mathbf{k}g} &=& \bra{\mathbf{k}+\mathbf{A}(t'')}V\ket{\psi_{1g}^{(\mathcal{C})}}\\ &= &\frac{1}{(2\pi)^{3/2}}\int d^3r_1e^{-i[\mathbf{k} +\mathbf{A}(t'')]\cdot\mathbf{r}_1}V(\mathbf{r}_1)\psi_{1g}^{(\mathcal{C})}(\mathbf{r}_1), \notag
\end{eqnarray}
where $V(\mathbf{r}_1)$ is the neutral atom's binding potential, and $\psi_{1g}^{(\mathcal{C})}(\mathbf{r}_1)=\braket{\mathbf{r}_1|\psi_{1g}^{(\mathcal{C})}}$ is the ground-state wave function for the first electron. The above-stated matrix element takes the first electron from the initial state $\ket{\psi_{1g}^{(\mathcal{C})}}$ to an intermediate continuum state $\ket{\psi^{(L)}_{\mathbf{k}}}=\ket{\mathbf{k}+\mathbf{A}(t'')}$. One should note that Eq.~\eqref{eq:pre3} is written in the length gauge. The unitary transformation from the length to the velocity gauge is a translation in momentum space that removes the vector potential from the intermediate state, so that it is given by $\ket{\psi^{(V)}_{\mathbf{k}}}=\ket{\mathbf{k}}$ instead. For details, we refer to \cite{Shaaran2010}.

The prefactor 
\begin{eqnarray}
\label{eq:Vp1ekg}V^{(\mathcal{C})}_{\mathbf{p}_1e,\mathbf{k}g}\hspace*{-0.1cm}&=& \hspace*{-0.1cm} \bra{\mathbf{p}_1,\psi_{2e}^{(\mathcal{C})}}V_{12}
\ket{\mathbf{k},\psi_{2g}^{(\mathcal{C})}}
    \\
    &=&\hspace*{-0.1cm} \frac{V_{12}(\mathbf{p}_1-\mathbf{k})}{(2\pi)^{3/2}}\hspace*{-0.1cm}\int \hspace*{-0.1cm}d^3r_2e^{-i(\mathbf{p}_1-\mathbf{k})\cdot \mathbf{r}_2}\psi_{2e}^{*(\mathcal{C})}(\mathbf{r}_2)\psi_{2g}^{(\mathcal{C})}(\mathbf{r}_2),  \notag 
\end{eqnarray}
where
\begin{equation}
    V_{12}(\mathbf{p}_1 - \mathbf{k}) =  \frac{1}{(2\pi)^{3/2}}\int d^3rV_{12}(\mathbf{r})\exp[-i(\mathbf{p}_1-\mathbf{k})\cdot\mathbf{r}]
\end{equation}
is the electron-electron interaction in momentum space and
$\mathbf{r} = \mathbf{r}_1-\mathbf{r}_2$, while $V_{12}(\mathbf{r})$ is taken to be of contact type, and it is associated with the excitation process. The interaction takes the first electron from the intermediate Volkov state $\ket{\mathbf{k}+\mathbf{A}(t')}$ to a final state with momentum $\ket{\mathbf{p}_1+\mathbf{A}(t')}$, and excites the second electron from $\ket{\psi_{2g}^{(\mathcal{C})}}$ to $\ket{\psi_{2e}^{(\mathcal{C})}}$. In Eq.~\eqref{eq:Vp1ekg},  $\braket{\mathbf{r}_2|\psi_{2e}^{(\mathcal{C})}}=\psi_{2e}^{(\mathcal{C})}(\mathbf{r}_2)$ and $\braket{\mathbf{r}_2|\psi_{2g}^{(\mathcal{C})}}=\psi_{2g}^{(\mathcal{C})}(\mathbf{r}_2)$.

Finally, the second electron being released in the continuum from an excited state is described by \begin{eqnarray}
    V^{(\mathcal{C})}_{\mathbf{p}_2e} &=& \bra{\mathbf{p}_2+\mathbf{A}(t)}V_{\mathrm{ion}}\ket{\psi_{2e}^{(\mathcal{C})}}  \\ &=& \frac{1}{(2\pi)^{3/2}}\int d^3r_2V_{\mathrm{ion}}(\mathbf{r}_2)e^{-i[\mathbf{p_2}+\mathbf{A}(t)]\cdot\mathbf{r}_2}\psi_{2e}^{(\mathcal{C})}(\mathbf{r}_2),\notag
    \label{eq:Vp2e}
\end{eqnarray}
where $V_{\mathrm{ion}}(\mathbf{r}_2)$ is the potential of the singly ionized target, and it describes the ionization of the second electron. Similarly to the prefactor given by  Eq.~\eqref{eq:pre3}, $ V^{(\mathcal{C})}_{\mathbf{p}_2e} $ is written in the length gauge, the final state being $\ket{\psi^{(L)}_{\mathbf{p}_2}}=\ket{\mathbf{p}_2+\mathbf{A}(t)}$. In the velocity gauge, the vector potential is removed by a unitary transformation, so that, instead, $\ket{\psi^{(V)}_{\mathbf{p}_2}}=\ket{\mathbf{p}_2}$.
Because ionization occurs most probably around field maxima, for monochromatic fields and few-cycle pulses $|\mathbf{A}(\tau)|\ll 1$, with $\tau=t'',t$, one may neglect the vector potential in Eqs. \eqref{eq:pre3} and \eqref{eq:Vp2e} when calculating length-gauge prefactors.  This approximation has been investigated in \cite{Shaaran2010} and used in our subsequent publications. However, as shown in \cite{Hashim2024b}, this fails for two-color fields with comparable strengths, requiring prefactors in both gauges. Here, we use hydrogenic bound-state prefactors from \cite{Shaaran2010,Maxwell2015,Hashim2024} and apply the Gaussian basis from \cite{Hashim2024b} in length-gauge calculations to avoid bound-state singularities.

For bichromatic fields, including prefactors narrows all distributions, confining them to smaller momentum regions \cite{Hashim2024b}. Thus, it is crucial to map prefactor nodes and phase shifts from the $\mathbf{p}_{n}$ $(n=1,2)$ space onto the $p_{1\parallel}p_{2\parallel}$ plane, as detailed in \cite{Hashim2024}. In the velocity gauge, nodes and phase shifts from the $V_{\mathbf{p}_{2}e}$ prefactor appear in the momentum distributions: radial nodes produce minima parallel to the $p_{n\parallel}$ axes, $p$-state angular nodes suppress the signal along $p_{n\parallel}=0$, and $d$-state nodes cause suppression parallel to the momentum axes, with possible additional phase shifts. The influence of the recollision prefactor $V_{\mathbf{p}_1e,\mathbf{k}g}\hspace*{-0.1cm}$ is more difficult to predict \cite{Hashim2024}, but will also cause narrowing and potentially phase shifts.  
In the length gauge, nodes from $V_{\mathbf{p}_{2}e}$ may blur if the vector potential at the second electron's ionization time $t$ is large \cite{Hashim2024b}, or new nodes and phase shifts may appear.

\subsection{Saddle-point equations}
 The transition amplitude \eqref{eq:Mp} is calculated using the saddle-point method \cite{Milosevic2024TopRev}. This procedure seeks variables $t''$, $\mathbf{k}$, $t'$ and $t$ such that the action is stationary, and leads to the saddle-point equations 
\begin{equation}\label{eq:sphati1}
[\vk+\vA(t'')]^2=-2E_{1g},
\end{equation}  
\begin{equation}\label{eq:sphati2}
\vk=-\frac{1}{t'-t''}\int_{t''}^{t'}d\tau\vA(\tau),
\end{equation}
\begin{equation}\label{eq:sphati3}
[\vp_1+\vA(t')]^2=[\vk+\vA(t')]^2-2(E_{2g}-E_{2e}),
\end{equation}
and
\begin{equation}\label{eq:spati}
[\vp_2+\vA(t)]^2=-2E_{2e},
\end {equation}
where the indices $(\mathcal{C})$ have been dropped for simplicity. This can be done without loss of generality as we focus on intra-channel interference.
Equations \eqref{eq:sphati1} and \eqref{eq:spati} give the energy conservation conditions at the time of ionization for the first and second electron, respectively. Those equations have no real solution, resulting from tunneling having no classical counterpart. Equation~\eqref{eq:sphati2} restricts the intermediate momentum $\vk$ of the first electron so that it can return to the parent ion. Moreover, Eq.~\eqref{eq:sphati3}  states that energy is conserved at the time $t'$ for which the first electron rescatters. Explicitly,  it provides a relation between the first electron's final momentum $\mathbf{p}_1$ and its intermediate momentum $\mathbf{k}$. 

Assuming that the field is linearly polarized allows us to rewrite the above equations in terms of the momentum components $p_{n\parallel}$, 
$\mathbf{p}_{n\perp}$, $n=1,2$, parallel and perpendicular to the driving-field polarization. This formulation sheds light on the RESI momentum regions. Equation~\eqref{eq:spati} reads 
\begin{equation}\label{eq:sp4ppar}
[p_{2\parallel}+A(t)]^2=-2E_{2e}-\mathbf{p}^2_{2\perp},
\end {equation}
which describes a sphere of complex radius in the $\mathbf{p}_{2}$ space, centered at $(p_{2x},p_{2y},p_{2\parallel})=(0,0,-A(t))$, where $\mathbf{p}_{n\perp}=p_{nx}\hat{e}_x+ p_{ny}\hat{e}_y$, $(n=1,2)$ and $\hat{e}_x$ and $\hat{e}_y$ are vectors spanning the plane perpendicular to the driving field polarization. If we assume $\mathbf{p}_{2\perp}$ to be constant, its role is to effectively shift the second ionization potential, so that for $\mathbf{p}_{2\perp}=\mathbf{0}$ this quantity is minimal. Equation~\eqref{eq:spati} is formally equal to that obtained for direct ATI, so that the maximal energy obtained by the second electron is given by the corresponding cutoff \cite{Shaaran2010,Shaaran2010a}. 

Equation~\eqref{eq:sphati3} describes a sphere centered at $(p_{1x},p_{1y},p_{1\parallel})=(0,0,-A(t'))$, whose radius is proportional to the difference between the electron's kinetic energy upon return and the energy gap between the ground and excited states' energies of the singly ionized target.  If its right-hand side is positive, rescattering has a classical counterpart, and the momentum region for which this holds is known as the classically allowed region. Similarly to what happens to the second electron, if $\mathbf{p}_{1\perp}$ is kept fixed, it effectively adds a term to the energy gap $E_{2g}-E_{2e}$. This will shrink the classically allowed region, so that an upper bound is determined for $\mathbf{p}_{1\perp}=\mathbf{0}$ \cite{Shaaran2010a}.  Apart from the energy gap, Eq.~\eqref{eq:sphati3} resembles that obtained for rescattered ATI. Thus, for high enough laser intensities, the energy of the first electron at the detector approaches the rescattered ATI cutoff, which is much higher than that of the direct  \cite{MilosReviewATI}.
If the above-stated information is combined, the width and the length of the momentum regions for which the RESI probability densities will be physically relevant will be determined by the first and second electron, respectively \cite{Shaaran2010,Shaaran2010a}. 
\subsection{Correlated ionization probabilities}
The RESI transition amplitude is then expressed as an asymptotic expansion over saddle-point solutions representing relevant events. For the second electron, we apply the standard saddle-point approximation, as its saddles are well separated. For the first electron, closely spaced saddles are treated in pairs using the uniform approximation from \cite{Faria2002}. The final transition amplitude is symmetrized with regard to momentum exchange, to account for electron indistinguishability \cite{Shaaran2010}.

The quantity of interest is the RESI two-electron momentum density as a function of the parallel components $p_{n\parallel}$ ($n=1,2$), given by
\begin{align}
	\mathcal{P}(p_{1\parallel},p_{2\parallel})= \int\int d^2 p_{1\perp}d^2 p_{2\perp}\mathcal{P}(\mathbf{p}_{1},\mathbf{p}_{2}), \label{Eq:Channels}
\end{align}
where $\mathcal{P}(\mathbf{p}_{1},\mathbf{p}_{2})$ is the fully resolved two-electron momentum probability density for a single channel, and the transverse momentum components have been integrated over.  
Here, $\mathcal{P}(\mathbf{p}_{1},\mathbf{p}_{2})$  is written differently, depending on the question at hand. 
A coherent and incoherent sum over events $\varepsilon$ and symmetrization leads to
\begin{equation}
\mathcal{P}_{(\mathrm{cc})}(\mathbf{p}_{1},\mathbf{p}_{2})=\left|\sum_{\varepsilon}\left[M_{\varepsilon}(\mathbf{p}_1,\mathbf{p}_2)+M_{\varepsilon}(\mathbf{p}_2,\mathbf{p}_1)\right]\right|^2,
\label{eq:1coherent}
\end{equation}
and
\begin{equation}
\mathcal{P}_{(\mathrm{ii})}(\mathbf{p}_{1},\mathbf{p}_{2})=\sum_{\varepsilon}\left[\left|M_{\varepsilon}^{(\mathcal{C})}(\mathbf{p}_1,\mathbf{p}_2)\right|^2\hspace*{-0.2cm}+\hspace*{-0.1cm}\left|M_{\varepsilon}^{(\mathcal{C})}(\mathbf{p}_2,\mathbf{p}_1)\right|^2\right], 
\label{eq:1ii}
\end{equation}
respectively. 
We also employ the partly coherent sums
 \begin{equation}
\mathcal{P}_{(\mathrm{ci})}(\mathbf{p}_{1},\mathbf{p}_{2})=\sum_{\varepsilon}\left|M_{\varepsilon}^{(\mathcal{C})}(\mathbf{p}_1,\mathbf{p}_2)+M_{\varepsilon}^{(\mathcal{C})}(\mathbf{p}_2,\mathbf{p}_1)\right|^2. 
\label{eq:1ci}
\end{equation}
and
 \begin{equation}
\mathcal{P}_{(\mathrm{ic})}(\mathbf{p}_{1},\mathbf{p}_{2})=\left|\sum_{\varepsilon}M_{\varepsilon}^{(\mathcal{C})}(\mathbf{p}_1,\mathbf{p}_2)\right|^2\hspace*{-0.2cm}+\hspace*{-0.1cm}\left|\sum_{\varepsilon}M_{\varepsilon}^{(\mathcal{C})}(\mathbf{p}_2,\mathbf{p}_1)\right|^2. 
\label{eq:1ic}
\end{equation}

In Eq.~\eqref{eq:1ci}, the symmetrization is done coherently, but the events are summed over incoherently, while Eq.~\eqref{eq:1ic} sums events coherently and symmetrizes incoherently. Often in this work, it is necessary to carry out partial sums in which specific events are considered pairwise, or in which we look at an individual symmetrized event. Furthermore, in this work, we also consider the interference of a specific event with the symmetrized counterpart of a time-delayed event. In this case, assuming two events $\varepsilon$ and $\varepsilon '$, $\varepsilon \neq \varepsilon'$, and summing these events pairwise, the corresponding coherent sum reads
\begin{equation}
    \mathcal{P}_{(\mathrm{ci, \varepsilon\varepsilon'})}(\mathbf{p}_{1},\mathbf{p}_{2})=\sum_{\varepsilon, \varepsilon'}\left|M_{\varepsilon}^{(\mathcal{C})}(\mathbf{p}_1,\mathbf{p}_2)+M_{\varepsilon'}^{(\mathcal{C})}(\mathbf{p}_2,\mathbf{p}_1)\right|^2,  \varepsilon \neq  \varepsilon'
    \label{eq:combinedP}
    \end{equation}

We indicate differences of two-electron probability densities computed by different means by $\mathcal{P}_{\mathrm{diff}}(p_{1\parallel},p_{2\parallel})$ for simplicity in the figure axes, but are more specific about what differences we consider in the captions and discussions. 

\subsection{Field properties and dominant events}
\label{sec:fieldmapping}

Here, we employ linearly polarized bichromatic fields with commensurate frequencies $r\omega$ and $s\omega$, where $r,s$ are co-prime integers. The corresponding vector potential reads
\begin{equation}
   \mathbf{A}_{r, s, \xi, \phi}(t)= \frac{2 \sqrt{U_p} }{\zeta}\left[\frac{\xi  \cos (\phi+s\omega t)}{s}+\frac{\cos (r \omega t)}{r}\right]\hat{e}_z,
   \label{eq:Afield}
\end{equation}
where 
\begin{equation}
    \zeta=\sqrt{\frac{1}{r^2}+\frac{\xi ^2}{s^2}}
\end{equation}
and
\begin{equation}
    U_p=\frac{E^2_{r\omega}} {4\omega^2} \zeta^2
    \label{eq:epond}
\end{equation}
is the ponderomotive energy and the field is polarized along $\hat{e}_z$. The intensity  $E^2_{r\omega}$ is associated with the wave of frequency $r\omega$, $\xi=E_{s\omega}/E_{r\omega}$ is the field-strength ratio, and $\phi$ is the relative phase between the two driving waves.  

\begin{figure*}
 \hspace*{-1.1cm}\includegraphics[width=1.1\textwidth]{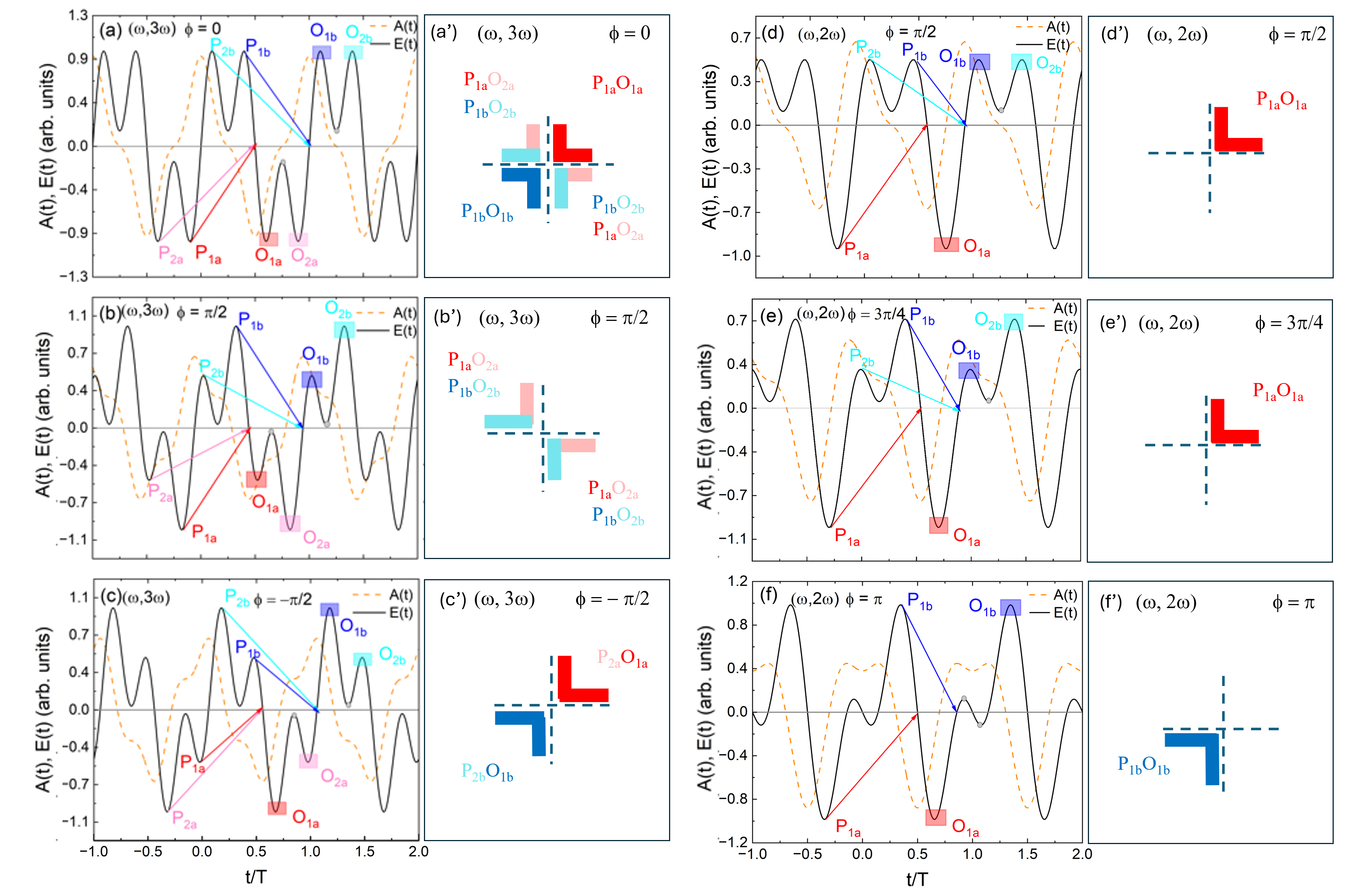}
    \caption{Electric field (black solid lines), corresponding vector potential (orange dashed lines) as functions of time, and diagrammatic representation of the dominant events for the ($\omega$,$3\omega$) [panels (a, a'), (b, b'), and (c, c')], and ($\omega$,$2\omega$) [panels (d, d'), (e, e'), and (f, f')] bichromatic linearly polarized field.  The ratio of the field amplitudes is $\xi=0.8$, and the relative phase is indicated in the panels.  The approximate values of the real part of the ionization and rescattering times of the first electron associated with the saddle-point pairs which lead to the most significant contributions to the photoelectron yield are indicated by the arrows, while the shaded rectangles correspond to the approximate values of the real part of the ionization time of the second electron. The subscript $n=1,2$ in the pairs $P_{n\mu}$ classifies them in increasing order of excursion times in the continuum, i.e., the excursion time for the pairs  $P_{1\mu}$ are smaller than those for pairs  $P_{2\mu}$. The index $\mu=a,b$ refers to the first and second half cycle taken into consideration, respectively. The red and pink, and the blue and cyan arrows indicate rescattering events populating the positive and negative momentum regions, respectively. The colors of the shaded rectangles match those of the arrow, but, rather, indicate that the orbits $O_{jk}$ of the second electron are associated with a specific pair for the first electron, instead of referring to the momentum region they populate. The subscript $j=1,2$ refers to how close the event is regarding the time of rescattering of the first electron, and the indices $a$ and $b$ refer to the first and second half cycle considered, respectively. The gray dots indicate irrelevant ionization events. The electric fields have been normalized to their maximum amplitude in each panel. 
     The thick straight lines in the figure provide a schematic of what momentum regions of the $p_{1\parallel}p_{2\parallel}$ plane the correlated RESI distributions occupy. The colors associated with the events are matched to those marked on the field and are colored according to the convention adopted for the orbits $O_{k\nu}$ rather than the total event.}
    \label{fig:fields}
\end{figure*}

We consider $(\omega,2\omega)$ and $(\omega, 3\omega)$ fields [$r=1$ and $s=2$ or $3$ in Eqs.~\eqref{eq:Afield} and \eqref{eq:epond}, respectively], whose relative phases are chosen to enforce or break specific symmetries. These fields have been employed in our publication \cite{Hashim2024b}, and are presented in Fig.~\ref{fig:fields}, together with diagrammatic representations of the expected RESI distributions (left and right columns, respectively). These representations have been introduced in \cite{Hashim2024} to indicate the momentum regions occupied by probability densities associated with particular events. Thereby, the dashed lines indicate the parallel momentum axes, and the thick solid lines give the approximate regions associated with specific events. If the events and symmetrization are summed incoherently, the distributions mirror these schematics.  If quantum interference is incorporated, we expect these symmetries to be broken. Studies of the incoherently summed RESI distributions and dominant events are presented in \cite{Hashim2024b}.

In Fig.~\ref{fig:fields}, different events are represented by different colors. The mapping is classical, but the approximate times indicated in Fig.~\ref{fig:fields} correspond to the real parts of the saddle-point solutions  $t'',t'$ associated with the first electron [Eqs.~\eqref{eq:sphati1}--\eqref{eq:sphati3}], and of the saddle-point solutions of Eq.~\eqref{eq:spati} giving the ionization time $t$ of the second electron. The solutions associated with the first electron occur in pairs \cite{Faria2002}, and $\mathrm{Re}[t'']$ and $\mathrm{Re}[t']$ are close to the extrema and zero crossings of $E(t)$ respectively. The classical times are indicated by arrows in Fig.~\ref{fig:fields}, where we have employed the tangent construction \cite{Faria1999}.

Pairs $P_{n\mu}$, associated with the first electron, are classified as follows. The subscript $n=1,2$ refers to the length of the orbit in the continuum, starting from the shortest, and the subscript $\mu=a,b$ indicates the half cycle in which ionization occurs, starting with that associated with events $a$. The red and pink (blue and cyan) arrows correspond to pairs leading to positive (negative) momenta $p_{1\parallel}$.
Due to the rescattering event at $\mathrm{Re}[t']$, the second electron is excited and leaves around the subsequent field maxima, indicated by the shaded regions in the figure. The ionization events of the second electron are classified by $O_{k\nu}$, where the subscript $k=1,2$ increases with the temporal distance from the rescattering time of the first electron.
The subscript $\nu=a,b$ gives the half-cycle with which the orbits should be associated to construct a two-electron event. For example, the event $P_{1a}O_{1a}$ means that we combine the shortest pair for the first electron, starting in the first half cycle we consider, with the ionization event $O_{1a}$ closest to the rescattering time. Picking up $P_{1a}O_{2a}$ implies that, instead, we take the second ionization event after rescattering, and so forth. The sign associated with the most probable momentum $p_{2\parallel}$ can be read from the instantaneous vector potential using the mapping $p_{2\parallel}=-A(t)$. One expects quantum interference to be substantial if different events occupy the same momentum region, outlined in the mappings provided in the right column of Fig.~\ref{fig:fields}. 

Before examining Fig.~\ref{fig:fields}, one should note the hierarchy of criteria determining the prevalence of a specific event. For the first electron, the key factors, in order of importance, are: (i) \textit{high ionization probability}; (ii) \textit{a short time delay between ionization and rescattering} to minimize wave-packet spreading in the continuum; and (iii) \textit{a large classically allowed momentum region} to prevent exponential decay of the transition amplitude. The probability of ionization depends on the instantaneous electric field $|E(t'')|$ at ionization, with higher fields increasing the probability. For the second electron, the key contributor is the ionization probability at $t$, which, once more, is associated with the instantaneous absolute value of the electric field. Another issue is bound-state depletion, which suppresses the events for $\mathrm{Re}[t]\gg \mathrm{Re}[t']$. In the present work, due to wave-packet spreading, we consider up to the second shortest pair for the first electron and, due to bound-state depletion, we only take the ionization events for the second electron occurring in the half-cycle after rescattering.

The $(\omega,3\omega)$ fields [Figs.~\ref{fig:fields}(a), (b), and (c)] are half-cycle symmetric, i.e., $E(t \pm T/2)=-E(t)$ and $A(t \pm T/2)=-A(t)$, where $T$ is a field cycle. Therefore, the resulting RESI distributions will exhibit reflection symmetry about the diagonals $p_{1\parallel}=\pm p_{2\parallel}$, as they should be invariant upon $(p_{1\parallel},p_{2\parallel}) \rightarrow (-p_{1\parallel},-p_{2\parallel}) $. For  $(\omega,2\omega)$ fields [Fig.~\ref{fig:fields}(d), (e), and (f)], the half-cycle symmetry is broken, so that only the reflection symmetry associated with the main diagonal $p_{1\parallel}= p_{2\parallel}$
is retained. 

 For $(\omega,3\omega,\phi=0)$, Figs.~\ref{fig:fields}(a) and (a') map key events in time and momentum space, respectively. Each half cycle features two equal-magnitude extrema, symmetric about $t = n\pi/\omega$. These extrema determine the ionization times for pairs $P_{1a,b}$ and $P_{2a,b}$ [arrows in Fig.~\ref{fig:fields}(a)] and, after rescattering, the orbits $O_{1a,b}$ and $O_{2a,b}$ [shaded regions]. These times are symmetric about a zero crossing of $A(t)$, so the contributions of $O_{1a,b}$ and $O_{2a,b}$ are mirror images peaked at symmetric, nonzero momenta. The shorter excursion time makes the contributions of $P_{1a,b}$ dominant, leading to the fourfold symmetry in Fig.~\ref{fig:fields}(a'): events from different half cycles occupy the second and fourth quadrants of the $p_{1\parallel}p_{2\parallel}$ plane, with their symmetrized counterparts in the first and third. 

Changing $\phi$ renders the above-mentioned field peaks unequal, although the half-cycle symmetry is preserved. For instance, taking $\phi=\pi/2$ will weaken the contributions of $O_{1a,b}$ for the $(\omega,3\omega)$ field [Fig.~\ref{fig:fields}(b)], moving the to RESI distributions to the second and fourth quadrants [Fig.~\ref{fig:fields}(b')]. For $(\omega,3\omega,\phi=-\pi/2)$, the contributions of $O_{2a,b}$ are suppressed, shifting the RESI distributions to the first and third quadrants [Figs.~\ref{fig:fields}(c) and (c')]. Furthermore, the dominant pairs for the first electron are now $P_{2 a,b}$, because a higher ionization probability at $t''$ supersedes a shorter excursion time. 

For $(\omega,2\omega)$ fields, similar features arise, although the half-cycle symmetry is broken. For $\phi=\pi/2$ [Fig.~\ref{fig:fields}(d)], a double peak in $E(t)$ appears during one half cycle, but the stronger field at $P_{1a}$ and $O_{1a}$ [red arrow and shaded area] dominates, producing RESI distributions on the positive half-axes of the parallel momentum plane [Fig.~\ref{fig:fields}(d')]. For $\phi=3\pi/4$ [Fig.~\ref{fig:fields}(e)], $P_{1a}O_{1a}$ remain dominant, but $P_{1b}$ and $O_{2b}$ gain strength, contributing to the second and fourth quadrants. 
Finally, for  $(\omega,2\omega, \phi=\pi)$, the contributions from $P_{1b}$ and $O_{2b}$ prevail over those of $P_{1a}$ and $O_{2a}$ due to similar ionization probabilities and a shorter excursion time for $P_{1b}$. In Figs.~\ref{fig:fields}(f) and (f'), $O_{2b}$ is labeled as $O_{1b}$ since the original $O_{1b}$ contribution is negligible.

\section{Expected Interference Patterns}\label{sec:expinterf}
Next, we adapt the interference conditions from \cite{Hashim2024} to cases with multiple second-electron ionization events per half-cycle. This covers the fields considered here, which have a single dominant first-electron pair of solutions \cite{Hashim2024b}. A similar generalization for the first electron is possible, but it will not be performed here. We then apply these conditions to the $(\omega, 2\omega)$ and $(\omega, 3\omega)$ fields in Fig.~\ref{fig:fields}.
  
\subsection{Generalized Interference Conditions}
\label{sec:generalizedinterf}
\begin{figure*}[!htbp]
\centering
\includegraphics[width=0.97\textwidth]{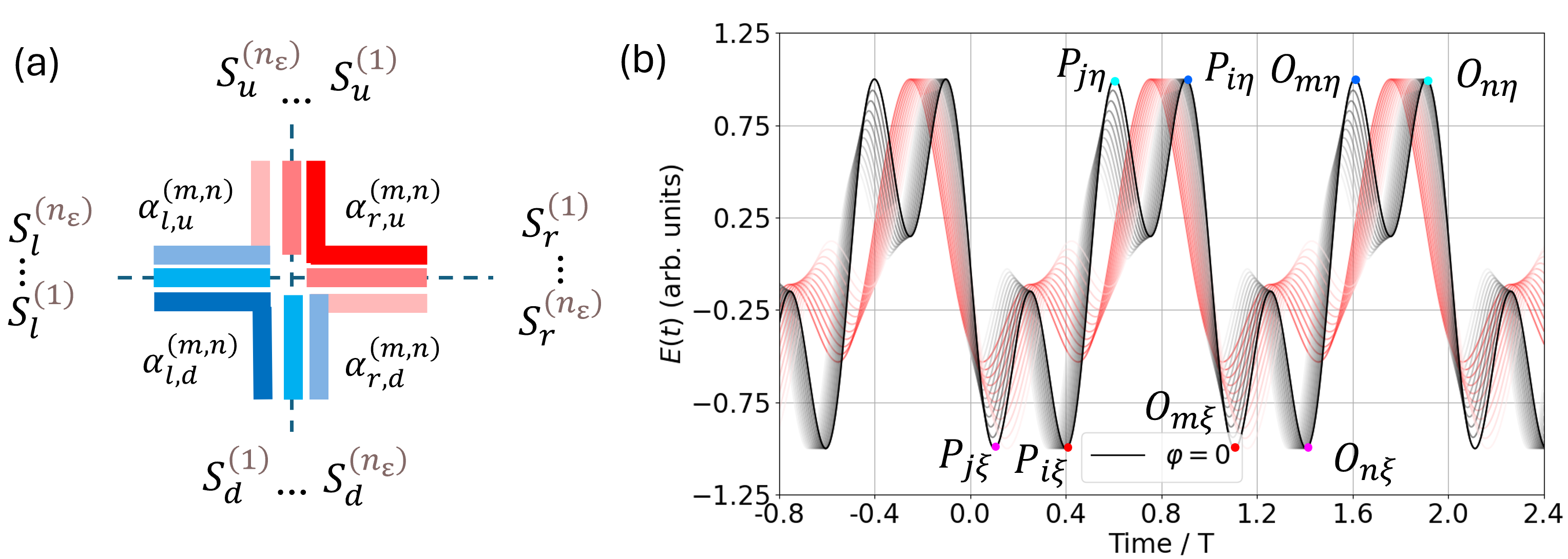}
    \caption{Generic schematics in momentum [panel (a)] and time [panel (b)] of interfering events for RESI in a two-color field. Panel (a) shows a schematic representation of the regions of the $p_{1\parallel}p_{2\parallel}$ plane occupied by the transition amplitudes and their symmetrized counterparts for RESI. The same (different) colors mean that the events occur in the same (different) half cycle. Different shades of the same color indicate multiple events occurring within the same half cycle. The actions of each event are denoted by $S_{\mu}^{(n)}$ where $\mu=l, d, r, u$ gives the momentum region in which the event is localized and $n=1,...,n_{\epsilon}$ where $n_\epsilon$ is the total number of events within a half cycle. Any generic event, $P_{i\xi}O_{m\xi}$ may interfere with any other event $P_{i\eta}O_{n\eta}$. While $\mu,\nu$ provides a good approximation of where the interference pattern will be localized, the interference may spill over the axes if $m-n$ is large. For simplicity, in the plot,  $n_{\varepsilon}$ is the same for $l$, $u$, $r$, and $d$, although this is not necessarily the case for $l$ and $r$ (or $d$ and $u$). Panel (b) aims to illustrate possible generic events in $2T$ for a bichromatic field with more than one maximum per half cycle.  The solid black line corresponds to the $(\omega,3\omega,\phi=0)$ field. This choice was made because this specific field provides the maximal number of relevant events in the parallel momentum plane for the parameter range considered in this work. However, other $(\omega,3\omega)$ fields with increasing phase are presented in gray along with $(\omega,2\omega)$ fields in red. The colors of the dots in (b) are kept consistent with Fig.~\ref{fig:fields} to allow for easier comparison. 
    }
    \label{fig:interfconditions}
\end{figure*}

Figure~\ref{fig:interfconditions} displays a momentum [Fig.~\ref{fig:interfconditions}(a)] and temporal [Fig.~\ref{fig:interfconditions}(b)] mapping for RESI in an arbitrary field, assuming a single dominant ionization event for the first electron and at least two relevant ionization events for the second electron per half cycle. In Fig.~\ref{fig:interfconditions}(a), we consider $n_\epsilon$ relevant ionization events for the second electron per half cycle e.g., $O_{1\xi}, O_{2\xi}, ..., O_{n_\epsilon \xi}$, with $\xi = a, b$ denoting negative and positive electric field amplitudes, respectively [see Fig.~\ref{fig:fields}]. Any transition amplitude may interfere with any other amplitude. An amplitude associated with an $m$-th event is related to its symmetrized counterpart by $M_\mu^{(m)} (\textbf{p}_1, \textbf{p}_2) = M_\nu^{(m)} (\textbf{p}_2, \textbf{p}_1)$, where $\mu, \nu = l, d$ or $r, u$. The specific indices $\mu, \nu$ depend on the momentum regions occupied by the event in question (left, down, right, or up). For fields with half-cycle symmetry, events associated with $M_l$ and $M_r$ are displaced a half-cycle apart and satisfy $|M_l|=|M_r|$. Their symmetrized counterparts thus satisfy $|M_d|=|M_u|$. 
A generic phase difference associated with interference between two arbitrary events is given by $\alpha_{\mu, \nu}^{(m, n)} = S_\mu^{(m)} - S_\nu^{(n)}$ where $S_\mu^{(m)}$ corresponds to a generic action related to the photoelectron yield in the $\mu$ (= left, down, up or right) part of the $p_{1\parallel}p_{2\parallel}$ plane for the $m$-th event as depicted in Fig.~\ref{fig:interfconditions}(a). 

The amplitudes of any two generic events, $P_{i\xi}O_{m\zeta}$ and $P_{j\eta}O_{n\upsilon}$, may interfere. Since only the dominant pair of solutions per half cycle is taken for the first electron, $i=j$. Depletion restricts relevant second-electron ionization events to the half cycle following rescattering, yielding $\xi = \zeta$ and $\eta = \upsilon$. Thus, interfering events reduce to $P_{i\xi}O_{m\xi}$ and $P_{i\eta}O_{n\eta}$. Figure~\ref{fig:interfconditions}(b) exemplifies such events for  $(\omega,2\omega)$ and $(\omega, 3 \omega)$ fields [red and black curves, respectively] with the field-strength ratio as in Fig.~\ref{fig:fields}.

In \cite{Hashim2024}, we identified three types of interference, associated with: (1) pure exchange phases from symmetrizing individual events, (2) pure temporal phases from time-displaced events without symmetrization, and (3) combined exchange-temporal phases from an event and the symmetrized counterpart of a time-displaced event. Time displacements are denoted by $\Delta \tau = (\Delta t'', \Delta t', \Delta t)$, where $t''$, $t'$, and $t$ are the ionization, rescattering, and second ionization times, respectively (see Sec.~\ref{sec:backgd}). For monochromatic fields, $\Delta t'' = \Delta t' = \Delta t$. However, this no longer holds for arbitrary fields, which may have multiple ionization events per half-cycle and lack half-cycle symmetry. While our prior studies considered events displaced by a full cycle $T$, we neglect them here, though the formalism remains applicable. For displacements within a half-cycle, $\Delta \tau = (0, 0, \Delta t)$ with $\Delta t < 0.5T$. More generally, $\Delta t < T$, so events in different half-cycles are temporally adjacent. We now discuss the phase differences for each interference type.

\subsubsection{Exchange-only interference}
\label{sec:exchange}
Pure exchange interference occurs between an event $P_{i\xi}O_{m\xi}$ and its symmetrized counterpart, with phase difference $\alpha^{(m,m)}_{\mu \nu}$.
For an event mapping to the third quadrant of the $p_{1\parallel}p_{2\parallel}$ plane, this phase is
\begin{eqnarray}
    \alpha_{l,d}^{(m, m)}&=&S_l^{(m)}-S_d^{(m)} \nonumber  \\ &=&\alpha_{\mathbf{p}_1, \mathbf{p}_2}^{(\text {exch) }}(t^{(m)}, t')+\alpha_{\mathbf{p}_1, \mathbf{p}_2}(t^{(m)}, t^{\prime}),
    \label{eq:phaseleftdown}
\end{eqnarray}
where
\begin{equation}
\alpha^{(\mathrm{exch})}_{\mathbf{p}_1,\mathbf{p}_2}(t,t')=\frac{1}{2}  \left(\mathbf{p}_{2 }^2 -\mathbf{p}_{1}^2\right)(t-t'),
\label{eq:alphaexch}
\end{equation}
\begin{equation}
\alpha_{\mathbf{p}_1,\mathbf{p}_2}(t,t')=  (\mathbf{p}_2-\mathbf{p}_1) \cdot \left[\mathbf{F}_{A}(t')-\mathbf{F}_{A}(t)\right],
  \label{eq:alphaexch2}
\end{equation}
and 
\begin{equation}
    \mathbf{F}_A(t)=\int^t \mathbf{A}(\tau)d\tau
    \label{eq:IntegralA}
\end{equation}
is the temporal integral of the vector potential.
Since $m=n$ here, the interfering amplitudes are represented by stripes of the same color and shade in Fig.~\ref{fig:interfconditions}(a). For interference in the left-down quadrant, the relevant stripes are parallel to the negative half-axes. Here, the times $t', t$ are those associated with the event in question. For clarity, we denote the ionization time of the second electron as $t^{(m)}$ - the time associated with the $m$-th event. This distinction shall become more useful when considering multiple second electron events within a half-cycle.

An event occurring roughly half a cycle before (or after) $P_{i\xi}O_{m\xi}$ can be denoted as $P_{i\eta}O_{m\eta}$ for which $\eta$ may be smaller (or larger) than $\xi$ [see Fig.~\ref{fig:interfconditions}(b)] and yields probability densities and amplitudes localized in the first quadrant. This event will have phase difference $\alpha^{(m,m)}_{r,u}$
\begin{align}
    \alpha_{r,u}^{(m, m)}=&S_r^{(m)}-S_u^{(m)} = - \alpha_{\mathbf{p}_1, \mathbf{p}_2}^{(\text {exch) }}(t^{(m)}, t')+\\&\alpha_{\mathbf{p}_1, \mathbf{p}_2}(t^{(m)} + \Delta t^{(m)}, t^{\prime} + \Delta t^{\prime}).
    \label{eq:phaserightup}
\end{align}
This phase is equal to Eq.~\eqref{eq:phaseleftdown} but with the times displaced by $\Delta \tau = (0, \Delta t^{\prime}, \Delta t^{(m)})$. In Fig.~\ref{fig:interfconditions}(a), the interfering amplitudes are represented by stripes of the same color and shade, parallel to the positive half-axes. 

Constructive exchange interference occurs when the total phase difference $\alpha_{\mu,\nu} = 2n\pi$, with $n$ an integer and $\mu, \nu = r, u$ or $l, d$. The simplest case, $n = 0$, requires each building block [Eqs.~\eqref{eq:alphaexch} and \eqref{eq:alphaexch2}] to vanish, as studied in \cite{Hashim2024} for few-cycle pulses and \cite{Maxwell2015} for monochromatic fields. Here, we examine both $n = 0$ and $n \neq 0$. Exchange-only interference produces \textit{hyperbolic fringes} and \textit{spine-like structures} along the diagonal $p_{1\parallel} = p_{2\parallel}$. The field-independent phase $\alpha^{(\mathrm{exch})}_{\mathbf{p}_1,\mathbf{p}_2}$ [Eq.~\eqref{eq:alphaexch}] describes a one-sheet hyperboloid proportional to the time delay $(t - t')$ between rescattering and second electron ionization. Constructive interference produces hyperbolae in the $p_{1\parallel}p_{2\parallel}$ plane with asymptotes at $p_{1\parallel} = \pm p_{2\parallel}$ and eccentricity $\sqrt{2}$. These match the hyperbolae from \cite{Maxwell2015} for $n = 0$, though asymptotes and axes may shift due to transverse momentum integration. The hyperbolae have vertices at $(p_{1\parallel}, p_{2\parallel}) = (0, \sqrt{p_{2\perp}^2 - p_{1\perp}^2 + 4n\pi/(t - t')})$.
There will be as many hyperbolae as there are events, for a given field and phase. For $n = 0$, hyperbolae from each event are identical and overlap exactly, making them the most prominent for all fields. For $n \neq 0$, overlapping but distinct hyperbolae may produce more complex patterns. In bichromatic fields (e.g., Fig.~\ref{fig:interfconditions}(b)), for a given phase, events with $m = i$ share the same $(t - t')$, reinforcing the hyperbolae, while events with $m > i$ have larger $(t - t')$. This has been numerically confirmed for all cases studied. Overall, electron exchange creates an intricate mix of hyperbolae with varying brightness and fringe spacing. Although the $n \neq 0$ fringes are generally weak, they are included here for completeness.

The building block $\alpha_{\mathbf{p}_1,\mathbf{p}_2}(t,t')$ is directly proportional to the integral of the vector potential at $t'$ and $t$, $F_A(t')-F_A(t)$.  For $n = 0$ and linear field polarization, constructive interference appears along the diagonal $p_{1\parallel} = p_{2\parallel}$, forming the `spine' identified in \cite{Hashim2024}. 
For integer $n \neq 0$, the spine generalizes to the lines 
\begin{equation}
  p_{2\parallel} = p_{1\parallel} - \frac{2n\pi}{F_A(t')-F_A(t)}
   \label{eq:spine}
\end{equation}
running parallel to the diagonal with field-dependent intercepts that vary with $n$ and $F_A(t') - F_A(t)$. The intercept’s sign depends on field symmetry, flipping with alternating first-electron solutions ($P_{1\xi}$ yields negative, $P_{2\xi}$ positive). While individual lines may shift above or below the diagonal, overall symmetry is preserved across all $n$. Though difficult to isolate in the total exchange interference pattern, spine lines for each $n$ can be identified by analyzing each event separately. 

The total exchange interference pattern results from the combined effect of parallel spine lines and hyperbolic fringes. Near the origin, $n = 0$ hyperbolae dominate and distort the spine lines, as they are centered at $(0,0)$ for vanishing perpendicular momenta. Hyperbolae from $n \neq 0$ can also skew spine lines near their vertices if within the momentum range.  If half-cycle symmetry breaks ($\alpha_{r,u}^{(m,m)} \neq -\alpha_{l,d}^{(m,m)}$), hyperbolae remain symmetric about the diagonal, but the $(p_{1\parallel}, p_{2\parallel}) \leftrightarrow (-p_{1\parallel}, -p_{2\parallel})$ symmetry is lost. Interference features associated with $n\neq0$ are expected to be much weaker. Contributions from less dominant events may blur in the total pattern but remain visible when the events are isolated.

\subsubsection{Temporal-only interference}
\label{sec:temporal}

Temporal-only shifts give rise to another key type of interference. When two interfering events, \( P_{i\xi}O_{m\xi} \) and \( P_{i\eta}O_{n\eta} \), occur in different half-cycles (\( \xi \neq \eta \)), the interference manifests in distinct ways depending on the relative timing of the second electron events. 
If the second electron events are separated by approximately \( 0.5T \) (\( m = n \)), the interference is characterized by a phase difference \( \alpha_{\mu,\nu}^{(m,m)} \), where \( \mu,\nu = r, l \) or \( u, d \). This interference is represented in Fig.~\ref{fig:interfconditions}(a) by combining any two stripes of different colors but the same shade.  
For events still in different half-cycles (\( \xi \neq \eta \)) but with second electron events separated by less than \( 0.5T \) (\( m \neq n \)), the interference is associated with the phase difference \( \alpha_{\mu,\nu}^{(m,n)} \).
Notably, $\alpha_{\mu,\nu}^{(m,n)} \neq \alpha_{\mu,\nu}^{(n,m)}$ as these phase differences represent distinct combinations of events since $\xi \neq \eta$. 
Their amplitudes are represented in Fig.~\ref{fig:interfconditions}(a) by stripes of different colors and shades.  
When both interfering events occur within the same half-cycle (\( \xi = \eta \)), but the second electron events are separated in time (\( m \neq n \) i.e. by less than $0.5T$), interference is described by phase differences \( \alpha_{\mu,\nu}^{(m,n)} \), where \( \mu,\nu \) = \( r,r \); \( l,l \); \( d,d \); or \( u,u \) depending on the event combination. This type of interference is unique to fields with multiple ionization events per half-cycle, such as bichromatic fields. It is represented in Fig.~\ref{fig:interfconditions}(a) by combining stripes of the same color but different shades.

One can compute temporal-only phase differences using $S^{(m)}_r(\mathbf{p}_1,\mathbf{p}_2,t^{(m)},t',t'')=S^{(m)}_l(\mathbf{p}_1,\mathbf{p}_2,t^{(m)}+\Delta t^{(m)},t'+\Delta t',t''+\Delta t'')$. This gives 
\begin{align}
         \alpha^{(m,n)}_{r,l}&=S^{(m)}_r-S^{(n)}_l\nonumber\\
         &=\alpha^{(A^2)}_{\Delta \tau}(t',t'')+ \frac{1}{2}\alpha^{(\mathrm{pond})}_{\Delta \tau}(t'',t^{(m)},t^{(n)})\nonumber\\
         &+\alpha^{(\mathrm{ene})}_{\Delta \tau}+\alpha^{(\mathbf{p}_1,\mathbf{p}_2)}_{\Delta \tau}(t',t^{(m)},t^{(n)}),\nonumber\\
     \label{eq:alpharightleft}
\end{align}
where 
\begin{align}
    \alpha^{(A^2)}_{\Delta\tau}(t',t'')= &\frac{\left[\mathbf{F}_A(t'+\Delta t')-\mathbf{F}_A(t''+\Delta t'')\right]^2}{2(t'-t''+\Delta t'-\Delta t'')}\nonumber \\ &-\frac{\left[\mathbf{F}_A(t')-\mathbf{F}_A(t'')\right]^2}{2(t'-t'')},
    \label{eq:alphaA2}
\end{align}
\begin{align}
    \alpha^{(\mathrm{pond})}_{\Delta \tau}(t'', t^{(m)}, t^{(n)})=&F_{A^2}(t^{(m)}+\Delta t^{(m)})\nonumber\\ &+F_{A^2}(t''+\Delta t'')\nonumber\\ &-F_{A^2}(t^{(n)})-F_{A^2}(t''),
     \label{eq:alphapond}
\end{align}
\begin{eqnarray}
\alpha^{(\mathrm{ene})}_{\Delta\tau}&=&\text{E}_{1 g}\Delta t''+\text{E}_{2 g}\Delta t'\notag\\&&+\text{E}_{2 e}(\Delta t^{(m)}-\Delta t' + t^{(m)} - t^{(n)})\notag\\&&+\frac{\mathbf{p}_{1 }^2\Delta t'}{2}+\frac{\mathbf{p}_{2 }^2(\Delta t^{(m)} + t^{(m)} - t^{(n)})}{2}
 \label{eq:alphaene}
\end{eqnarray}
and 
 \begin{align}
\alpha^{(\mathbf{p}_1,\mathbf{p}_2)}_{\Delta \tau}(t,t')=&\mathbf{p}_1\cdot\left[\mathbf{F}_A(t'+\Delta t') - \mathbf{F}_A(t') \right] \nonumber \\&+ \mathbf{p}_2 \cdot \left[\mathbf{F}_A(t+\Delta t) - \mathbf{F}_A(t) \right],
   \label{eq:alphap1p2tau}  
  \end{align}
with 
\begin{equation}
    F_{A^2}(t)=\int^t \mathbf{A}^2(\tau)d\tau.
    \label{eq:FA2}
\end{equation}
These expressions are more general than their counterparts in \cite{Hashim2024}.  
For $m=n$, and $\Delta t' = \Delta t''=\Delta t^{(m)}$, the expressions in \cite{Hashim2024} are recovered. 

Equation~\eqref{eq:alphaene} is linear in the time differences $\Delta t^{(m)}$, $\Delta t'$, and $\Delta t''$,  and contains the bound state and kinetic energies, i.e., it is target-dependent. 
In contrast, Eqs.~\eqref{eq:alphaA2}, \eqref{eq:alphapond} and \eqref{eq:alphap1p2tau} have temporal arguments $F_{A^2}(\tau)$ and $\mathbf{F}_A(\tau)$ and are thus highly dependent on field symmetry.  The phase difference in Eq.~\eqref{eq:alphapond} gives ponderomotive terms, which will add to those in Eq.~\eqref{eq:alphaene}.

Computing the symmetrized counterpart to Eq.~\eqref{eq:alpharightleft}, $ \alpha^{(n,m)}_{u,d}$, yields identical expressions but with electron momenta swapped in $\alpha^{(\mathbf{p}_2,\mathbf{p}_1)}_{\Delta \tau}(t,t')$.

 For a monochromatic wave, the effect of these temporal-only phases is minimal \cite{Shaaran2010,Maxwell2015}. However, for few-cycle pulses, these terms lead to noticeable interference patterns, particularly when events are detangled \cite{Hashim2024}. For bichromatic fields, multiple relevant ionization events per half-cycle require accounting for the additional phase shifts $\alpha^{(n,m)}_{\mu,\mu}$. Explicitly, these read
\begin{equation}
    \alpha^{(m,n)}_{l,l}=S^{(m)}_l-S^{(n)}_l=\alpha^{\mathbf{p}_2}_{\delta t}(t^{(m)},t^{(n)}),
\label{eq:alphall}
\end{equation}
where 
\begin{eqnarray}
    \alpha^{\mathbf{p}_2}_{\delta t}(t^{(m)},t^{(n)}) &=&\left( E_{2e}+\frac{\mathbf{p}^2_2}{2}\right)(t^{(m)}-t^{(n)})\notag\\&&+\frac{1}{2}\left[ F_{A^2}(t^{(m)})-F_{A^2}(t^{(n)}) \right]\notag \\&& + \mathbf{p}_2 \cdot \left[ \mathbf{F}_{A}(t^{(m)})-\mathbf{F}_{A}(t^{(n)})\right]. 
\label{eq:deltaphase}
\end{eqnarray}
Similarly, 
\begin{equation}
    \alpha^{(m,n)}_{r,r}=\alpha^{\mathbf{p}_2}_{\delta t} (t^{(m)} + \Delta t^{(m)},t^{(n)} + \Delta t^{(n)}),
    \label{eq:alpharr}
\end{equation}
\begin{equation}
    \alpha^{(m,n)}_{d,d}=\alpha^{\mathbf{p}_1}_{\delta t} (t^{(m)},t^{(n)}),
\label{eq:alphadd}
\end{equation}
and 
\begin{equation}
 \alpha^{(m,n)}_{u,u}=\alpha^{\mathbf{p}_1}_{\delta t} (t^{(m)} + \Delta t^{(m)},t^{(n)} + \Delta t^{(n)})
 \label{eq:alphauu}.
\end{equation}

Our previous study \cite{Hashim2024} shows that the above building blocks produce \textit{circularly- and linearly shaped fringes} on the $p_{1\parallel}p_{2\parallel}$ plane, with intricate interplay. By analyzing these building blocks and interference conditions, we determine the patterns formed in bichromatic fields. In addition to the interfering processes identified in \cite{Hashim2024} for few-cycle pulses, time-delayed ionization events within the same half-cycle must also be considered.

For phases involving events in different half cycles ($\xi \neq \eta$), the total temporal shifts comprise of four composite phase differences as given in \eqref{eq:alpharightleft}. One 
building block is $\alpha^{(\mathrm{ene})}_{\Delta\tau}$ [Eq.~\eqref{eq:alphaene}], which, for fields with $\Delta t'' = \Delta t' = \Delta t$, describes a hypersphere. Thus, in the parallel momentum plane, this term is expected to produce circular fringes with radii $\sqrt{4n\pi/\Delta t - p_{1\perp}^2 - p_{2\perp}^2}$. 
For few-cycle pulses, circular fringes appeared only for large $\Delta\tau$, corresponding to intercycle events \cite{Hashim2024}. In the present study, where only intracycle events are considered, such circular substructures are unlikely to be present. Furthermore, for $\Delta t'' \neq \Delta t' \neq \Delta t$, the shape induced by this phase shift becomes difficult to determine.
Another building block is $\alpha_{\Delta\tau}^{(\mathbf{p_1}, \mathbf{p_2})}$, given by Eq.~\eqref{eq:alphap1p2tau}. For linearly polarized fields, imposing the condition for constructive interference leads to straight lines, with gradients and intercepts strongly dependent on the field symmetry. Interference of events with the same rescattering time, $P_{i\xi}O_{m\xi}$ and $P_{i\xi}O_{n\xi}$, will result in lines parallel to the $p_{2\parallel}$ axes only. 

The building block $\alpha^{(A^2)}_{\Delta\tau}(t',t'')$ (Eq.~\eqref{eq:alphaA2}) depends solely on the first electron's ionization times. It vanishes for fields with half-cycle symmetry, as, in this case, $\Delta t' = \Delta t''$ and the two terms in Eq.~\eqref{eq:alphaA2} cancel out. For fields without half-cycle symmetry, this phase is shaped by $F_A(\tau)$
and induces small shifts, potentially forming linear alternating fringes whose spacing and magnitude depend on field symmetry and event localization \cite{Hashim2024}. The role of the field shape on these phase differences is examined in greater detail in the Appendix. 
Lastly, the ponderomotive building block $\alpha_{\Delta\tau}^{(\text{pond})}$  [Eq.~\eqref{eq:alphapond}] involves the integral of the square of the vector potential, 
a monotonic smooth step function passing through the origin. It causes numerical shifts which do not directly affect interference patterns, and is therefore not discussed further.

The total temporal-shift interference results from the combined effects of all phase differences discussed, making it difficult to predict their hierarchy or their interplay. For few-cycle pulses, observed features included a ``box” shape near the origin due to the absence of fringes from $\alpha_{\Delta\tau}^{(\mathbf{p}_1, \mathbf{p}_2)}$,  fringes with varying gradients causing `V'-shaped structures that combine to form criss-crossed patterns, wing-like structures along the anti-diagonal, straight-line fringes from $\alpha^{(A^2)}$ leading to checkerboard patterns upon electron symmetrization, and circular substructures from  $\alpha^{(\text{ene})}$.  However, many of these patterns are unlikely to appear in this study due to (a) fewer combined events, (b) strictly intracycle events with smaller $\Delta \tau$, reducing the influence of $\alpha^{(\mathrm{ene})}$, and 
(c) more symmetries in bichromatic fields compared to few-cycle pulses.

Interference between two distinct events within a single half-cycle gives rise to new patterns, characterized by temporal phase shifts $\alpha_{\mu,\mu}^{(m,n)}$ where $\mu=l,r,u,d$ are given by Eqs.~\eqref{eq:alphall} and \eqref{eq:alphauu}. The underlying building block [Eq.~\eqref{eq:deltaphase}] is a special case of those previously discussed. 
The first term produces fringes parallel to the axes according to 
\begin{equation}
    p_{2\parallel} = \sqrt{\frac{4n\pi-2E_{2e}}{t^{(m)}-t^{(n)}} - p_{2\perp}^2},
    \label{eq:temporalspecial}
\end{equation}
which is maximized when perpendicular momentum vanishes, the events are well-separated within the half-cycle, or 
 $n$ is large. The second term introduces only numerical shifts. The last term causes axes-parallel fringes at $2n\pi/[F_A(t^{(m)})-F_A(t^{(n)})]$, with interference along the axes for $n=0$. 

\subsubsection{Combined  exchange-temporal interference}
\label{sec:interfcombined}

Finally, combined exchange-temporal interference is computed by exchanging $\mathbf{p}_1$ and $\mathbf{p}_2$ and displacing one of the interfering events by $\Delta \tau$. 
As with pure temporal shifts, three scenarios arise. The first,  with $m=n, \xi \neq \eta$, is represented by combining stripes of different colors but the same shade in Fig.~\ref{fig:interfconditions}(a). The second, with $m \neq n, \xi \neq \eta$, corresponds to stripes of different colors and shades. Both cases lead to phase differences in the right-down and left-up regions given by $\alpha_{\mu, \nu}^{(m, n)}$ where $m$ may equal $n$ and $\mu,\nu= r, d$ or $u, l$. 
The right-down phase reads
\begin{align}
     \alpha^{(m,n)}_{r, d}&=S^{(m)}_r-S^{(n)}_d\nonumber \\
     &=\alpha^{(A^2)}_{\Delta \tau}(t',t'')+ \frac{1}{2}\alpha^{(\mathrm{pond})}_{\Delta \tau}(t'',t^{(m)},t^{(n)})\nonumber \\&+\alpha^{(\mathrm{ene})}_{\Delta \tau}+\alpha^{(\mathbf{p}_2 \leftrightarrow \mathbf{p}_1)}_{\Delta \tau}(t',t^{(m)},t^{(n)})+\alpha_{\mathbf{p}_1,\mathbf{p}_2}^{(\mathrm{exch})} (t^{(n)}, t'), 
     \label{eq:alphard}
\end{align}
where 
 \begin{align}
\alpha^{\mathbf{p}_2\leftrightarrow \mathbf{p}_1}_{\Delta \tau}(t^{(n)},t')=&\mathbf{p}_1\cdot\left[\mathbf{F}_A(t'+\Delta t') - \mathbf{F}_A(t^{(n)}) \right] \nonumber \\&+ \mathbf{p}_2 \cdot \left[\mathbf{F}_A(t^{(m)}+\Delta t^{(m)}) - \mathbf{F}_A(t') \right].
   \label{eq:alphaexctau}  
  \end{align}
Equation~\eqref{eq:alphaexctau} contains a momentum dependence, and the double arrow in the superscript indicates that the momenta in the second row are exchanged about the temporal arguments  $t$ and $t'$ of $\mathbf{F}_A$. We have verified that $|\alpha_{d, r}^{(m,n)}| = -|\alpha_{r, d}^{(n,m)}|$.

The left-up phase is given by
\begin{align}
     \alpha^{(m,n)}_{l, u}&=S^{(m)}_l-S^{(n)}_u\nonumber \\
     &=-\alpha^{(A^2)}_{\Delta \tau}(t',t'')- \frac{1}{2}\alpha^{(\mathrm{pond})}_{\Delta \tau}(t'',t^{(m)},t^{(n)})\nonumber \\&-\alpha^{(\mathrm{ene})}_{\Delta \tau}-\alpha^{(\mathbf{p}_1 \leftrightarrow \mathbf{p}_2)}_{\Delta \tau}(t',t^{(m)},t^{(n)})+\alpha_{\mathbf{p}_1,\mathbf{p}_2}^{(\mathrm{exch})} (t^{(m)}, t'), 
     \label{eq:alphalu}
\end{align}
where all phases except the exchange term switch signs, and the momenta in each term swap. For bichromatic fields $|\alpha_{r, d}^{(m,n)}| = |\alpha_{l, u}^{(m,n)}|$ still holds. This implies a symmetry about the diagonal when interference from both phase differences is included.  

The third scenario involves events with $m \neq n$ but $\xi = \eta$, represented by stripes of the same color but different shades in Fig.~\ref{fig:interfconditions}(a). These phase differences read
\begin{align}
     \alpha^{(m,n)}_{l, d}&=S^{(m)}_l-S^{(n)}_d\nonumber \\
     &=\alpha^{(A^2)}_{\Delta \tau=0}(t',t'')+ \frac{1}{2}\alpha^{(\mathrm{pond})}_{\Delta \tau=0}(t'',t^{(m)},t^{(n)})\nonumber \\&+\alpha^{(\mathrm{ene})}_{\Delta \tau=0}+\alpha^{(\mathbf{p}_2 \leftrightarrow \mathbf{p}_1)}_{\Delta \tau=0}(t',t^{(m)},t^{(n)})+\alpha_{\mathbf{p}_1,\mathbf{p}_2}^{(\mathrm{exch})} (t^{(n)}, t'). 
     \label{eq:alphaldcombined}
\end{align}
Equation~\eqref{eq:alphaldcombined} is merely a simplification of Eq.~\eqref{eq:alphard}. The exchange phase term remains unchanged as it does not depend on $\Delta \tau$. Meanwhile, the energy, ponderomotive, and $A^2$ phase terms simplify to the special case associated with $\Delta \tau = 0$.  
This causes the $A^2$ term to vanish. 
Notably, unlike exchange-only phases, $\alpha^{(m,n)}_{l, d} \neq \alpha^{(m,n)}_{d, l}$, with signs flipping for some phases and swapping of $\mathbf{p}_1$ and $\mathbf{p}_2$. Explicitly,
\begin{align}
     \alpha^{(m,n)}_{d, l}&=S^{(m)}_d-S^{(n)}_l\nonumber \\
     &=\alpha^{(A^2)}_{\Delta \tau=0}(t',t'')+ \frac{1}{2}\alpha^{(\mathrm{pond})}_{\Delta \tau=0}(t'',t^{(m)},t^{(n)})\nonumber \\&+\alpha^{(\mathrm{ene})}_{\Delta \tau=0}+\alpha^{(\mathbf{p}_1 \leftrightarrow \mathbf{p}_2)}_{\Delta \tau=0}(t',t^{(m)},t^{(n)})-\alpha_{\mathbf{p}_1,\mathbf{p}_2}^{(\mathrm{exch})} (t^{(n)}, t'). 
     \label{eq:alphadlcombined}
\end{align}
The phase shift $\alpha_{r, u}^{(m,n)}$ is calculated similarly, also resulting in $\alpha^{(A^2)}=0$ and simplified remaining terms.

To predict the interference from the combined temporal-exchange shifts $\alpha_{\mu,\nu}^{(m,n)}$ ($\mu,\nu = l,u; r,d$), one must once more investigate their building blocks. They are five in total, as outlined in Eq.~\eqref{eq:alphalu}, 
and behave 
as previously discussed. 
Constructive interference for 
the combined temporal-exchange phase difference, $\alpha_{\Delta\tau}^{(\mathbf{p_2}\leftrightarrow \mathbf{p_1})}$[Eq.~\eqref{eq:alphaexctau}] leads to linear fringes described by
\begin{equation}
    p_{2\parallel} = \frac{-p_{1\parallel}[F_A(t^{(n)}+\Delta\tau)-F_A(t')] + 2n\pi}{F_A(t'+\Delta\tau)-F_A(t^{(m)})},
     \label{eq:combinedfringes}
\end{equation}
which are highly dependent on the field shape and symmetry. Similar to temporal-only patterns, these lines are expected to be skewed by other building blocks. However, one can predict their individual behavior by inspecting the vector potential integrals [Eq.~\eqref{eq:IntegralA}] for specific fields and phases. 

\subsection{Interference mapping}
\label{sec:interfmapping}

\begin{table*}[!htbp]
\vspace*{-0.4cm}
\includegraphics[height=1.1\paperwidth, keepaspectratio, trim=0.5cm 6.5cm 0cm 0cm, clip]{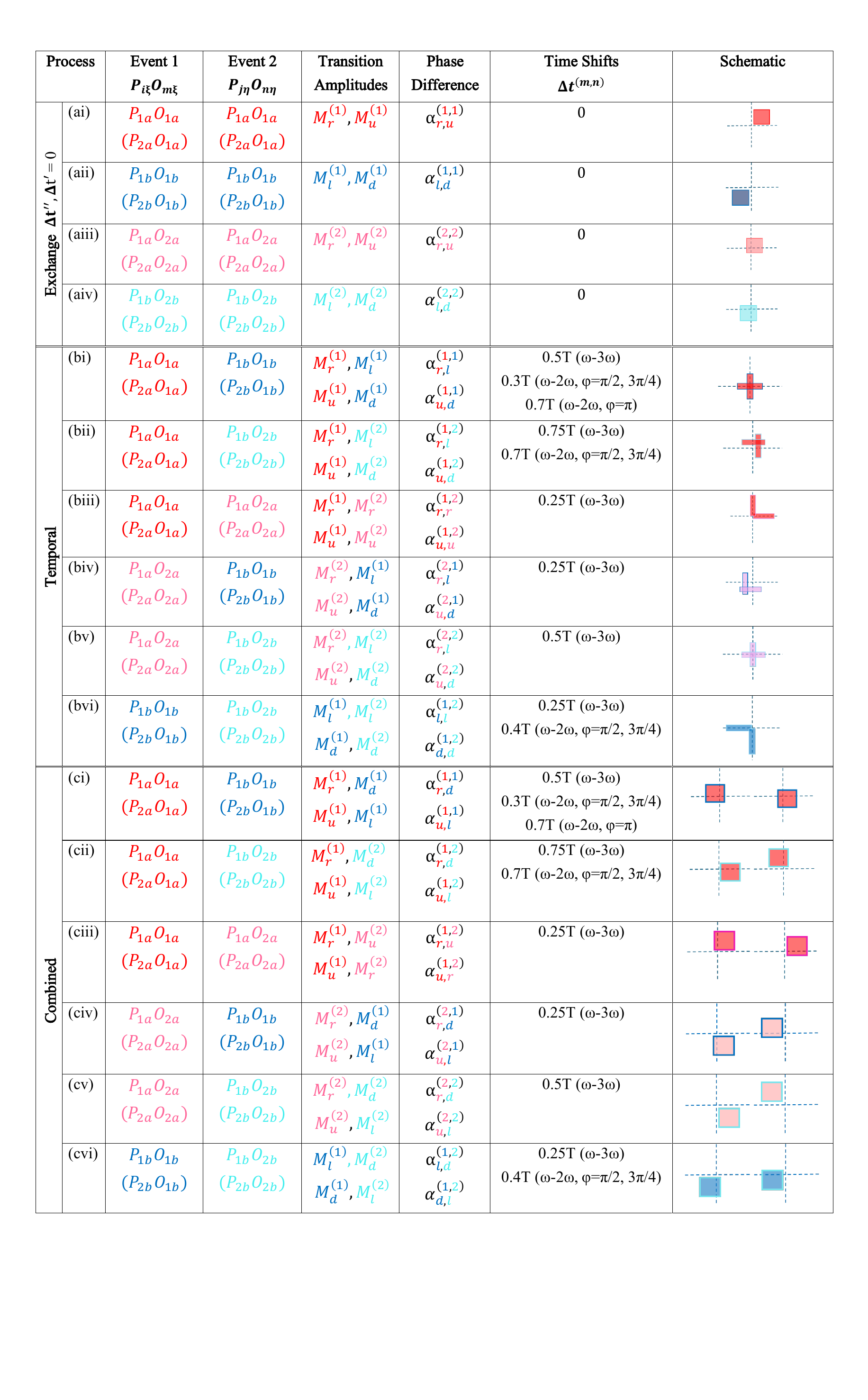}
\vspace*{-0.4cm}
   \caption{All possible exchange, temporal and combined phase differences for bichromatic fields studied in this work.
   From left to right, the columns give the process, the constituent interfering pair of events, the transition amplitudes, relative temporal displacement between events, and a schematic of where such interference is expected to be significant in the parallel momentum plane. We have used the same colors as in Fig.~\ref{fig:fields} to refer to specific events. Momentum regions with a single (two) colors indicate the interference of the same (different) event(s).}
   \label{fig:phasetab}
\end{table*}

Here, we map the different types of interference to their corresponding momentum regions using the fields and diagrams in Fig.~\ref{fig:fields}. A summary of this mapping is provided in Table \ref{fig:phasetab}, which gives the events involved, the quantum phase differences and the regions of the $p_{1\parallel}p_{2\parallel}$ plane expected to be occupied. Besides Fig.~\ref{fig:fields}, the generic interference schematics in Fig.~\ref{fig:interfconditions}, together with the field illustrations, are good references to understand the table.
In all cases, except for $(\omega,3\omega,\phi=-\pi/2)$ fields, the contributions of pairs $P_{1a,b}$ dominate the first electron dynamics. In that specific case, one should consider $P_{2a,b}$ instead [see schematic in Fig.~\ref{fig:fields}(c')], as indicated by the events in parentheses in Table \ref{fig:phasetab}. 

Exchange-only shifts and their contributions are represented in the four upper rows of Table \ref{fig:phasetab}, and are called processes (a). The momentum regions for which interference is significant differ according to the ionization pathway of the second electron.  
For events $P_{ia,b}O_{1a,b}$ [processes (ai) and (aii)], the dominant orbit pairs for the first electron are matched with the first ionization event after the field zero crossing for the second electron. Interference occurs predominantly in the first and third quadrants of the $p_{1\parallel}p_{2\parallel}$ plane and is associated with phase shifts $\alpha^{(1,1)}_{r,u}$ and $\alpha^{(1,1)}_{l,d}$, respectively.  
The relevant momentum regions are highlighted as the red and blue shaded rectangles in the last column of Table \ref{fig:phasetab}, respectively.
The other interfering events, denoted by  $P_{i\xi}O_{2\xi}$  [processes (aiii) and (aiv), whose phase shifts are $\alpha^{(2,2)}_{r,u}$ and $\alpha^{(2,2)}_{l,d}$, respectively], comprise the dominant pair for the first electrons and the second ionization events after rescattering for the second electron. In this case, exchange interference occurs near the origin and may extend into the second and fourth quadrants. 
The pink and cyan rectangles in the last column of Table \ref{fig:phasetab} indicate the primary momentum region for this type of interference.  

The interference processes arising from the coherent sum of events separated in time by $\Delta\tau$ (temporal-only interference) are outlined in the next six rows of Table \ref{fig:phasetab} [processes (b)].  
Process (bi) represents the interference of the events $P_{ia}O_{1a}$ and $P_{ib}O_{1b}$, where the rescattering events for the first electron are separated by half a cycle, combined with the orbits $O_{1a}$ of the second electron consecutive to rescattering. These events are associated with the phase differences $\alpha^{(1,1)}_{r,l}$ and $\alpha^{(1,1)}_{u,d}$. A similar type of interference is that of $P_{ia}O_{2a}$ and $P_{ib}O_{2b}$ [process (bv)], for which, instead, we consider the second orbits after rescattering for the second electron. The corresponding phase differences are $\alpha^{(2,2)}_{r,l}$ and $\alpha^{(2,2)}_{u,d}$. The patterns arising from this interference occur near the axes $p_{n\parallel}=0$. They are strongest near the origin, but there are also fainter secondary patterns at slightly larger momenta.

Processes (bii) and (biv) account for rescattering pairs occurring in different half cycles combined with different orbits for the second electron, i.e., $P_{ia,b}O_{2a,b}$ and $P_{ib,a}O_{1b,a}$. The corresponding phase differences are $\alpha^{(m,n)}_{rl}$ and $\alpha^{(m,n)}_{ud}$, with $m \neq n$ and $m, n=1,2$. In this case, interference is also appreciable close to $p_{n\parallel}=0$. However, because of the slightly different most probable parallel momenta for each event, the regions for which the patterns are the strongest are slightly mismatched. Finally, the interference type labelled processes (biii) and (bvi) involves a single rescattering event for the first electron and two consecutive ionization events for the second electron, for example, $P_{ia,b}O_{1a,b}$ and $P_{ia,b}O_{2a,b}$. These processes are associated with the phase shifts $\alpha^{(1,2)}_{m,m}$, and are unique to fields with more than one prominent ionization event per half cycle. They lead to notable patterns on the positive [(biii)] and negative [(bvi)] half axes.

The interference processes involving electron symmetrization \textit{and} temporal shifts are outlined in the bottom six rows of Table \ref{fig:phasetab} [processes (c)]. 
First, we consider the interference of the amplitude associated with $P_{ia,b}O_{m a,b}$ and the transpose of the amplitude related to $P_{ib,a}O_{m b,a}$, \textit{both} taking the first or second ionization event for the second electron after rescattering [processes (ci) and (cv), with $m=1,2$, respectively]. These processes are associated with phase differences $\alpha^{(m, m)}_{r,d}$ and $\alpha^{(m, m)}_{u,l}$, and exhibit substantial interference mostly in the second and fourth quadrants of the $p_{1\parallel}p_{2\parallel}$ plane. Furthermore,  the amplitudes related to $P_{i a,b}O_{m a,b}$ and the transpose of the contributions from $P_{ib,a}O_{n b,a}$, with $n \neq m$ may also interfere. This interference is signposted as processes (cii) and (civ), associated with $\alpha^{(m, n)}_{r,d}$ and $\alpha^{(m, n)}_{u,l}$, $n \neq m$. The resulting patterns are expected to be visible in the second and fourth quadrants of the parallel momentum plane as well.   
Finally, there are different pathways associated with a single rescattering event for the first electron and two consecutive ionization events for the second electron, which interfere [processes (ciii) and (cvi)]. However, one must take the transpose of one of the transition amplitudes. The interference patterns arise in the first and third quadrants of the parallel momentum plane and are associated with $\alpha^{(1,2)}_{\mu ,\nu}$ and $\alpha^{(1,2)}_{\nu ,\mu}$, with $(\mu,\nu)=(r,u)$ or $(\mu,\nu)=(l,d)$. It should be noted that the phase differences associated with swapped indices $\mu, \nu$ are no longer equal unlike for temporal-only shifts. This is because of the momentum and sign swapping.

Further to the regions in Table \ref{fig:phasetab}, the equations in Sec.~\ref{sec:generalizedinterf} have remarkable predictive power, if analyzed together with the integral of the vector potential. Below, we provide a summary of what to expect. For explanations, and field- and event-specific equations, we refer to the Appendix. 

Exchange-only fringes consist of spine lines and hyperbolic fringes. For both $(\omega, 2\omega)$ and $(\omega, 3\omega)$ fields, there are alternating spine lines along the diagonal. 
The spines are reflection-symmetric regarding the anti-diagonal only for $(\omega, 3\omega)$ fields.  The spine is skewed by hyperbolae along the diagonal at larger momenta, driven closer by events with large $(t-t')$.

Event interference produces a myriad patterns, including prominent wing shapes from $P_{ia}O_{1a} + P_{ib}O_{1b}$ and $P_{ia}O_{2a} + P_{ib}O_{2b}$, criss-cross and chequerboard patterns in the first and third quadrants due to overlapping horizontal and vertical straight-line fringes, and faint circular substructure at larger momenta. Overall, interference is brightest at the origin, with patterns skewed unpredictably by various building blocks. Fringe overlap is more pronounced in the $(\omega, 2\omega)$ field than in $(\omega, 3\omega)$ due to half-cycle symmetry breaking. For both fields, interference from events involving two second-electron solutions in the same half-cycle yields simpler fringes parallel to the axes, as their building block [Eq.\eqref{eq:alphaA2}] involves only one momentum. These fringes may be partially or fully obscured by other interference effects.  Shifts and cuts may occur, altering fringe direction and width. 

Combined temporal-exchange interference produces patterns similar to temporal-only shifts, including faint circular substructures, hyperbolae, and straight-line fringes. If the contributing events spill into different quadrants of the momentum plane, the patterns from the temporal-exchange shifts become more pronounced. Linear fringes dominate, forming fine feathery or fishbone-like patterns whose size and intensity depend on the localization of the contributing events' momentum distribution. Feathery fringes are expected to be clearest when the PMD primarily occupies the second and fourth quadrants, for the  $(\omega, 3\omega,\phi=\pi/2)$ and the $(\omega, 2\omega,\phi=\pi)$ fields. For $(\omega, 3\omega,\phi=0)$, these fringes are also expected to be prominent, as the RESI probability density is appreciable in all quadrants.

\begin{figure}[!htbp]
\centering
\includegraphics[width=1.1\columnwidth]{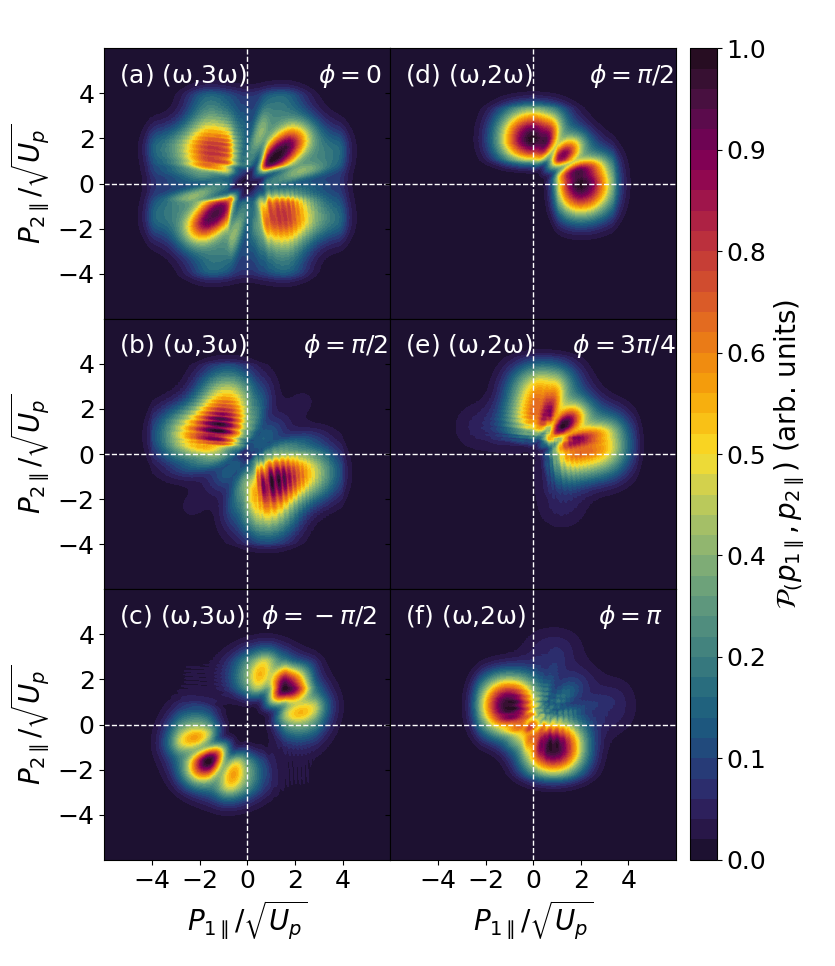}
    \caption{Fully coherent momentum distributions $\mathcal{P}_{(cc)}(p_{1\parallel},p_{2\parallel})$ without prefactors calculated for the ($\omega$,$3\omega$) (left column) and ($\omega$,$2\omega$) (right column) driving fields with the relative phase as indicated in the panels. The intensity of the $\omega$ field component is $E_{\omega}^2=6\times 10^{13}$W/cm$^2$, the ratio of the field amplitudes is $\xi=0.8$, and the relative phase is indicated in the panels. The fundamental wavelength is 800~nm. Other driving-field parameters are the same as in the corresponding panels of Fig.~\ref{fig:fields}. }
    \label{fig:coherent}
\end{figure}

\section{Two-electron momentum distributions}\label{sec:coherent}

Here, we analyze the quantum interference types given above, for the same parameters as in \cite{Hashim2024b}. As a target, we choose argon, for which the first and second ionization potentials are $E^{(\mathcal{C})}_{1g}=0.58$ a.u.,  $(\mathcal{C}=1...6)$\footnote{There are six relevant excitation channels in Argon, which have been used in our previous publications \cite{Maxwell2015,Maxwell2016,Hashim2024}.} and $E^{(1)}_{2g}= E^{(2)}_{2g}= 1.016$ a.u., respectively, where the superscripts indicate the ground state $3s$ or $3p$. To minimize field gradient effects from loosely bound states \cite{Hashim2024}, most results have been calculated for the $3s \rightarrow 3p$ excitation channel (electron configuration $3s3p^6$), with $E^{(1)}_{2e}=0.52$ a.u. This is the deepest excited bound state for this specific target. In Sec.~\ref{sec:prefac}, we also employ the $3p \rightarrow 4s$ $(3p^54s)$ excitation pathway, with $E^{(2)}_{2e}=0.40$ a.u., to assess the influence of bound-state geometry. As shown in \cite{Hashim2024b}, the prefactor for $d$ excited states is of relatively little importance; hence, they are not investigated here.
 
Figure~\ref{fig:coherent} displays the fully coherent RESI distributions $\mathcal{P}(p_{1\parallel},p_{2\parallel})$ without prefactors for the two investigated fields.  The momentum regions occupied by the distributions follow the mappings in Fig.~\ref{fig:fields}, where the associated events are indicated. The fourfold symmetry of the distribution obtained with the $(\omega$,$3\omega, \phi=0)$ field is broken compared to the fully incoherent sum in \cite{Hashim2024b}. All distributions are reflection-symmetric about the diagonal, but only that in Fig.~\ref{fig:coherent}(c) appears to show reflection symmetry regarding the anti-diagonal. For $(\omega,2\omega)$ fields, this symmetry is broken by the lack of half-cycle symmetry, while, for $(\omega,3\omega)$ fields [Figs.~\ref{fig:coherent}(a) and (b)], it is destroyed by feather-like fringes in the second and fourth quadrants. The fringes in the first and third quadrant are thicker and, for  $(\omega,3\omega)$ fields, symmetric about $(p_{1\parallel},p_{2\parallel}) \leftrightarrow (-p_{1\parallel},-p_{2\parallel})$.
Contributions from each interference type will now be detangled.

\subsection{Exchange Interference}
\label{sec:exchangediscussion} 

\begin{figure}[!htbp]
\centering
\includegraphics[width=1.1\columnwidth]{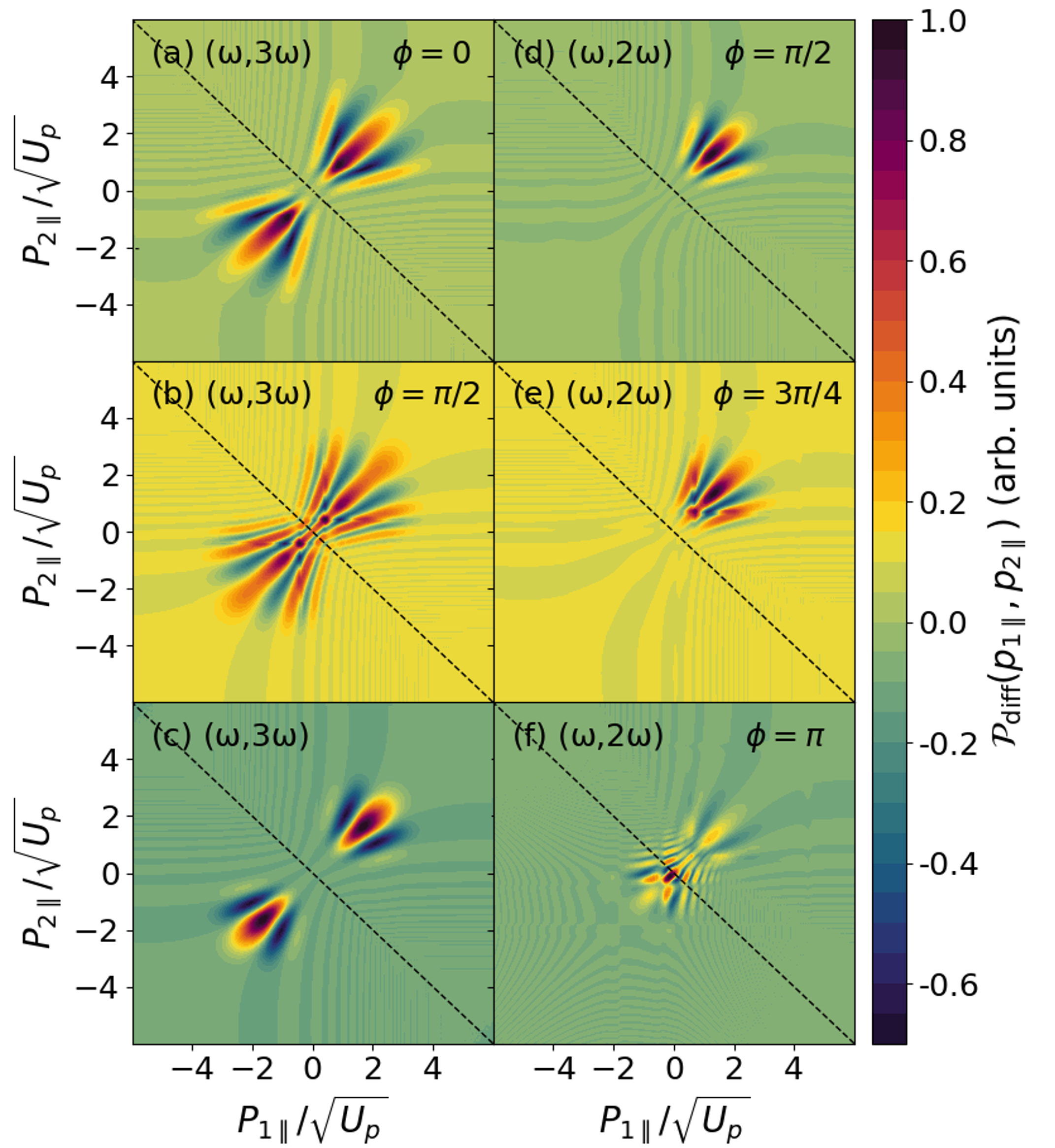}
    \caption{Difference between the RESI distributions $\mathcal{P}_{(ci)}(p_{1\parallel},p_{2\parallel})$ symmetrized coherently but with events added incoherently, and the fully incoherent probability densities   $\mathcal{P}_{(ii)}(p_{1\parallel},p_{2\parallel})$, calculated without any prefactors for the fields and phases employed in the corresponding panels of Fig.~\ref{fig:fields}. The field parameters are given in Fig.~\ref{fig:coherent}. The signal in each panel has been normalized with regard to its maximum. The anti-diagonals $p_{1\parallel}=-p_{2\parallel}$ are indicated with dashed lines. }
\label{fig:exchange}
\end{figure}

Figure~\ref{fig:exchange} isolates exchange interference, schematically represented by processes (a) in Table~\ref{fig:phasetab}, confirming key predictions from Sec.~\ref{sec:expinterf}. First, all distributions remain reflection-symmetric about the diagonal, featuring a central spine and alternating parallel spine lines (given by $n=0$  and $n\neq 0$ in Eq.~\eqref{eq:spine}, respectively). Second, hyperbolic fringes from $\alpha^{(\mathrm{exch})}_{\mathbf{p}_1,\mathbf{p}_2}$ extend outward from the diagonal. Third, these interference patterns are primarily confined to the first and third quadrants. Figures~\ref{fig:exchange}(a)-(c) exhibit reflection symmetry regarding the anti-diagonal, making the first and third quadrants mirror images. This happens because,  for the $(\omega,3\omega)$ field, contributions from pairs with $p_{1\parallel}>0$ are identical to those with $p_{1\parallel}<0$. However, this symmetry breaks in the $(\omega,2\omega)$ field [Figs.~\ref{fig:exchange}(d)-(f)] since events mapping to $p_{1\parallel}>0$ and $p_{1\parallel}<0$ differ. In Fig.~\ref{fig:exchange}(b), the spines are more convoluted, and in Fig.~\ref{fig:exchange}(f) it is even suppressed, suggesting that it may compete with the hyperbolic fringes. Next, we perform an in-depth analysis of these structures for selected field parameters.

\begin{figure}[!htbp]
\centering
\includegraphics[width=1.05\columnwidth]{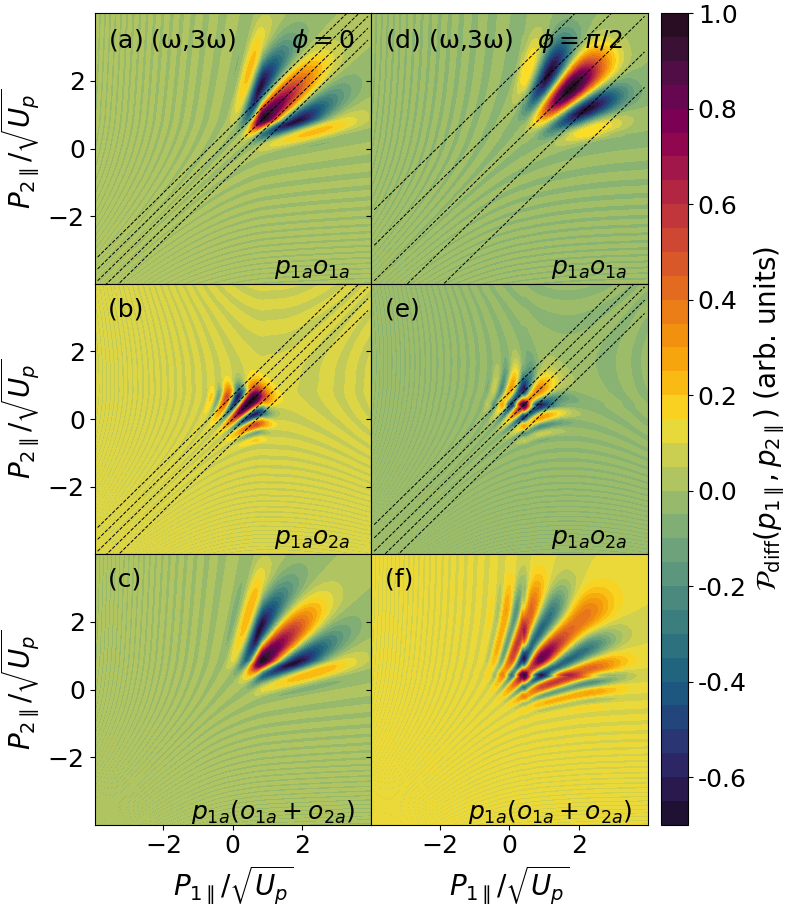}
    \caption{ Difference between RESI distribution $\mathcal{P}_{(\mathrm{ci})}(p_{1\parallel},p_{2\parallel})$, in which symmetrization is performed coherently, and fully incoherent RESI probability densities $\mathcal{P}_{(\mathrm{ii})}(p_{1\parallel},p_{2\parallel})$, calculated without prefactor for individual events $P_{1a}O_{1a}$ and $P_{1a}O_{2a}$ (first and second rows), and for $P_{1a}O_{1a}+P_{1a}O_{2a}$ (third row), using $(\omega,3\omega, \phi=0)$ and $(\omega,3\omega, \phi=\pi/2)$ fields (left and right columns, respectively). These symmetrized coherent sums are related to the phase differences $\alpha_{r,u}^{(1,1)}$, $\alpha_{r,u}^{(1,2)}$, $\alpha_{r,u}^{(1,1)}+\alpha_{r,u}^{(1,2)}$, respectively. 
   Other driving-field parameters are the same as in the corresponding panel of Fig.~\ref{fig:fields} and  Fig.~\ref{fig:coherent}. The contributing events are indicated in the lower-right corner of each panel. The dashed lines in the figure indicate the predicted analytical spine fringes with $n=-2$ to $2$. The signal in each panel has been normalized to its maximum value.}
    \label{fig:exchange0}
\end{figure}

Figure~\ref{fig:exchange0} focuses on exchange interference cases with clear fringes around the diagonal, occupying the first quadrant of the $p_{1\parallel}p_{2\parallel}$ plane. We consider the $(\omega, 3\omega)$ field, for which the contributing events considered are $P_{1a}O_{1a}$ and $P_{1a}O_{2a}$. Interference patterns involving $O_{2\xi}$ ($\xi=a,b)$  are 
distributed close to the origin, spilling into other quadrants, whilst those involving $O_{1\xi}$ remain contained.  Due to half-cycle symmetry, the patterns in the third quadrant, associated with events $P_{1b}O_{1b}$, $P_{1b}O_{2b}$, are its mirror images, rendering their discussion redundant [see Figs.~\ref{fig:coherent} and \ref{fig:exchange}].  Approximate expressions for the spine lines are calculated with Eq.~\eqref{eq:spine} for values of $n$ between $\pm2$ and indicated by dashed lines in the figure. These expressions are more accurate for the central fringe (associated with $n=0$) and near the origin for all fields and phases \cite{Maxwell2015}. Away from the origin, the spine lines are skewed by the hyperbolae arising from the field-independent exchange phase. The hyperbolic fringes have an asymptote along the central diagonal and are associated with different values of $n$. As predicted, hyperbolae are not visible for all events. With all fields and phases, they can only be observed for $P_{1\xi}O_{2\xi}$, for which the quantity $(t-t')$ is largest.

For $(\omega, 3\omega,\phi=0)$ [Figs.~\ref{fig:exchange0}(a)-(c)], the amplitudes of the electric field associated with $P_{1a}O_{1a}$, $P_{1a}O_{2a}$ are identical [Fig.~\ref{fig:fields}(a)]. Thus, the predicted intercepts of the spine lines are equal for both events. Spine lines are visible for the interference of $P_{1a}O_{1a}$ [Fig.~\ref{fig:exchange0}(a)], with hyperbolae also visible for $P_{1a}O_{2a}$ [Fig.~\ref{fig:exchange0}(b)]. Combining both events [Fig.~\ref{fig:exchange0}(c)] shows that $P_{1a}O_{1a}$ prevails. Whilst the predicted spine lines hold well for $P_{1a}O_{2a}$, the spacing for the $P_{1a}O_{1a}$ event is wider than expected. This is because the main hyperbola for this event is centered at larger positive momenta (out of frame in the first quadrant), which skews the spine lines significantly. Near the origin, we verify that this spacing still holds. In this scenario, the exchange building blocks compete very strongly. Whilst the spine line equation provides a useful guideline for predicting the interference, it is crucial to consider the hyperbolae to fully understand the pattern. In contrast, since the $P_{1a}O_{2a}$ event [Fig.~\ref{fig:exchange0}(b)] is concentrated around the origin with the main hyperbola centered at smaller momenta, the skewing effect is more subtle. The interplay of the building blocks is particularly important when interfering events have identical $F_A(t)$ values -- we have verified this effect also occurs for the $(\omega, 2\omega, \phi=\pi/2)$ field - see Appendix.

For ($\omega$,$3\omega$, $\phi=\pi/2$) fields [Figs.~\ref{fig:exchange0}(d)-(f)], the events $O_{2\xi}$ gain dominance compared to $O_{1\xi}$ as shown in Fig.~\ref{fig:fields}(b'). Thus, the spine lines are spaced further apart for $O_{1\xi}$ [see Figs.~\ref{fig:exchange0}(d) and (e) for $P_{1a}O_{1a}$ and $P_{1a}O_{2a}$, respectively]. This diminishes the interplay between the exchange building blocks. Breaking certain field symmetries may make it easier to detangle the effects arising from different phase differences. Furthermore, when events are combined, this leads to finer patterns [Fig.~\ref{fig:exchange0}(f)]. For these field parameters, combining both events results in spine lines essentially obfuscating the hyperbolae.

\begin{figure}[!htbp]
%\centering
\hspace*{-0.2cm}\includegraphics[width=1.1\columnwidth]{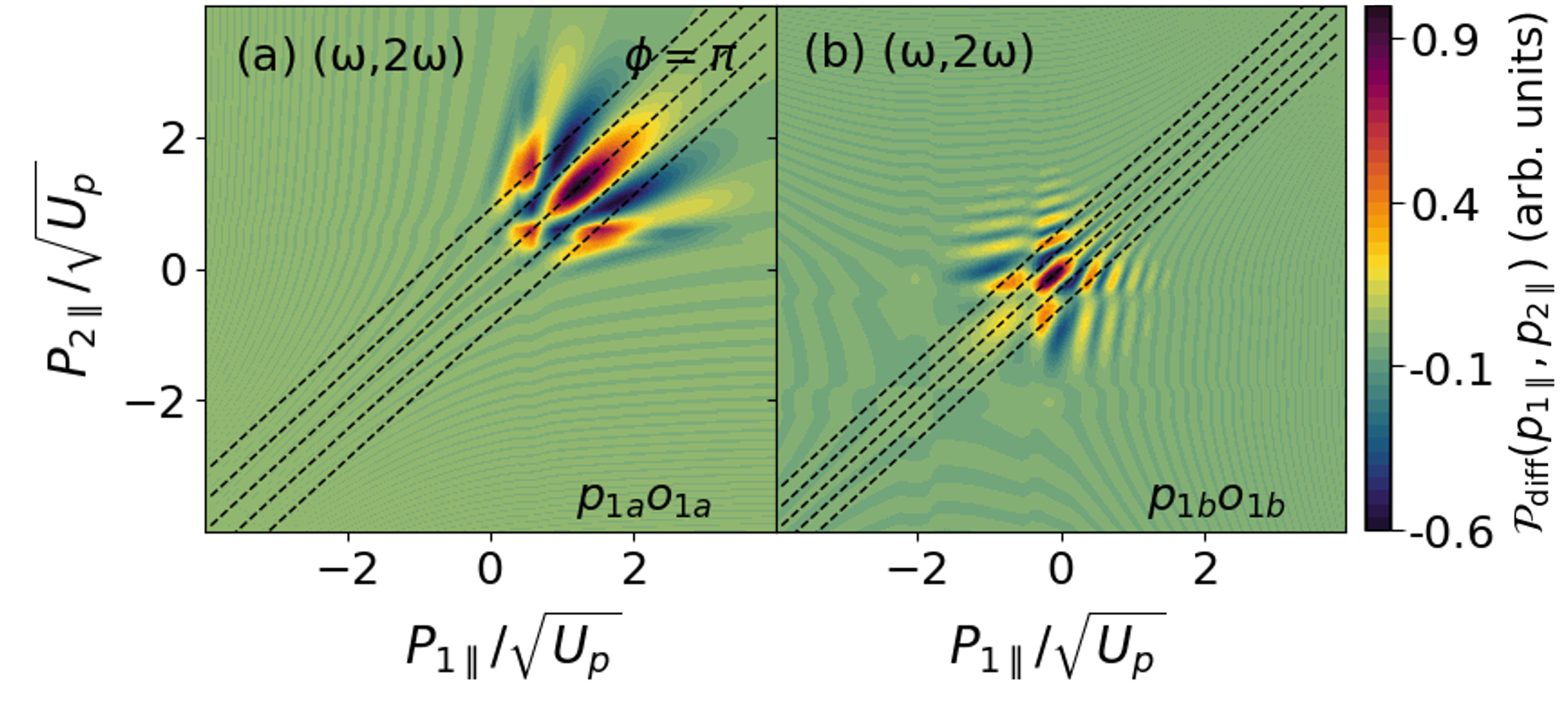}
    \caption{Difference between RESI distribution $\mathcal{P}_{(\mathrm{ci})}(p_{1\parallel},p_{2\parallel})$, in which symmetrization is performed coherently, and fully incoherent RESI probability densities $\mathcal{P}_{(\mathrm{ii})}(p_{1\parallel},p_{2\parallel})$, calculated without prefactor for individual events $P_{1a}O_{1a}$ and $P_{1b}O_{1b}$ [panels (a) and (b), respectively] and a $(\omega$,2$\omega$ $\phi = \pi$) driving field. The coherent sums are related to the $\alpha_{r,u}^{(1,1)}$ (left) and $\alpha_{l,d}^{(1,1)}$ (right) phase differences. Other driving-field parameters are the same as in the corresponding panel of Fig.~\ref{fig:fields} and Fig.~\ref{fig:coherent}. Dashed lines indicate locations of predicted spine lines arising from the pure-exchange phase term with varying $n$. Each panel has been normalized to its maximum probability density. The maximum probability density in panel (a) is around 3.65 times smaller than that in panel (b). }
    \label{fig:exchange180}
\end{figure}

Still, sometimes the hyperbolae may gain in dominance over the spine.
In Fig.~\ref{fig:exchange180}, we analyze the pairwise contributions to Fig.~\ref{fig:exchange}(f) for which the hyperbolae appear to be more prominent than the spine. For the $(\omega$,2$\omega$ $\phi = \pi$) field, there is a single second-electron ionization event per half cycle. The amplitudes of the first electron are much larger for $p_{1\parallel}<0$, i.e. $P_{1b}$ making $\alpha_{l,d}^{(1,1)}$ more significant than $\alpha_{r,u}^{(1,1)}$ (associated with $P_{1a}$) and thereby dominating the total distribution in Fig.~\ref{fig:exchange}(f) [see the mapping in Figs.~\ref{fig:fields}(f) and (f')]. Figure~\ref{fig:exchange180}(a) shows the spine arising from the interference of the symmetrized $P_{1a}O_{1a}$ event, while Fig.~\ref{fig:exchange180}(b) depicts the hyperbolae from $P_{1b}O_{1b}$. The RESI probability density for the dominant event is primarily located near the origin, which increases the relevance of the hyperbolae. A spine would form at higher momenta, for which the probability density is highly suppressed. 

For all fields and phases studied other than $(\omega$,2$\omega, \phi=\pi)$, there are two second-electron events per half-cycle, which are summed coherently before symmetrization. Therefore, the total exchange interference patterns in the first or third quadrant will be a combination of the exchange interference from these individual events.

\subsection{Event Interference}
\label{sec:temporaldiscussion}

\begin{figure}[!htbp]
%\centering
\hspace{0.5cm}\includegraphics[width=1.05\columnwidth]{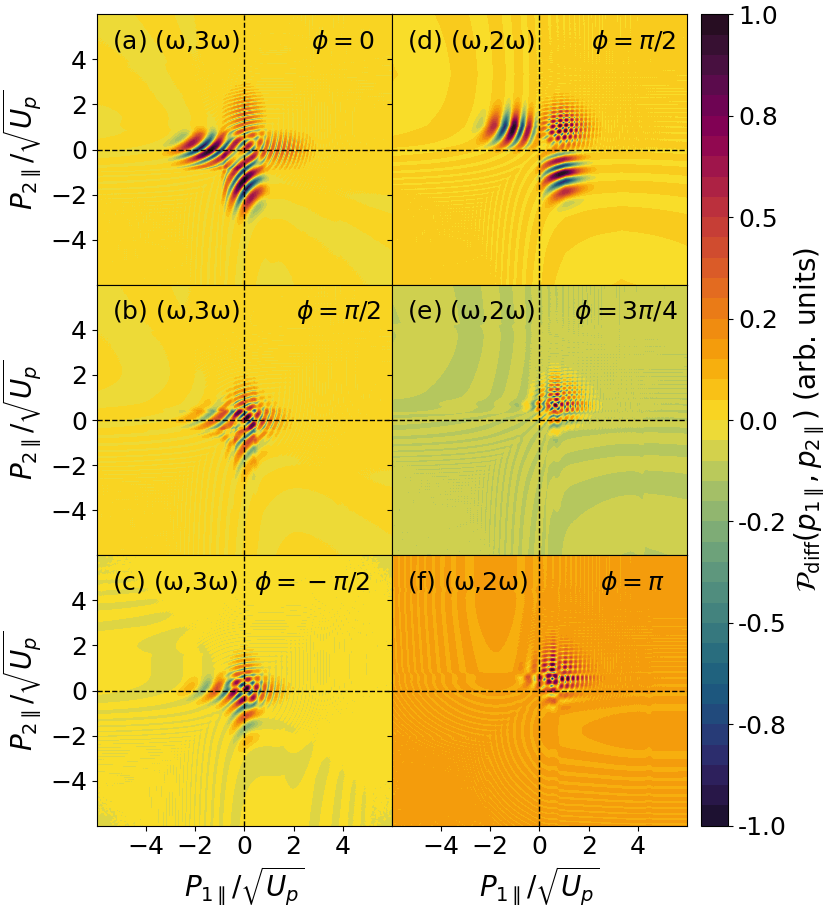}
    \caption{Difference between the RESI distributions $\mathcal{P}_{(ic)}(p_{1\parallel},p_{2\parallel})$ symmetrized incoherently but with events added coherently, and the fully incoherently probability densities  $\mathcal{P}_{(ii)}(p_{1\parallel},p_{2\parallel})$, without any prefactors for the fields and phases employed in the corresponding panels of Fig.~\ref{fig:fields}. The field parameters are given in Fig.~\ref{fig:coherent}. The signal in each panel has been normalized with regard to its maximum. The axes are indicated with dashed lines. }
    \label{fig:temporal}
\end{figure}

\begin{figure}[!htbp]
\centering
\hspace*{-0.2cm}\includegraphics[width=1.1\columnwidth]{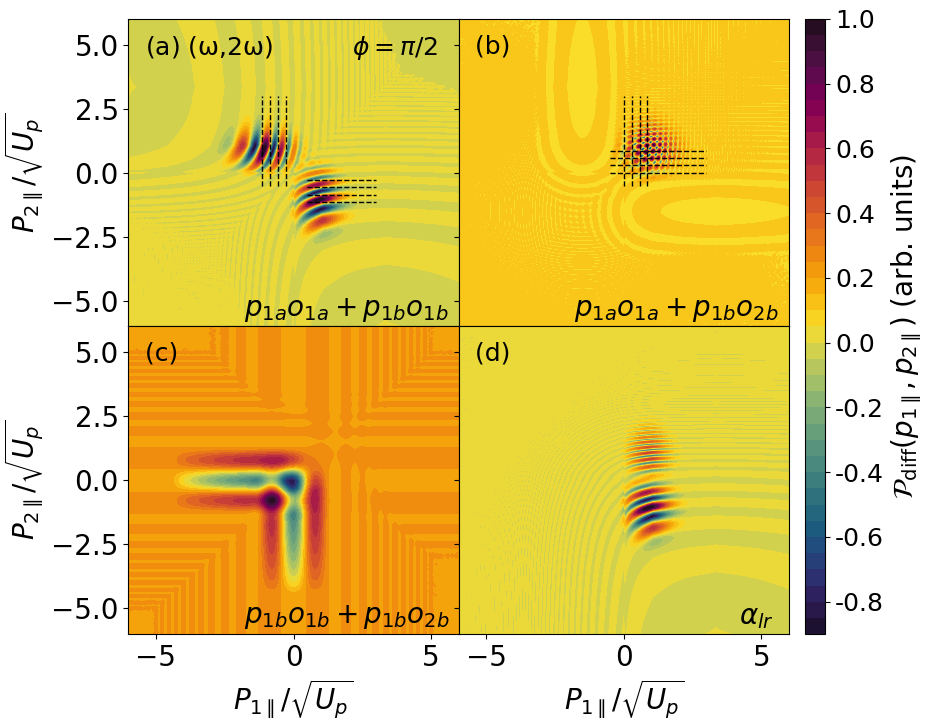}
   \caption{Difference between RESI distributions $\mathcal{P}_{(\mathrm{ic})}(p_{1\parallel},p_{2\parallel})$, in which events are summed coherently, and fully incoherent RESI probability densities $\mathcal{P}_{(\mathrm{ii})}(p_{1\parallel},p_{2\parallel})$, calculated without prefactors for all possible combinations of events with the $(\omega$,2$\omega$ $\phi = \pi/2$) driving field.    
   The coherent sums are related to the  $\alpha_{r,l}^{(1,1)} + \alpha_{u,d}^{(1,1)}$, $\alpha_{r,l}^{(1,2)}+ \alpha_{u,d}^{(1,2)}$ (top row) and $\alpha_{l,l}^{(1,2)} + \alpha_{d,d}^{(1,2)}$, $\alpha_{r,l}^{(1,1)}+\alpha_{r,l}^{(1,2)}$ (bottom row) phase differences. 
   Other driving-field parameters are the same as in the corresponding panel of Fig.~\ref{fig:fields} and Fig.~\ref{fig:coherent}. Dashed lines indicate locations of some predicted linear fringes arising from the temporal building blocks. Each panel has been normalized to its maximum probability density. Panel (c) is two orders of magnitude smaller than panels (a) and (b).}
   \label{fig:temporal90}
\end{figure}

Figure~\ref{fig:temporal} presents the total temporal-only interference across all fields and phases, obtained by subtracting the RESI distributions with events summed coherently and incoherently [Eqs.~\eqref{eq:1ic} and \eqref{eq:1ii}, respectively]. For $(\omega$,3$\omega)$ fields, this interference consists of fringes that are most pronounced along the axes forming a cross shape, in agreement with Table \ref{fig:phasetab} [see processes (b)]. For ($\omega$, $2\omega$) fields, a chequerboard pattern emerges in quadrant one and along the positive half-axes, with wide fringes (`wings') in the second and fourth quadrants.  Notably, the resulting patterns, though always reflection-symmetric about the diagonal, do not necessarily mirror the field symmetry. In particular, for the $(\omega$,3$\omega)$ field the reflection symmetry regarding the anti-diagonal and the $(p_{1\parallel},p_{2\parallel}) \leftrightarrow (-p_{1\parallel},-p_{2\parallel})$ symmetry are broken. 

Detangling the interference of time-delayed events sheds light on a few key questions: what are the patterns emerging from the four composite temporal-shift phase differences? To what extent does the dominance of events affect pure temporal interference? How does the interference arising from second electron events within the same half-cycle manifest? 

Figs.~\ref{fig:temporal90}(a)-(c)
show the event-wise breakdown for $(\omega$,2$\omega$, $\phi=\pi/2$) [Fig.~\ref{fig:temporal}(d)], while Fig.~\ref{fig:temporal90}(d) presents the combined unsymmetrized contributions of all events. Alternating fringes parallel to the axis associated with $\alpha_{r,l}^{(1,1)} + \alpha_{u,d}^{(1,1)}$  [panel (a)], and with $\alpha_{r,l}^{(1,2)} + \alpha_{u,d}^{(1,2)}$ [panel (b)] arise, in agreement with Table \ref{fig:phasetab}(bi) and (bii), respectively. In panel (a), fringes widen away from the origin, with wing shapes appearing along the axes. The interference is localized along the positive half-axes since $P_{1a}O_{1a}$ is the dominant event, as shown in Fig.~\ref{fig:fields}(d'). In panel (b), a chequerboard pattern emerges. As discussed in \cite{Hashim2024}, this is a symptom of incoherent symmetrization occurring when the linear fringes (coming from the $\alpha^{(A^2)}_{\Delta \tau}(t',t'')$ and/or $\alpha_{\Delta\tau}^{(\mathbf{p}_1, \mathbf{p}_2)}$ building blocks) associated with $\alpha_{l,r}^{(m,n)}$ and $\alpha_{u,d}^{(m,n)}$ overlap. Figure~\ref{fig:temporal90}(d) shows the total interference associated with $\alpha_{lr}$. The narrower set of fringes in the first quadrant is the unsymmetrized version of Fig.~\ref{fig:temporal90}(b). This effect occurs for all fields and phases where $\alpha_{l,r}^{(m,n)}$, $\alpha_{u,d}^{(m,n)}$ overlap. The dashed lines indicate the predicted fringes from the $\alpha_{\Delta\tau}^{(\mathbf{p}_1, \mathbf{p}_2)}$ building block which agree reasonably well for this phase. However, predictions break down for other phases, due to the pattern being skewed by competing building blocks. It is difficult to determine the interplay of the building blocks, but it depends on the field shape indirectly through the temporal shift between events, and directly through $F_A(\tau)$, which change the relevance of certain building blocks depending on the shape. Finally, Fig.~\ref{fig:temporal90}(c) shows interference associated with $\alpha_{l,l}^{(1,2)} + \alpha_{d,d}^{(1,2)}$, and shown schematically in Table \ref{fig:phasetab}(biii), (bvi). This interference consists of linear fringes in the third quadrant parallel to the axes, and behaves as expected, as the momentum dependence and the times in  Eq.~\eqref{eq:deltaphase} are only associated with the second electron. This type of interference, for events with $\xi=\eta$ and $m\neq n$, is specific to bichromatic fields.  However, the resulting patterns are roughly two orders of magnitude smaller than the outcome of other interference processes. Thus, the linear fringes will not be visible in the total interference patterns in Fig.~\ref{fig:temporal}.

\begin{figure}[!htbp]
\centering
\includegraphics[width=1.05\columnwidth]{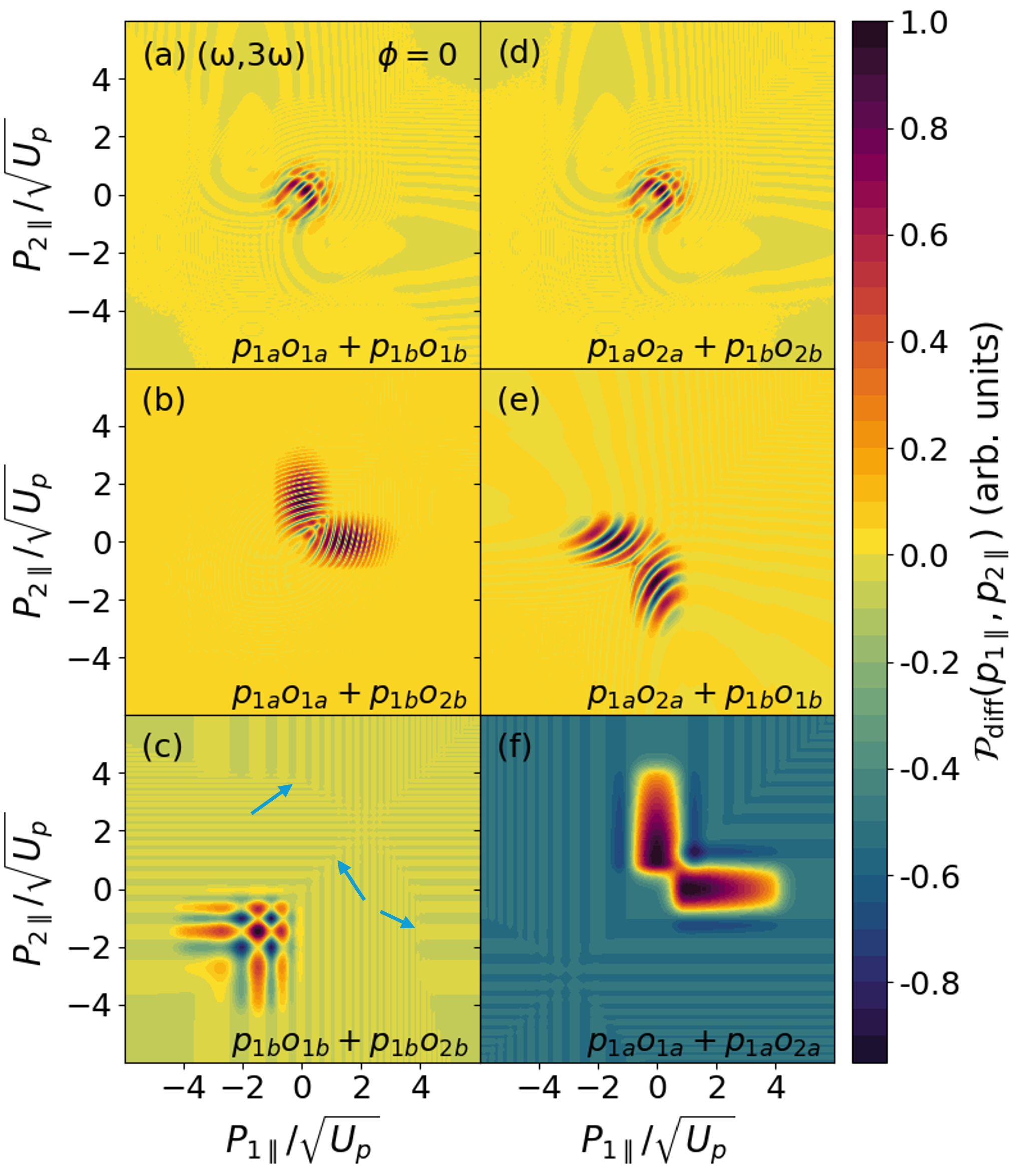}
    \caption{Difference between RESI distributions $\mathcal{P}_{(\mathrm{ic})}(p_{1\parallel},p_{2\parallel})$, in which events are summed coherently, and fully incoherent RESI probability densities $\mathcal{P}_{(\mathrm{ii})}(p_{1\parallel},p_{2\parallel})$, calculated without prefactors for all possible combinations of events with the $(\omega$,3$\omega$ $\phi = 0$) driving field.    
   The coherent sums are related to the $\alpha_{r,l}^{(1,1)} + \alpha_{u,d}^{(1,1)}$, $\alpha_{r,l}^{(1,2)}+ \alpha_{u,d}^{(1,2)}$, $\alpha_{l,l}^{(1,2)} + \alpha_{d,d}^{(1,2)}$ (left column) and  $\alpha_{r,l}^{(2,2)} + \alpha_{u,d}^{(2,2)}$, $\alpha_{r,l}^{(2,1)}+\alpha_{r,l}^{(2,1)}$, $\alpha_{r,r}^{(1,2)} + \alpha_{u,u}^{(1,2)}$ (right column)  phase differences. 
   Other driving-field parameters are the same as in the corresponding panel of Fig.~\ref{fig:fields} and Fig.~\ref{fig:coherent}. The arrows in panel (c) indicate `cuts' at which the direction or width of fringes abruptly changes.    Each panel has been normalized to its maximum probability density. Panels (c) and (f) are two orders of magnitude smaller than the other panels.}
    \label{fig:temporal0}
\end{figure}

\begin{figure}[!htbp]
\centering
\includegraphics[width=\columnwidth]{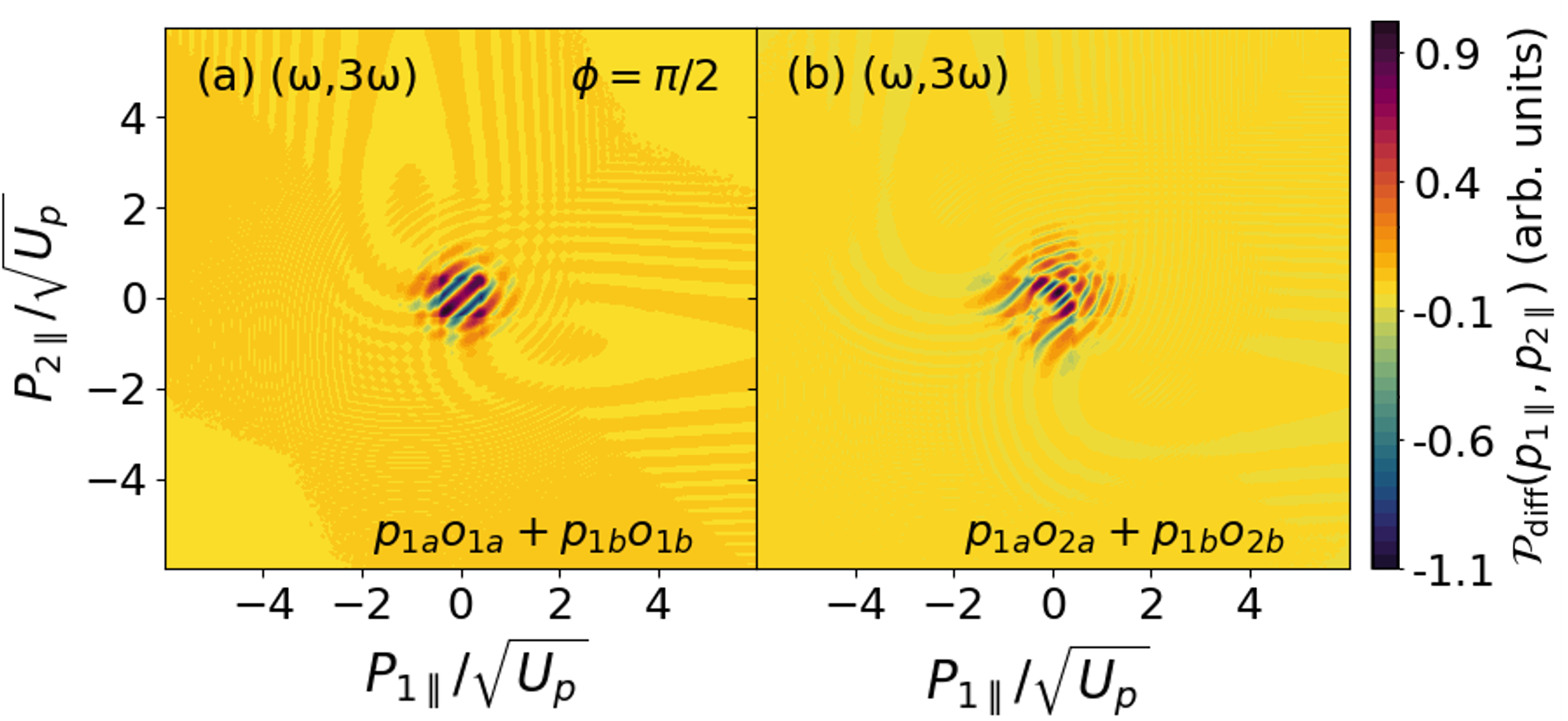}
    \caption{Difference between RESI distributions $\mathcal{P}_{(\mathrm{ic})}(p_{1\parallel},p_{2\parallel})$, in which events are summed coherently, and fully incoherent RESI probability densities $\mathcal{P}_{(\mathrm{ii})}(p_{1\parallel},p_{2\parallel})$, calculated without prefactors for two possible events with the $(\omega$,3$\omega$ $\phi = \pi/2$) driving field, related to the $\alpha_{r,l}^{(1,1)} + \alpha_{u,d}^{(1,1)}$ and $\alpha_{r,l}^{(1,2)}+ \alpha_{u,d}^{(1,2)}$ phase differences [panels (a) and (b), respectively]. Other driving-field parameters are the same as in the corresponding panel of Fig.~\ref{fig:fields} and Fig.~\ref{fig:coherent}. Each panel has been normalized to its maximum probability density.}
    \label{fig:temporal90w3w}
\end{figure}

Figure~\ref{fig:temporal0} displays the event-wise breakdown for $(\omega$,$3\omega$, $\phi=0$). Figures~\ref{fig:temporal0}(a) and (d), showing the effect of $\alpha_{r,l}^{(1,1)} + \alpha_{u,d}^{(1,1)}$ and $\alpha_{r,l}^{(2,2)} + \alpha_{u,d}^{(2,2)}$ respectively, are almost identical. This is expected since all four second-electron solutions are identical for this field. Thus, the event in panel (d) is just temporally shifted with regard to panel (a). Each of the times, $t''$, $t'$ and $t$ are shifted equally, so the pattern does not change. This property no longer holds if the ionization events of the second electron are made unequal, such as for $(\omega$,$3\omega$, $\phi=\pm \pi/2$) - see Fig.~\ref{fig:temporal90w3w}.
These fringes are predicted to occur along the axes by Fig.~\ref{fig:phasetab}(bi), (bv) but are most intense close to the origin. This is because the RESI probability densities of the events themselves are localized in the right-up and left-down quadrants, leading to interference in the middle (at the origin). These distributions exhibit wing shapes, with ``static" in the corners of the first and third quadrants, similar to what is observed with few-cycle pulses \cite{Hashim2024}. Figures~\ref{fig:temporal0}(b) and (e) show interference related to $\alpha_{r,l}^{(1,2)} + \alpha_{u,d}^{(1,2)}$ and $\alpha_{r,l}^{(2,1)} +\alpha_{u,d}^{(2,1)}$ (i.e. $m \neq n$). These interference patterns agree with the predicted schematics in Table~\ref{fig:phasetab}(bii), (biv) but the momentum regions to which the events $P_{1\xi}$ are mapped causes only the positive ($\xi=a$) and negative ($\xi=b$) half-axes to be occupied [see Figs. \ref{fig:fields}(a) and (a')].  Very faint wing shapes can be observed in panel (e). Finally, interference from $\alpha_{l,l}^{(1,2)} + \alpha_{d,d}^{(1,2)}$ and $\alpha_{r,r}^{(1,2)} + \alpha_{u,u}^{(1,2)}$ arising from two events within the same half-cycle and shown schematically in Table \ref{fig:phasetab}(biii), (bvi) are shown in Fig.~\ref{fig:temporal0}(c),(f). Both panels show fringes parallel to the axes. This is expected as the $\alpha^{(A^2)}_{\Delta \tau}(t',t'')$ temporal phase depends solely on the ionization and rescattering times of the first electron [see Sec.~\ref{sec:temporal}]. Symmetrization leads to a checkerboard pattern in Fig.~\ref{fig:temporal0}(c) but not in Fig.~\ref{fig:temporal0}(f) as the fringes are broader and farther apart in the latter. There also appear to be shifts or cuts where the direction of the fringes swaps, and in some cases, the width changes. These are indicated by the arrows in Fig.~\ref{fig:temporal0}(c).

The dominance of events plays a significant role in determining the total event interference pattern.  For the $(\omega,3\omega, \phi=0)$ field, Fig.~\ref{fig:temporal}(a) shows strong patterns along the axes and near the origin, as $O_{1\xi}$ and $O_{2\xi}$ are equally dominant. For other phases, however,  Fig.~\ref{fig:temporal}(b) and (c) show that the contributions around the axes weaken. This is due to the unequal dominance of $O_{1\xi}$ and $O_{2\xi}$ as the fringes along the axes stem from the interference involving these events in different cycles [see Figs. \ref{fig:temporal0}(b) and (e)]. 

\subsection{Exchange and Event Interference}
\label{sec:combineddiscussion}

\begin{figure}[!htbp]
\centering
\includegraphics[width=\columnwidth]{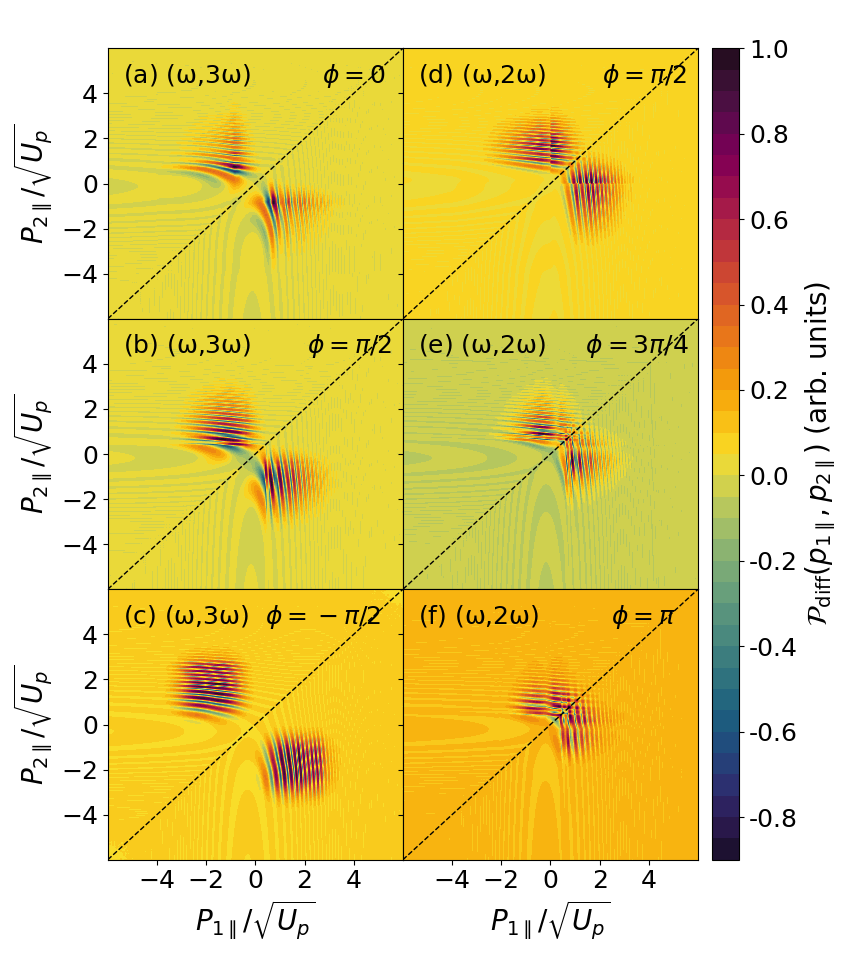}
    \caption{Difference between the RESI distributions given by Eq.~\eqref{eq:combinedP}, where the symmetrized and unsymmetrized counterparts of temporally-shifted events are summed coherently, and the fully incoherent distribution $\mathcal{P}_{(ii)}(p_{1\parallel},p_{2\parallel})$. No prefactors are included and the fields and phases are as in the corresponding panels of Figs.~\ref{fig:fields} and \ref{fig:coherent}. The signal in each panel has been normalized with regard to its maximum. The diagonals $p_{1\parallel}=p_{2\parallel}$ are indicated with dashed lines.}
    \label{fig:differencecombined}
\end{figure}

Figure~\ref{fig:differencecombined} presents the combined exchange and event interference, schematically represented by processes (c) in Table~\ref{fig:phasetab}. This type of interference arises from a coherent sum of the transition amplitude of one event and the symmetrized amplitude of a different (i.e., temporally-shifted) event. The distributions are reflection-symmetric about the diagonal, confirming that $|\alpha_{l,u}^{(m,n)}| = |\alpha_{r,d}^{(m,n)}|$. `Feathery' fringes parallel to the axes are located in regions where the constituent dominant events are localized, i.e., the locations of the brightest spots and their widths are dependent on the field itself. Fainter, thicker fringes are also visible along the negative half-axes along with faded partial circular substructure. This assortment of patterns stems from the five temporal-exchange building blocks, as in Eq.~\eqref{eq:alphalu}. The hyperbolic fringes expected from the exchange term, along with the four pure temporal phases lead to secondary faint features which occur in the first and third quadrants of Fig.~\ref{fig:differencecombined} and at large momenta. These features serve to skew the main visible features (the feathers) arising from the combined temporal-exchange phase difference building block, $\alpha^{(\mathbf{p_1 \leftrightarrow p_2})}_{\Delta \tau}(t, t')$. Similar patterns were observed for the few-cycle pulse for this excitation pathway \cite{Hashim2024}.  

\begin{figure}[!htbp]
\centering
\includegraphics[width=\columnwidth]{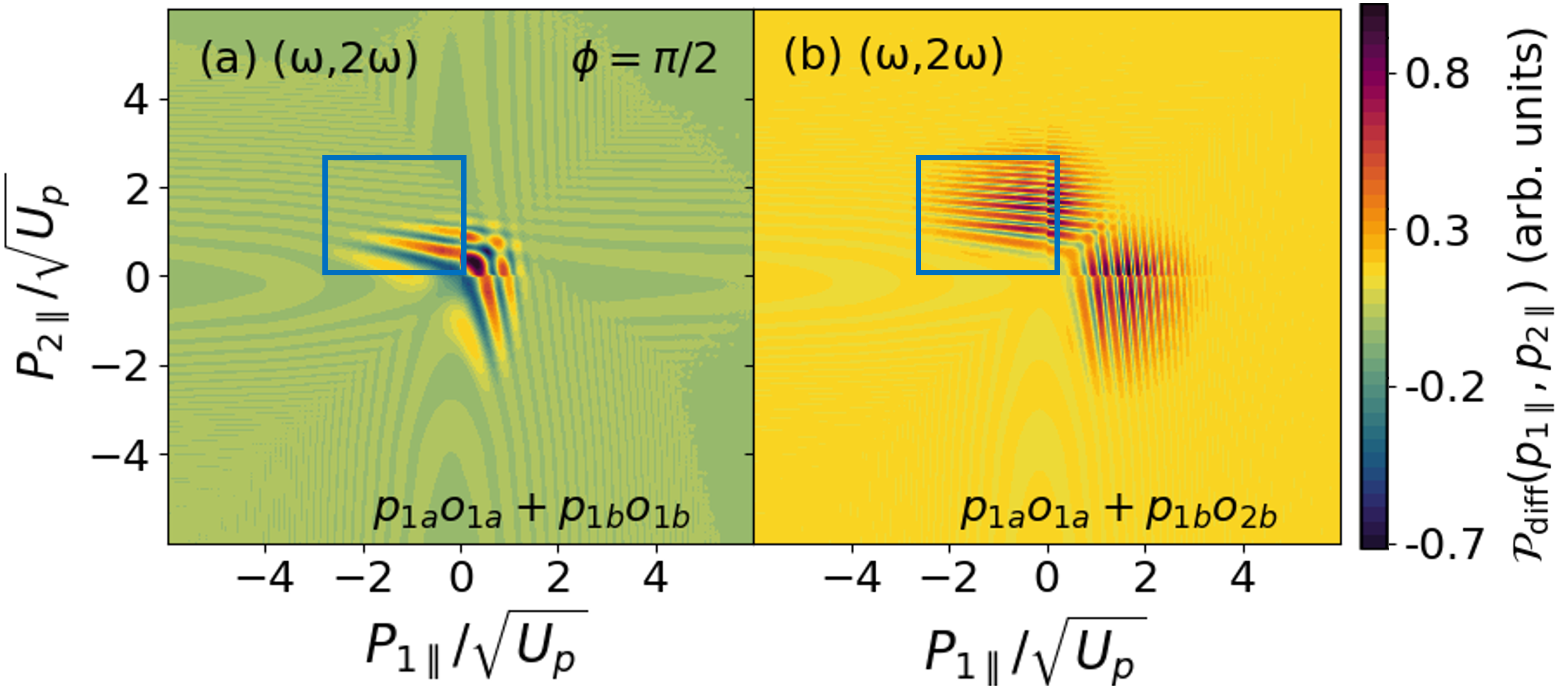}
    \caption{Difference between RESI distributions where the symmetrized and unsymmetrized counterparts of temporally-shifted events are summed coherently [Eq.~\eqref{eq:combinedP}] and the fully incoherent distribution calculated without prefactors for all possible combinations of events with the $(\omega,2\omega, \phi=\pi/2)$ driving field. The coherent sums are related to the $\alpha_{r,d}^{(1,1)} + \alpha_{u,l}^{(1,1)}$, $\alpha_{r,d}^{(1,2)}+ \alpha_{u,l}^{(1,2)}$ phase differences [panels (a) and (b), respectively]. Other driving-field parameters are the same as in the corresponding panels of Figs.~\ref{fig:fields} and \ref{fig:coherent}. The contributing events are indicated in the lower-right of each panel. Each panel has been normalized to its maximum probability density.}
    \label{fig:combined90}
\end{figure}

Figure~\ref{fig:combined90} displays the eventwise breakdown of combined interference for ($\omega$, 2$\omega$, $\phi=\pi/2$) and all possible pairs of events. Panels (a) and (b) depict interference of events in different half-cycles. The location of this interference, in the second and fourth quadrants with spilling into other regions, is in good agreement with predictions in Table~\ref{fig:phasetab}(ci) and (cii). Both panels consist of wider fringes near the origin that thin out with increasing absolute momentum, creating feathery fringes. Equation~\eqref{eq:combinedfringes} predicts fringes from these events to have the same gradient and intercept due to the symmetry of this field. This prediction has been verified. However, the feathers \textit{appear} to be much narrower for the event involving $O_{2b}$ [panel (b)] compared to $O_{1b}$ [panel (a)]. This discrepancy stems from the fact that these events occupy different regions in momentum space, meaning different parts of the fainter fringes are more intense for different events - see boxes on Fig.~\ref{fig:combined90}. 
\begin{figure}[!htbp]
\centering
\includegraphics[width=\columnwidth]{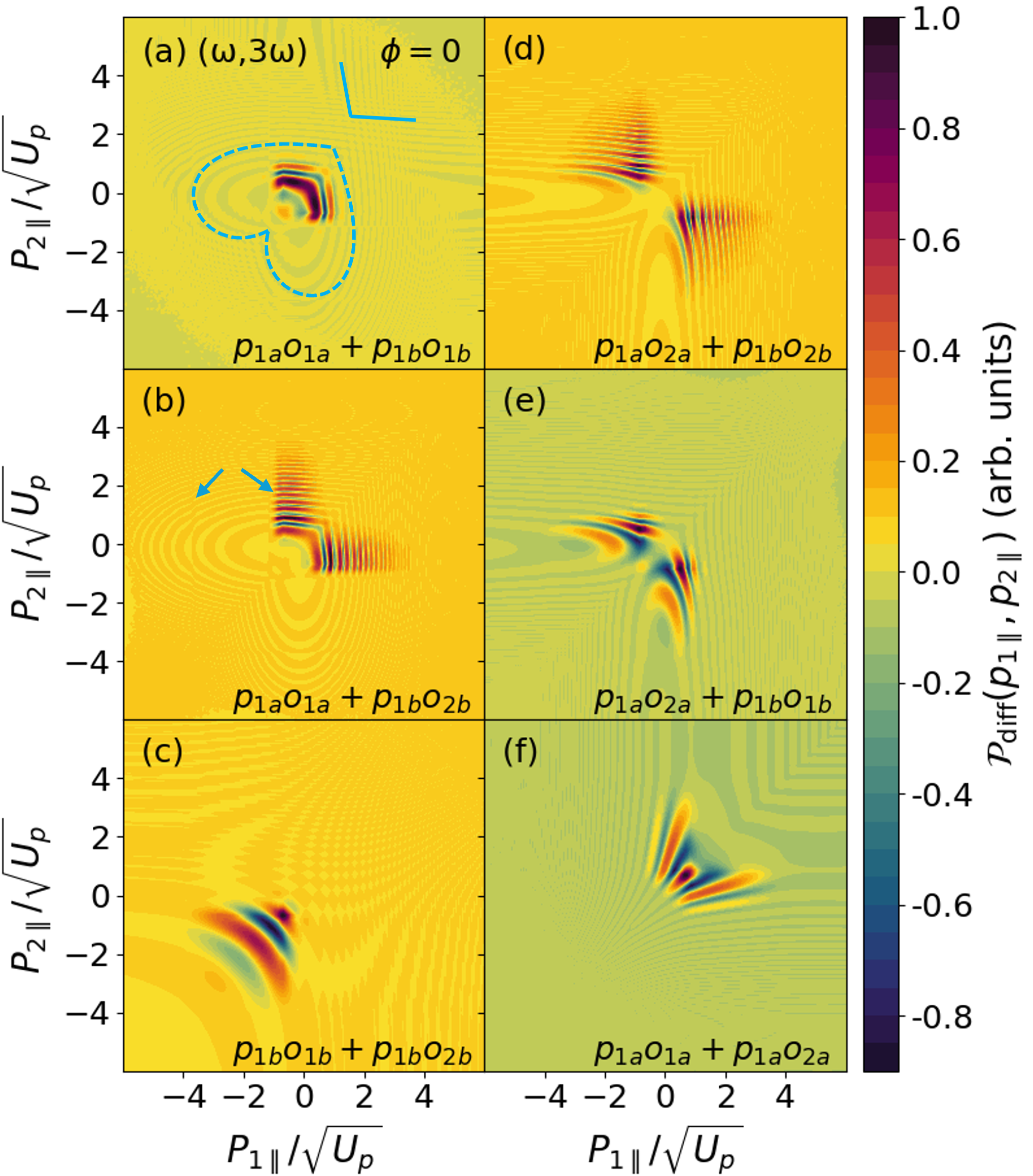}
    \caption{
    Difference between RESI distributions where the symmetrized and unsymmetrized counterparts of temporally-shifted events are summed coherently [Eq.~\eqref{eq:combinedP}] and the fully incoherent distribution calculated without prefactors for all possible combinations of events with the $(\omega,3\omega, \phi=0)$ driving field. The coherent sums are related to the   $\alpha_{r,d}^{(1,1)} + \alpha_{u,l}^{(1,1)}$, $\alpha_{r,d}^{(1,2)}+ \alpha_{u,l}^{(1,2)}$, $\alpha_{l,d}^{(1,2)} + \alpha_{d,l}^{(1,2)}$ (left column) and $\alpha_{r,d}^{(2,2)} + \alpha_{u,l}^{(2,2)}$, $\alpha_{r,d}^{(2,1)}+ \alpha_{u,l}^{(2,1)}$ and $\alpha_{r,u}^{(1,2)} + \alpha_{u,r}^{(1,2)}$ (right column)  phase differences. Other driving-field parameters are the same as in the corresponding panels of Figs.~\ref{fig:fields} and \ref{fig:coherent}. The contributing events are indicated in the lower-right of each panel. The `heart' and some of the linear fringes are indicated with blue annotations. Each panel has been normalized to its maximum probability density.}
    \label{fig:combined0}
\end{figure}

Figure~\ref{fig:combined0} illustrates event-wise temporal-exchange interference for ($\omega$, 3$\omega$ $\phi=0$). With this field, there are six pairs of events that may interfere when accounting for depletion and considering only the dominant first electron events $P_{i\xi}$ - see Fig.~\ref{fig:fields}(a'), which leads to a richer collection of interference patterns. All distributions agree with the locations predicted in Table \ref{fig:phasetab}(c). 

In Figs.~\ref{fig:combined0}(a) and (b), faint heart shapes are visible, with criss-crossed v-shaped fringes in the first quadrant and faint circular substructure at large momenta - see annotations. In panels (b) and (e), feathers are visible. The ends of the hearts are stretched out along the negative half-axes in panel (d). Interestingly, temporal-exchange interference occurring from $P_{1a}O_{1a}+P_{1b}O_{1b}$ [panel (a)] and from $P_{1a}O_{2a}+P_{1b}O_{2b}$ [panel (d)] are neither identical (as for temporal-only shifts) nor mirror images (as for exchange-only shifts). This is as predicted since $\alpha_{\mu,\nu}^{(m,m)}$ where $\mu,\nu=r,d;l,u$ and $m=1,2,...$ are distinct phase differences. Similarly, panels (b) and (e) are not identical: $\alpha_{\mu,\nu}^{(1,2)} \neq \alpha_{\mu,\nu}^{(2,1)}$ as expected. Cuts can also be seen in these plots indicated by arrows in panel (b). Such patterns are similar to those observed with few-cycle pulses.

 Figures~\ref{fig:combined0}(c) and (d) show combined interference of events within the same half-cycle. In agreement with the schematic representations of processes (cvi) and (ciii) in Table \ref{fig:phasetab}, these interference patterns occupy the first and third quadrants, respectively, with slight spilling into neighbouring quadrants. However, these interference fringes are orders of magnitude smaller than the others and thus do not contribute much to the total interference in Fig.~\ref{fig:differencecombined}.

What is perhaps most interesting about temporal-exchange shifts is that, unlike pure exchange and temporal shifts, dominant events are not the primary contributors to the combined interference in Fig.~\ref{fig:differencecombined}(a). Given that all second-electron solutions are equally dominant with $\phi=0$, one expects the total distribution to be a fairly equal mix of the patterns in Fig.~\ref{fig:combined0}. However, it appears to resemble only Fig.~\ref{fig:combined0}(d). This may be attributed to the exchange building block in Eq.~\eqref{eq:alphalu} which depends strongly on $(t-t')$. This value is largest for pairs of events involving $O_{2a}$ and $O_{2b}$. Thus, it appears that $O_{2b}$ plays a bigger role in the total temporal interference pattern, despite $O_{1b}$ being of equal dominance. This is also observed with the $(\omega, 2\omega, \phi=\pi/2)$ driving field in Fig.~\ref{fig:combined90} - panel (b) is the dominant contributor to Fig.~\ref{fig:differencecombined}(d). We have verified that this holds for other fields and phases, when the pairs of events involving $O_{1\xi}$ and $O_{2\xi}$ occupy the similar regions. Therefore, it is possible to control the intensity and pattern of the interference by choosing to incorporate certain events. 

Inspection of Fig.~\ref{fig:coherent} shows that pure-exchange and combined temporal-exchange interference features play the biggest role for all fields and phases. The spine lines along the diagonal, skewed by the hyperbolae away from the origin, along with the feathery fringes in the second and fourth quadrants are evident. This is unlike few-cycle pulses, where temporal-exchange shifts were not so visible in the total distribution. Pure temporal shifts remain least important and are barely visible in Fig.~\ref{fig:coherent}. Interference between events separated by half a cycle prevail for temporal-only phase differences, and thus are also not as influential in the total interference patterns.

\subsection{Interference with prefactors}
\label{sec:prefac}

Next, we investigate how the prefactors affect the previous results, focusing on the $3s\rightarrow 3p$ and $3p\rightarrow 4s$ transitions. Transitions involving $d$ excited states are less important \cite{Hashim2024b} and will not be discussed here.
\subsubsection{$3s\rightarrow3p$}

\begin{figure*}[!htbp]
\centering
\includegraphics[width=\textwidth]{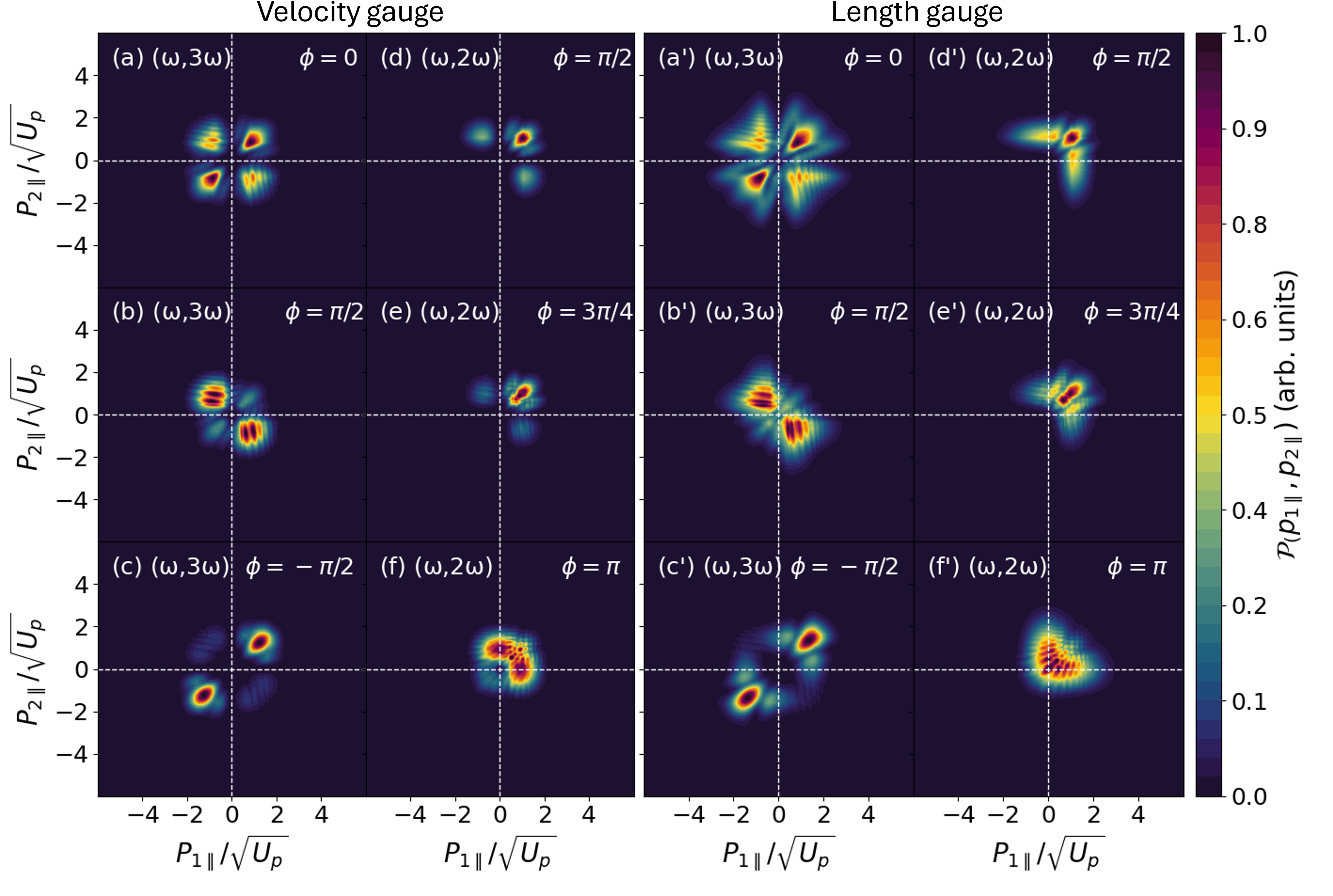}
    \caption{Fully coherent momentum distributions with prefactors in the velocity and length gauges (left and right blocks, respectively, as indicated), calculated for the ($\omega$,$3\omega$) (left columns) and ($\omega$,$2\omega$) (right columns) driving fields with the relative phase as indicated in the panels. Other driving-field parameters are the same as in the corresponding panels of Figs.~\ref{fig:fields} and \ref{fig:coherent}. The RESI distributions in each panel have been normalized to their maximum values. }
    \label{fig:coherentvgauge}
\end{figure*}

In Fig.~\ref{fig:coherentvgauge}, we present the fully coherent PMDs with prefactors incorporated in both gauges (left and right blocks, respectively) for the $3s\rightarrow 3p$  transition. The brightest fringes are now narrowed and occupy momentum regions much more localized around the origin, compared to the no-prefactor interference presented in Fig.~\ref{fig:coherent}.  This is expected given that the prefactor narrows the momentum distributions. In both cases, the symmetry along the diagonal is retained, and the key shapes of the interference patterns (spine lines, hyperbolae, alternating fringes, feathers) are retained. However, the velocity-gauge prefactor subtly enhances the signal in the second and fourth quadrants for $(\omega, 3\omega, \phi=-\pi/2)$[see Figs.\ref{fig:coherentvgauge}(c) and (c')] and $(\omega, 2\omega, \phi=3\pi/4)$ [see Figs.\ref{fig:coherentvgauge}(e) and (e')]. 

\begin{figure*}[!htbp]
\centering
\includegraphics[width=0.97\textwidth]{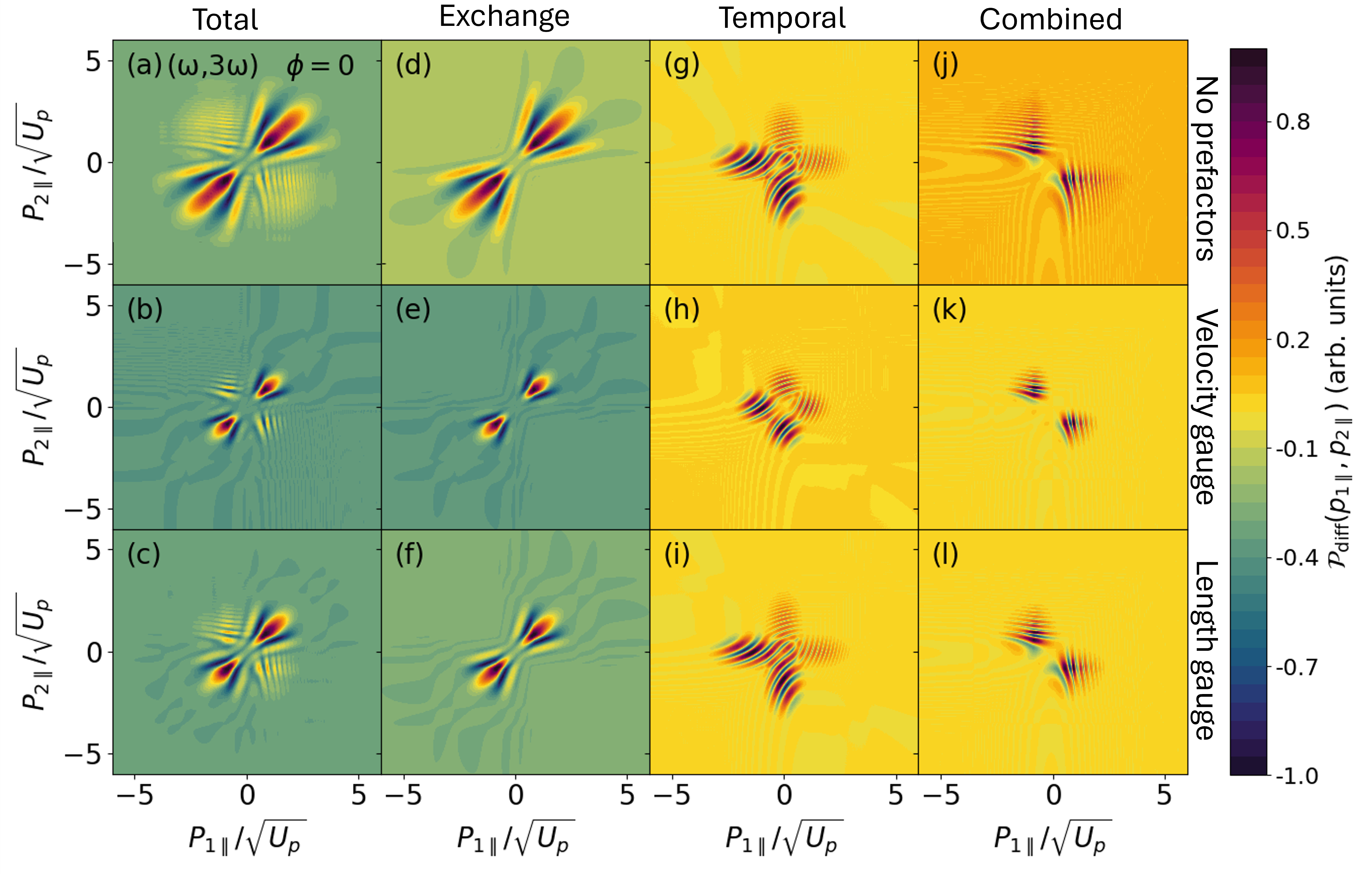}
    \caption{Full interference patterns for the $3p$ excited state calculated without prefactors (top row), with prefactors calculated in the velocity gauge (middle row), and with prefactors calculated in the length gauge (bottom row) for the ($\omega$, $3\omega$) field with the relative phase $\phi=0$. From left to right, we display the differences $\mathcal{P}_{\mathrm{diff}}(p_{1\parallel},p_{2\parallel})$, between different types of coherent RESI distributions and their fully incoherent counterparts. The left column considers the fully coherent and  RESI distributions, the second left column the coherently symmetrized distributions $\mathcal{P}_{(ci)}(p_{1\parallel},p_{2\parallel})$, to highlight exchange interference, the second right column displays the distributions $\mathcal{P}_{(ic)}(p_{1\parallel},p_{2\parallel})$ for which time-delayed events were summed coherently and the symmetrization was done incoherently, showcasing temporal interference, and in the right column we consider combined exchange-temporal interference, for which the symmetrized and unsymmetrized counterparts of temporally-shifted events are summed coherently [Eq.~\eqref{eq:combinedP}]}.  The field parameters are the same as in the previous figures.
    \label{fig:3p0}
\end{figure*}

\begin{figure*}[!htbp]
\centering
\includegraphics[width=0.97\textwidth]{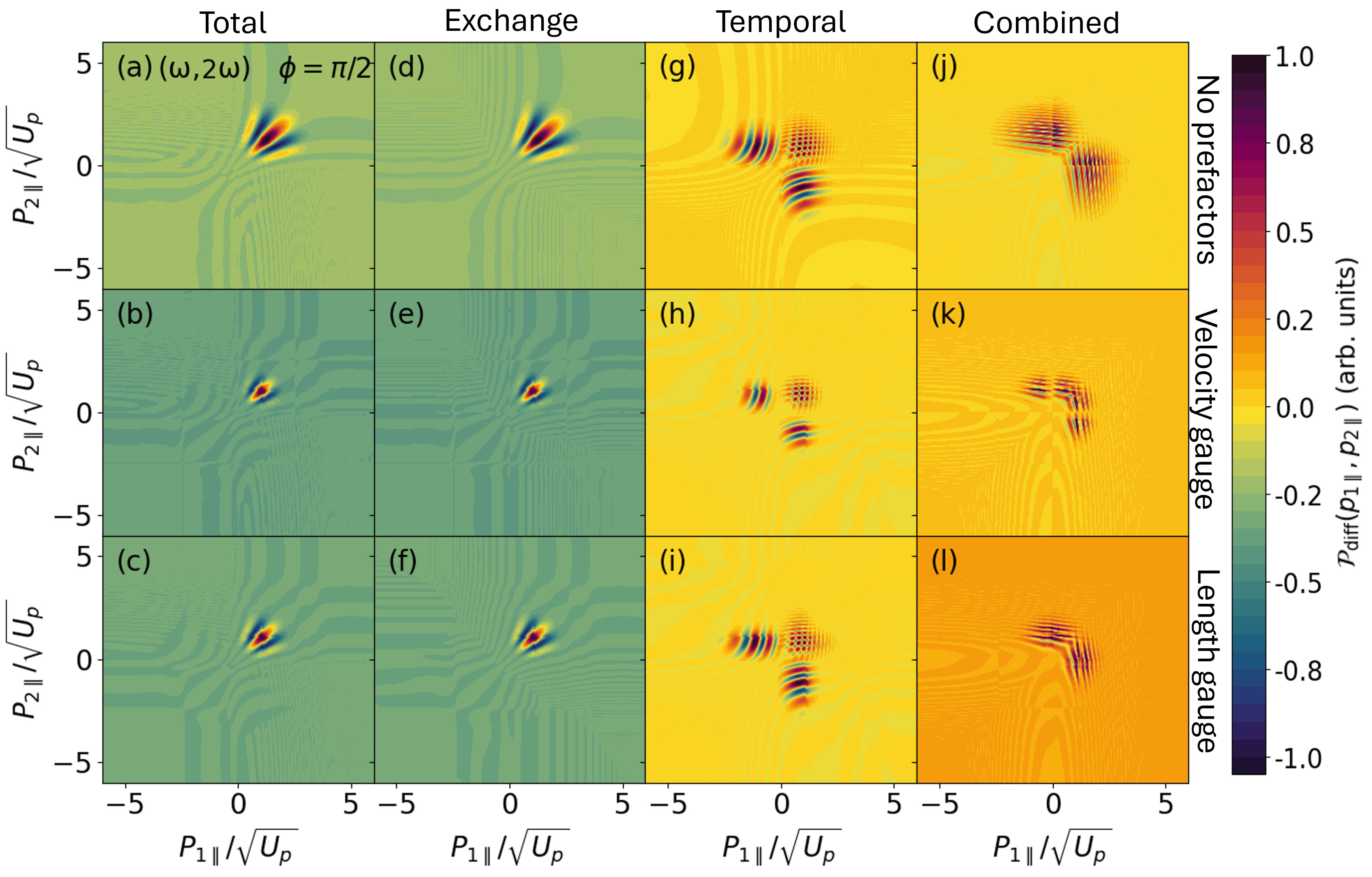}
\caption{Full interference patterns for the $3p$ excited state calculated without prefactors (top row), with prefactors calculated in the velocity gauge (middle row), and with prefactors calculated in the length gauge (bottom row) for the ($\omega$, $2\omega$) field with the relative phase $\phi=\pi/2$. From left to right, we display the differences $\mathcal{P}_{\mathrm{diff}}(p_{1\parallel},p_{2\parallel})$, between different types of coherent RESI distributions and their fully incoherent counterparts. The left column considers the fully coherent and  RESI distributions, the second left column the coherently symmetrized distributions $\mathcal{P}_{(ci)}(p_{1\parallel},p_{2\parallel})$, to highlight exchange interference, the second right column displays the distributions $\mathcal{P}_{(ic)}(p_{1\parallel},p_{2\parallel})$ for which time-delayed events were summed coherently and the symmetrization was done incoherently, showcasing temporal interference, and in the right column we consider combined exchange-temporal interference, for which the symmetrized and unsymmetrized counterparts of temporally-shifted events are summed coherently [Eq.~\eqref{eq:combinedP}].  The field parameters are the same as in the previous figures.}
    \label{fig:3p90}
\end{figure*}

\begin{figure}[!htbp]
\centering
\includegraphics[width=\columnwidth]{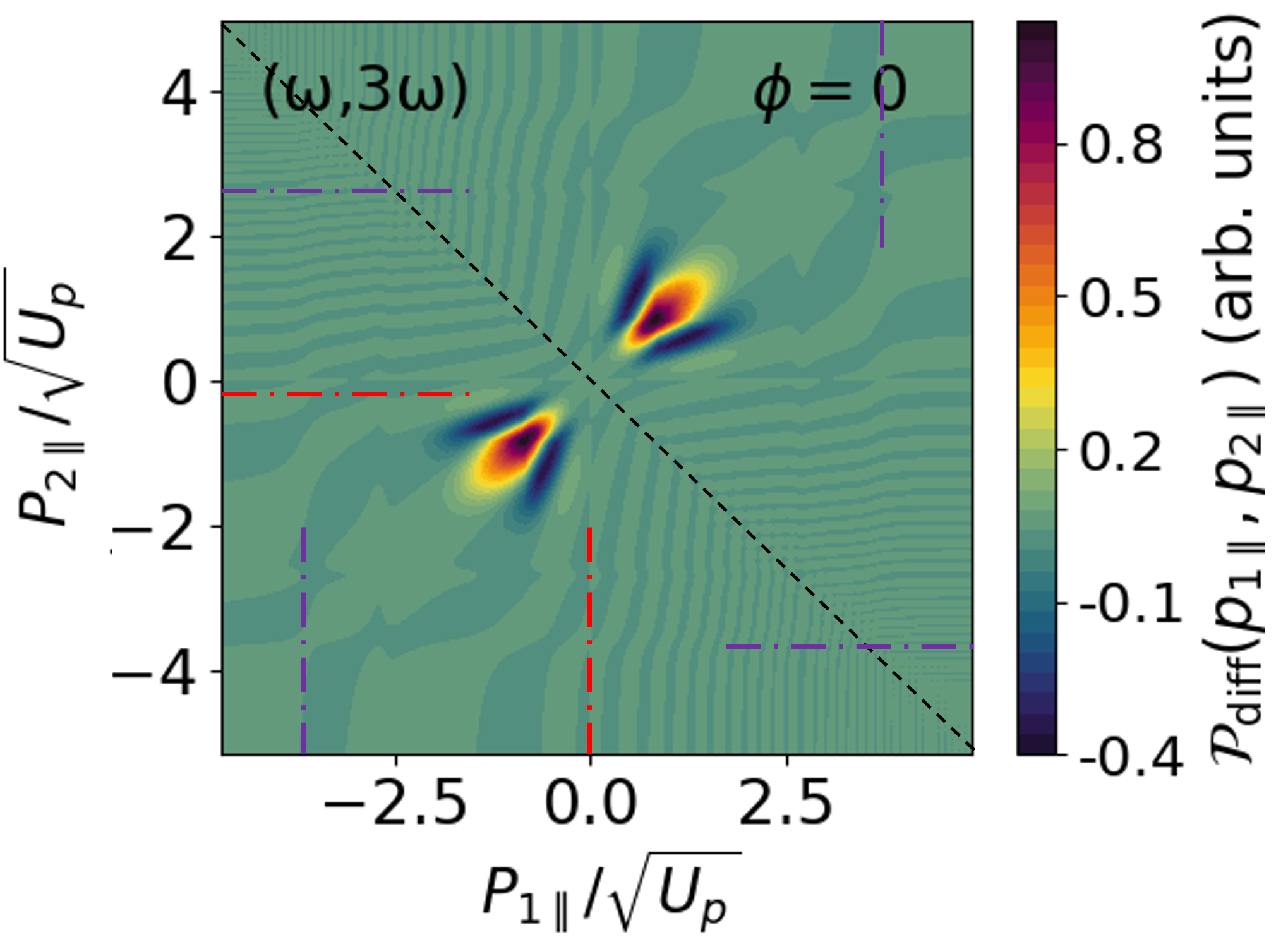}
    \caption{Difference between the fully coherent and incoherent RESI distributions for the ($\omega$,$3\omega$, $\phi=0$) driving field with the $3p$ prefactors in the velocity gauge. The black dashed line marks the anti-diagonal $p_{1\parallel}=-p_{2\parallel}$. The cuts are indicated by purple and red dashed-dot lines. The figure has been normalized with respect to its maximum. All other field parameters are the same as in the corresponding panel of Figs.~\ref{fig:fields} and \ref{fig:coherent}.}
    \label{fig:exchangevgauge}
\end{figure}

 We now investigate how exchange-only, temporal-only and temporal-exchange shifts are altered by the prefactor. A general feature is that specific types of interference keep their main characteristics, regardless of the bias introduced by the prefactors. Thus, in broad terms, exchange interference leads to hyperbolae and spines, temporal interference is strongest near the $p_{n\parallel}$ axes, and temporal-exchange interference leads to feather-like fringes. 
 This is shown in the difference plots presented in Figs.~\ref{fig:3p0}, \ref{fig:3p90} and \ref{fig:exchangevgauge},  for selected driving fields. As expected, the momentum ranges for which the probability density is appreciable are significantly reduced, and, in the velocity gauge, the signal tends to be suppressed near the axes. This is clearly seen in Figs.~\ref{fig:3p0}(b), (e) and (k), and in Figs.~\ref{fig:3p90}(b), (e), (h) and (k). This suppression is characteristic of $V_{\mathbf{p}_{2}e}$ for $p$ states \cite{Shaaran2010,Maxwell2015}. In Figs.~\ref{fig:coherentvgauge}(f)  and \ref{fig:3p0}(f) this suppression is washed out due to the prefactor $V_{\mathbf{p}_1e,\mathbf{k}g}$, which narrows the distribution and shifts it to the origin, in conjunction with the steep and asymmetrical partial momentum distribution for the second electron. Another noteworthy feature is the different ``bending'' of the RESI distributions in Figs.~\ref{fig:coherentvgauge}(f) and (f'), away from and towards the origin, respectively. These different shapes are caused by the different prefactors $V_{\mathbf{p}_2e}$ in the velocity and length gauges. In the velocity gauge, $V_{\mathbf{p}_2e}$ exhibits nodes along the axes for $p$ states, with an upper and lower lobe \cite{Shaaran2010,Maxwell2015}. For $(\omega, 2\omega, \phi=\pi)$ fields, the upper lobe is more prominent, bringing the distributions higher up around the positive $p_{n\parallel}$ half axes. In the length gauge, these nodes are blurred in $V_{\mathbf{p}_2e}$ due to the complex argument introduced by $A(t)$. Therefore, the RESI distribution is not shifted as high up along the axes, and, upon symmetrization, bends differently. However, this effect does not stem from quantum interference and is discussed in more detail in  Ref.~\cite{Hashim2024b}.

For exchange-only interference, there are additional phase shifts or cuts at locations determined by the $3p$ prefactor mapping, indicated by dash-dot purple lines in Fig.~\ref{fig:exchangevgauge}. At these points, the direction of the fringes changes, and the fringes are not continuous and appear disjointed beyond these points. In the velocity gauge, there are additional cuts at the axes arising from the nodes of the prefactor [dash-dot red lines in Fig.~\ref{fig:exchangevgauge}].

Temporal shifts are the least influential, even with the prefactor. We have checked that the length-gauge prefactor simply narrows the temporal interference-patterns slightly without altering the shapes, in comparison with the non-prefactor case presented in Fig.~\ref{fig:temporal}. It also blurs out the faint secondary interference fringes for the ($\omega$, 3$\omega$, $\phi=0$) field and ($\omega$, 2$\omega$, $\phi=\pi/2$). Meanwhile, the velocity-gauge prefactor results in significantly narrowed temporal interference patterns, with slightly more prominent secondary interference patterns at higher momenta. In both cases, the cuts and phase shifts are less obvious than for pure exchange shifts. Finally, the temporal-exchange shifts remain influential in the total interference pattern when prefactors are included. 

\subsubsection{$3p\rightarrow4s$}

\begin{figure*}[!htbp]
\centering
\includegraphics[width=0.97\textwidth]{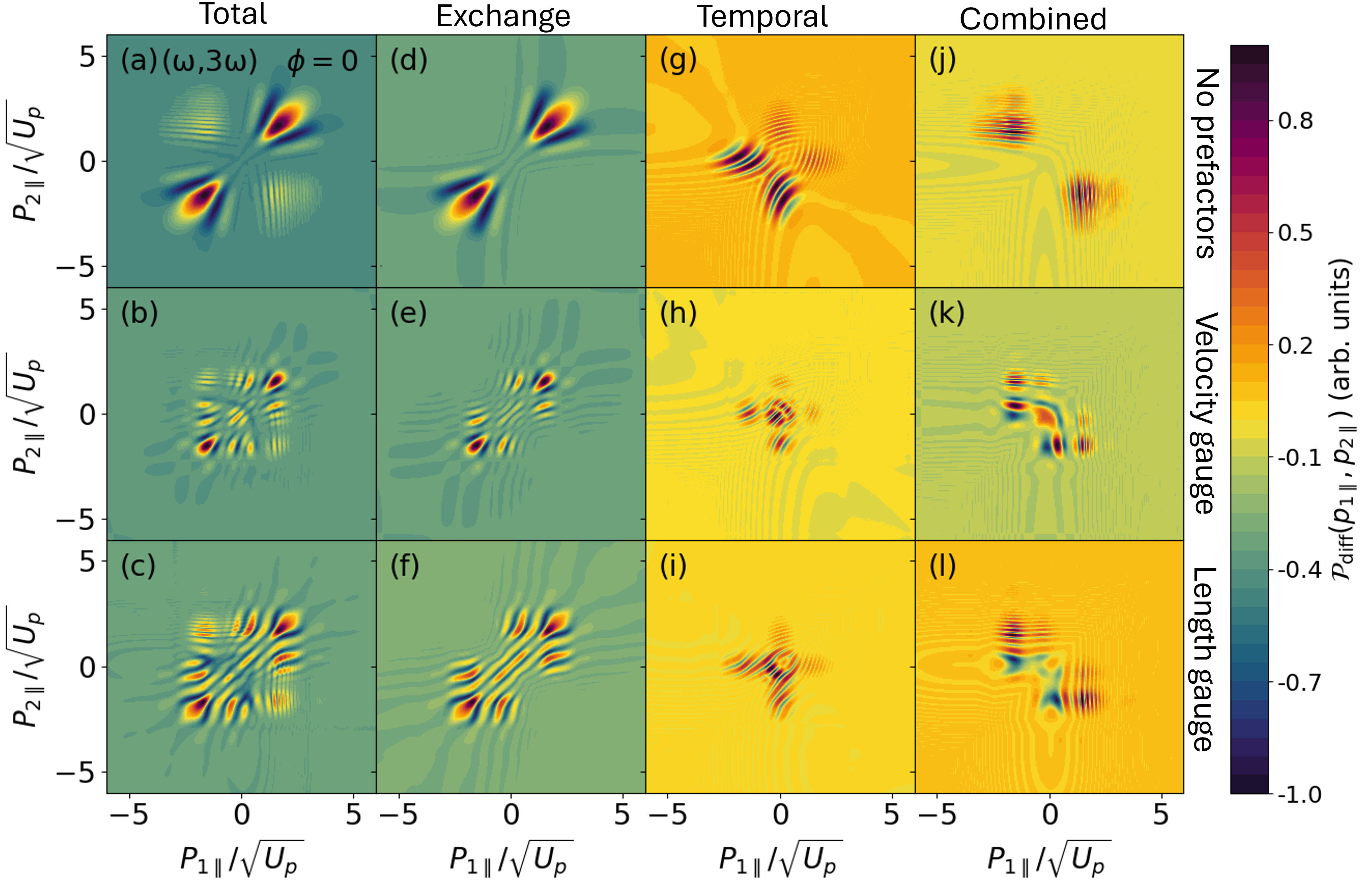}
    \caption{Full interference patterns for the $4s$ excited state calculated without prefactors (top row), with prefactors calculated in the velocity gauge (middle row), and with prefactors calculated in the length gauge (bottom row) for the ($\omega$, $3\omega$, $\phi=0$) field. From left to right, we display the differences $\mathcal{P}_{\mathrm{diff}}(p_{1\parallel},p_{2\parallel})$, between different types of coherent RESI distributions and their fully incoherent counterparts, following the same ordering from left to right as in Fig.~\ref{fig:3p0} (all types of interference, exchange-only interference, temporal interference and combined exchange-temporal interference). The remaining field parameters are the same as in the previous figures. The signal in each panel was normalized to its maximum value. }
    \label{fig:4s0}
\end{figure*}

\begin{figure*}[!htbp]
\centering
\includegraphics[width=0.97\textwidth]{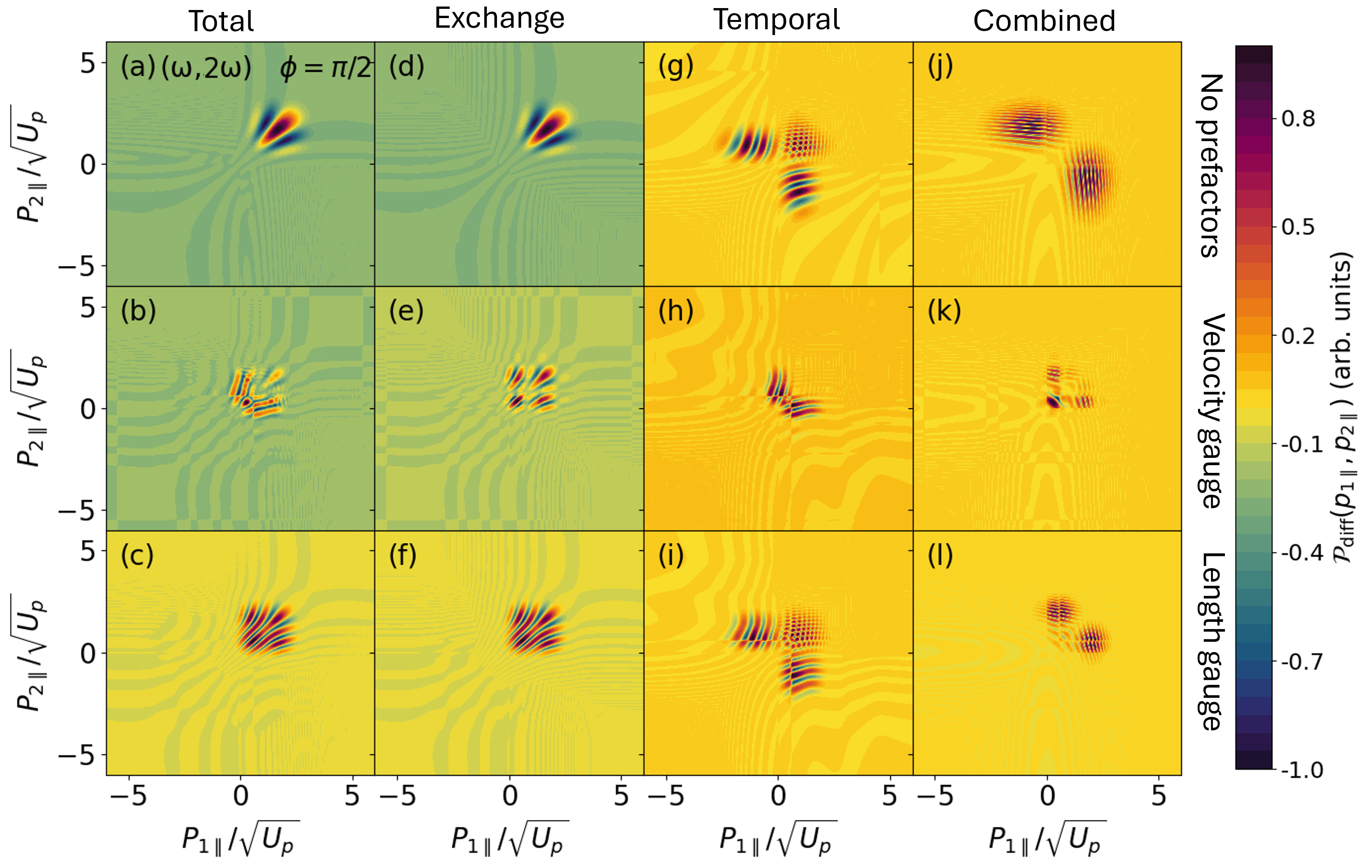}
    \caption{Full interference patterns for the $4s$ excited state calculated without prefactors (top row), with prefactors calculated in the velocity gauge (middle row), and with prefactors calculated in the length gauge (bottom row) for the ($\omega$, $2\omega$, $\phi=\pi/2$) field. From left to right, we display the differences $\mathcal{P}_{\mathrm{diff}}(p_{1\parallel},p_{2\parallel})$, between different types of coherent RESI distributions and their fully incoherent counterparts, from left to right as in  Figs.~ Fig.~\ref{fig:3p0} and  \ref{fig:3p90} (all types of interference, exchange-only interference, temporal interference and combined exchange-temporal interference). The remaining field parameters are the same as in the previous figures. The signal in each panel was normalized to its maximum value.}
    \label{fig:4s90}
\end{figure*}

Now we investigate the $3p\rightarrow4s$ transition, for which the nodes of the prefactor, and thus its mapping, will affect the interference differently. In Figures~\ref{fig:4s0} and \ref{fig:4s90} [computed for ($\omega$, $3\omega$, $\phi=0$) and ($\omega$,$2\omega$, $\phi=\pi/2)$ fields, respectively], we display different types of $\mathcal{P}_{\mathrm{diff}}(p_{1\parallel},p_{2\parallel})$, computed without prefactors (top row), and with prefactors in the velocity and length gauges (middle and bottom rows, respectively).  As in the $3s\rightarrow3p$ case, the interference patterns with prefactors are confined to smaller momentum regions, with the velocity-gauge prefactor leading to the narrowest distributions. In addition, the prefactors cause more intricate secondary maxima to be present for higher momenta. 
The change in shapes of Figs.~\ref{fig:4s0}(a) and \ref{fig:4s90}(a) can be better understood by considering the exchange-only shifts [panels (d)-(f)]. For both phases, there are now far more spine lines. The central spine also occupies the origin, as the $4s$ prefactor does not exhibit nodes at the axes. This is especially evident in Fig.~\ref{fig:4s0}(j). The amount of skewing experienced by the spine lines from the hyperbolae is unchanged in all cases. The length gauge spine lines are much ``smoother'', with fewer cuts in the interference pattern where fringe 
 directions or widths change. This is because, for this gauge, the prefactor is orbit-dependent due to being shifted by the vector potential $A(t)$ for each event. In contrast, the prefactor acts on the events as a whole in the velocity gauge, since there is no $A(t)$ shift. Thus, events localized entirely in the right-up quadrant (e.g., $P_{1a}O_{1a}$) or in the left-down quadrant (e.g., $P_{1b}O_{1b}$) will experience the mapping of the prefactor in those momentum regions. This may lead to additional cuts or phase shifts, if the event is influenced by, for example, the fainter outer radial node of the $4s$ prefactor. These cuts are visible in panel (f)  of Figs.~\ref{fig:4s0}, \ref{fig:4s90}.
 
The shape of the temporal interference is barely affected by the prefactors. However, they narrow the relevant interference region, with the effect being more pronounced in the velocity gauge. 
Thus, fringes are so narrow that they may no longer overlap to cause a chequerboard pattern when symmetrized incoherently - see Fig.~\ref{fig:4s90}(g). Alternatively, confining the interference pattern to a smaller momentum region may cause additional overlap of temporal fringes that were originally close but not touching. This leads to a new artificial chequerboard pattern as in Fig.~\ref{fig:4s0}(g). Cuts are present due to the prefactor mapping. It should be noted that temporal shifts still remain least influential in the total interference pattern in panels (a)-(c), even with the prefactor.

Finally, the momentum regions with appreciable temporal-exchange interference also narrow with the prefactor. Cuts may be present, particularly visible for ($\omega$, 2 $\omega$, $\phi=\pi/2$) in the length gauge because the prefactor shifts the featherlike structure into the first quadrant (see Fig.~\ref{fig:4s90}(l)), due to vanishing $A(t)$ at the ionization time for the dominant $P_{1a}O_{1a}$ event. This does not occur for ($\omega$, 3 $\omega$, $\phi=0$) since all events are identical. For both gauges, confinement to a narrower momentum range also leads to bright fringes near the origin and along the axes. 
The building blocks causing the interference shapes, such as the heart and the thick fringes along the negative half-axes, still act in the same way. The feathery fringes are skewed by these shapes, particularly evident with the blue fringes in Fig.~\ref{fig:4s0}(l). Similar to the $3s\rightarrow3p$ transition, temporal-exchange shifts play a significant role in the total interference pattern. 

\section{Conclusions}
\label{sec:conclusions}

In this paper, we investigate quantum interference in recollision excitation with subsequent ionization (RESI) for two-color linearly polarized fields, focusing on two-electron interference patterns from a single excitation channel. Previous analytical work derived quantum phase shifts assuming a single relevant ionization event for the second electron \cite{Hashim2024}. Here, we relax this assumption to generalize interference conditions for bichromatic fields, which leads to additional temporal-only and combined temporal-exchange phase differences. These arise from interfering events involving distinct second-electron ionization events within the same half-cycle, significantly enhancing the predictive power of analytical conditions. This refinement helps disentangle interference patterns by comparing fully coherent and incoherent distributions, revealing a rich variety of structures. Using the building blocks of such phase differences, we can accurately predict the shape and location of interference arising from each interference type. 
In bichromatic fields, dominant interfering events dictate momentum distribution regions \cite{Hashim2024b} and influence event contributions.

The interference patterns stem from phase differences related to electron exchange, time-delayed events, and from electron exchange \textit{and } temporally shifted events. While interference from temporal-only shifts is subtle, the other two types of interference may lead to prominent structures, easily identifiable in the RESI momentum distributions integrated over the transverse momenta. In contrast to our results in \cite{Hashim2024} for few-cycle pulses, their hierarchy is unclear. It can be manipulated by an appropriate choice of bichromatic-field parameters, such as the frequency ratio and the relative phase between the two driving waves. 

This is possible because different types of interference (exchange-only and exchange-temporal) occur mostly in specific regions of the $p_{1\parallel}p_{2\parallel}$ plane. If the two driving waves composing the bichromatic fields are comparable, one may control dominant events and confine the RESI distributions to specific momentum regions \cite{Hashim2024b}. For RESI distributions concentrated in the first and third quadrants of the parallel momentum plane, exchange-only interference prevails, while, for those located in the second and fourth quadrants, exchange-temporal interference is dominant. Examples of the former scenario are clear for $(\omega, 3\omega, \phi=-\pi/2) $ and $(\omega, 2\omega, \phi=3\pi/4) $  fields, while the latter situation is observed in, for instance, RESI with $(\omega, 3\omega, \phi=\pi/2) $ fields. A third possibility is to make both types of interference equally dominant using a RESI distribution located in the four quadrants, by using a field such as $(\omega, 3\omega, \phi=0) $. This specific field has all the symmetries associated with a monochromatic wave, which leads to a fourfold symmetric RESI distribution if interference is neglected \cite{Rook2022,Hashim2024b}.

Exchange-only interference forms alternating bright fringes along the main diagonal (``spine lines") skewed by hyperbolic patterns. 
Fringes along the anti-diagonal and the $(p_{1\parallel},p_{2\parallel}) \leftrightarrow (-p_{1\parallel},-p_{2\parallel})$ symmetry  are absent for $(\omega$,2$\omega)$ fields, but retained for $(\omega$,3$\omega)$. This is expected, as both are consequences of the half-cycle symmetry \cite{Hashim2024}. The main difference from previous studies is that, due to more than one event per half cycle, overlapping hyperbolic patterns may skew the spine or each other. 
Temporal-only processes remain least significant for bichromatic fields, consistent with results obtained for monochromatic waves \cite{Maxwell2015} and few-cycle pulses \cite{Hashim2024}. When isolated, they yield wing-shaped patterns and straight-line fringes in the second and fourth quadrants, as well as very faint ring-shaped patterns resembling those in \cite{Wang2012}.
Finally, combined temporal-exchange shifts gain relevance for two-color fields in comparison with few-cycle pulses. They lead to prominent 'feather-like' fringes in the second and fourth quadrants of the $p_{1\parallel}p_{2\parallel}$ plane skewed by temporal-only patterns.        

Bound-state prefactors introduce additional phase shifts and momentum biases, with  $V_{\mathbf{p}_{2}e}$ influencing the shapes of the distributions, and $V_{\mathbf{p}_{1}e,\mathbf{k}g}$ their centers and widths \cite{Hashim2024}. Care is needed when incorporating $V_{\mathbf{p}_{2}e}$ in the electron-momentum distributions peaked at non-vanishing momenta $p_{2||}$, mapped via 
$p_{2\parallel}=-A(t)$, to avoid bound-state singularities and account for complex tunnel-ionization times (see \cite{Hashim2024b} for details). The velocity-gauge prefactor leads to a greater number of `cuts' in the distributions where the direction, width, or spacing of fringes abruptly changes. The choice of gauge determines the region to which the interference is confined, although the patterns remain qualitatively similar.

Therefore, the key message is that one may confine, enhance, or suppress quantum interference in specific momentum regions by manipulating the driving-field parameters. For bichromatic fields, the relative phase between driving waves controls amplitude differences between same-half-cycle events, shifting event dominance and the location of momentum distributions. Moreover, changing the dominance of events alters the contrast of the interference fringes in certain regions and their skewing of other interference patterns. The field strongly influences the spacing, gradient, and intercept of straight-line fringes, and this can be foretold by inspecting the interference building blocks.  Thus, modifying the field can, in essence, alter the shapes of the total interference and, hence, the hierarchy of the interference types.  
Contributions from same-half-cycle events enhance patterns in specific quadrants; for instance, first- and third-quadrant patterns now include both exchange-only and combined temporal-exchange shifts. Adjusting amplitude differences or increasing event numbers per half-cycle can further enhance or suppress interference in certain regions. 

If using confinement in a specific momentum region to enhance quantum interference in RESI, or employing orbit-based approaches to model RESI distributions, several factors must be considered. First, the SFA neglects residual binding potentials in the continuum, which alters trajectory weights by modifying ionization probabilities \cite{Lai2015a} and introducing new trajectory classes \cite{Yan2010,Maxwell2017,rodriguez2023}. These new trajectories, if linked to quantum pathways, will carry phase differences, potentially altering interference patterns. For example, in photoelectron holography, the residual potential significantly affects near-threshold patterns \cite{Maxwell2017} but only slightly impacts high-energy, backscattered orbit interference \cite{Rook2024a}. Second, even within the SFA, interference between multiple excitation channels must be considered for shaped fields \cite{Maxwell2015,Maxwell2016}. Lastly, a more realistic model may need focal averaging and the inclusion of Gouy/Maslov phases \cite{Brennecke2020,Werby2021,Rook2024} to facilitate experimental comparisons, and different types of interference may be unequally affected by these features. Future investigations will address these aspects.

\textbf{Acknowledgements: } This work has been funded by the UK Engineering and Physical Sciences Research Council (EPSRC) (Grant No. EP/T019530/1) and by UCL. D. Habibovi\'{c} thanks UCL for its kind hospitality. 

\appendix
\section*{Appendix: The role of $F_{A}(\tau)$ on interference}

\label{Sec:integralAppendix}

\begin{figure*}[!htbp]
\centering
\includegraphics[width=0.97\textwidth]{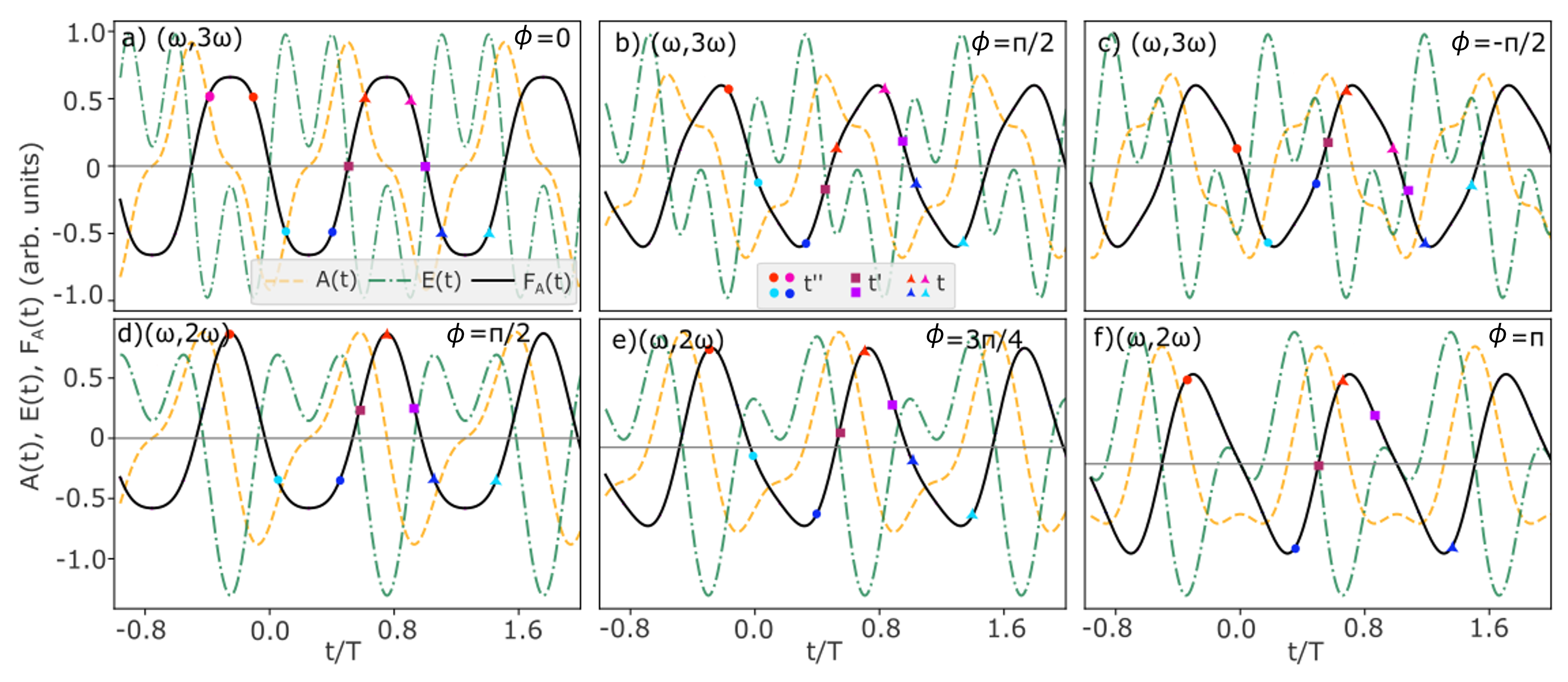}
  \caption{Integral of the vector potential (black solid line), vector potential (yellow dashed line) and electric field (green dash-dot line) for the ($\omega$,$3\omega$) (upper row) and ($\omega$,$2\omega$) (lower row) driving fields with the relative phase as indicated in the panels. The values of the integrated vector potential at the ionization and rescattering times of the first electron are denoted by circles and squares, respectively, whilst the values at the ionization time of the second electron are indicated with triangles. The colours of the $t'', t$ annotations are consistent with those of each $P_{i\xi} O_{m\xi}$ in the corresponding panels of Fig.~\ref{fig:fields}. Burgundy (purple) squares have been chosen to represent $t'$ associated with $P_{ia}$ ($P_{ib}$). All other driving-field parameters are consistent with those in the corresponding panels of Fig.~\ref{fig:fields}. Values of $F_A(t), E(t)$, and $A(t)$ must be extracted numerically to determine the slopes and intercepts of interference fringes, since arbitrary units are used here.}
   \label{fig:integral}
\end{figure*}

Figure~\ref{fig:integral} shows the integral of the vector potential for each of the fields and phases considered in this study. This term plays a significant role in many building blocks, and is highly dependent upon the field symmetry. In this Appendix, we illustrate its predictive power for the three different types of interference studied in this work.  

\subsection{Exchange Interference}

The field-dependent exchange building block [Eq.\eqref{eq:alphaexch2}] leads to `spines' parallel to the diagonals at $p_{1\parallel} = p_{2\parallel} + \delta_{\mathrm{exch}}$ [described by Eq.~\eqref{eq:spine}], the intercept of which is inversely proportional to $F_A(\tau)$.  One can compute the values of the intercept for a given $n$ by numerically obtaining the values of $F_A(\tau)$ at the rescattering and tunnelling times. For the fields and phases in this study, these values are provided in Table \ref{tab:exchangetable}. 

For all fields, there can be multiple fringes, each associated with a different value of $n$ which changes the intercept. The spacing of these fringes is given by 
\begin{equation}
 |\delta_{\mathrm{exch}}|=  \left|\frac{2\pi}{F_A(t+\Delta\tau)-F_A(t)}\right|.
   \label{eq:intercept}
\end{equation}

\begin{table}[h!]
   \centering
    \begin{tabular}{|c|c|c|c|c|c|}
       \hline
        \multicolumn{2}{|c|}{Field, $\phi$} & \multicolumn{4}{c|}{$F_A(t') - F_A(t)$} \\ \cline{1-6}
         &  & $P_{1a}O_{1a}$ & $P_{1a}O_{2a}$ & $P_{1b}O_{1a}$ & $P_{1b}O_{2b}$ \\ \hline
        \multirow{3}{*}{($\omega$, 3$\omega$)} 
        & 0        &       18      &     -18        &     18        &  -18           \\ \cline{2-6}
         & $\pi/2$  &        6     &      16       &  -6           &    -18         \\ \cline{2-6}
        & $-\pi/2$ &        8     &      0       &       -9      &   0          \\ \hline
       \multirow{3}{*}{($\omega$, 2$\omega$)} 
        & $\pi/2$  &        14     &       -      &     -11        & -11            \\ \cline{2-6}
         & $3\pi/4$ &      14       &     -        &    -8         &   -16          \\ \cline{2-6}
        & $\pi$    &       14      &    -         &      -21       &    -         \\ \hline
    \end{tabular}
   \caption{$F_A(t') - F_A(t)$ for each of the dominant events for all fields and phases considered. These values can be determined by numerically computing the difference between the $F_A(t)$ values at the positions of the rescattering and tunnelling times as exemplified in Fig.~\ref{fig:integral} by the squares and associated triangles for each event. Depending on the units at hand, some rescaling of the given values may be necessary. However, their ratios are a good indicator of the fringes' overall behavior, with specific signs dependent upon the field shape, and thus hold predictive power. }
    \label{tab:exchangetable}
\end{table}

\begin{figure}[!htbp]
\centering
\includegraphics[width=0.97\columnwidth]{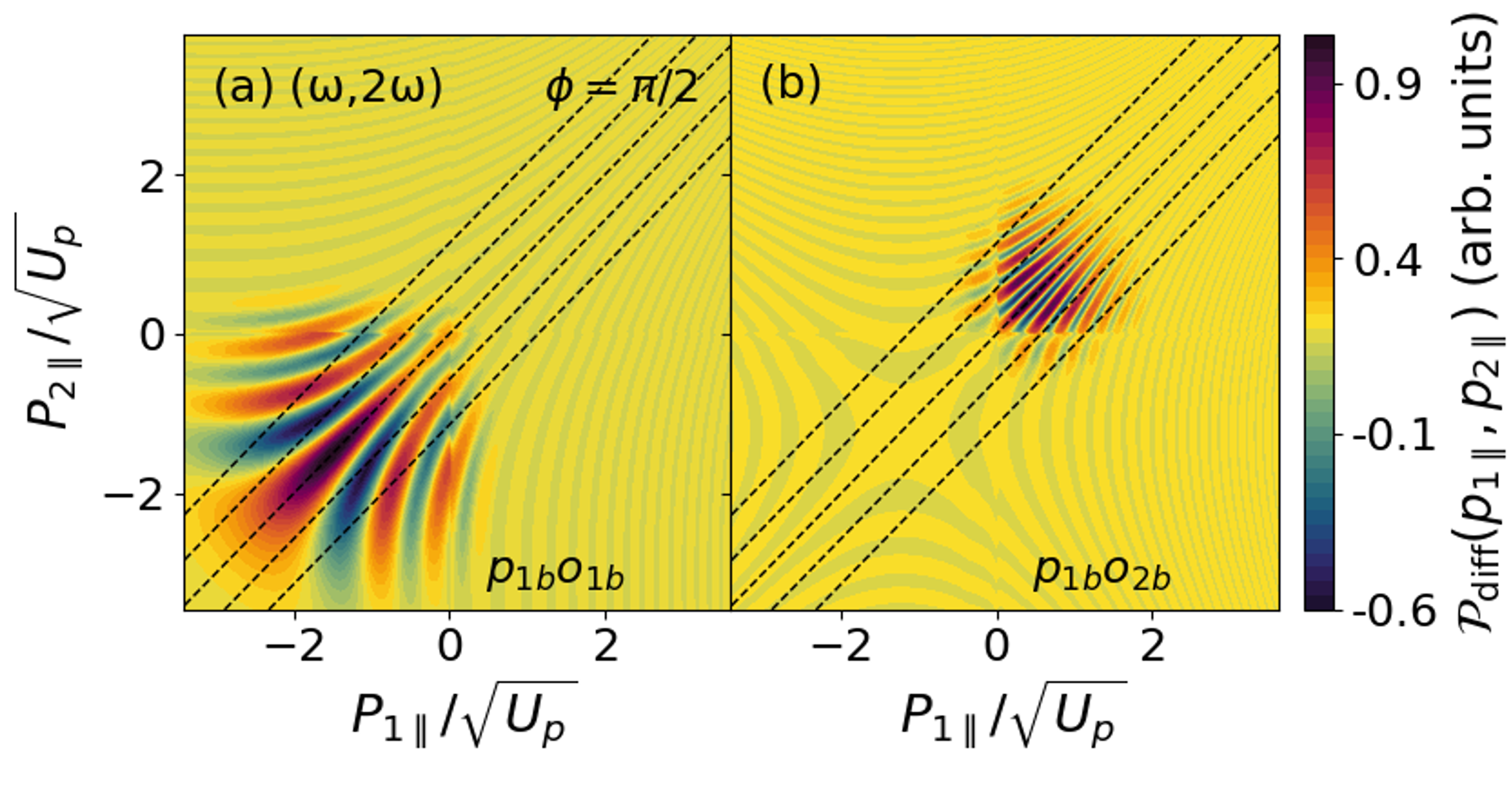}
    \caption{Difference between RESI distribution $\mathcal{P}_{(\mathrm{ci})}(p_{1\parallel},p_{2\parallel})$, in which symmetrization is performed coherently, and fully incoherent RESI probability densities $\mathcal{P}_{(\mathrm{ii})}(p_{1\parallel},p_{2\parallel})$, calculated without prefactor for individual events, $P_{1b}O_{1b}$ and $P_{1b}O_{2b}$ [panels (a), (b) respectively] with the $(\omega$,2$\omega$ $\phi = \pi/2$) driving field. Other driving-field parameters are the same as in the corresponding panel of Fig.~\ref{fig:fields} and Fig.~\ref{fig:coherent}. Dashed lines indicate locations of predicted spine lines arising from the pure-exchange phase term with varying $n$. Each panel has been normalized to its maximum probability density.}
    \label{fig:exchange90appendix}
\end{figure}

The interplay between the two exchange building blocks leads to spine lines being skewed by hyperbolae away from the origin. For some driving fields, the effect of the hyperbolae may be so strong that the spine line spacing predictions no longer hold. 
This deviation from the prediction is observed most prominently in fields where events are predicted to have the same $|\delta_{\mathrm{exch}}|$ i.e. for events with the same $F_A(t)$ value, which occur in the same half-cycle. This happens for the $(\omega, 3\omega, \phi=0)$ [Fig.~\ref{fig:exchange0}(a), (b)] and $(\omega, 2\omega, \phi=\pi/2)$ fields. Fig.~\ref{fig:exchange90appendix} shows the exchange fringes and identical predicted spacing for the $P_{1b}O_{1b}$ and $P_{1b}O_{2b}$ events respectively for the $(\omega, 2\omega, \phi=\pi/2)$ field. Whilst the predicted spines agree well in panels (a), the actual spacing in panel (b) is much narrower than expected. Closer to the origin, and hence closer to the center of the hyperbolae the fringes get thicker and the predicted spines are in better agreement.

\subsection{Event Interference}
The temporal building blocks $\alpha_{\Delta \tau}^{(A^2)}$ and $\alpha_{\Delta \tau}^{(\mathbf{p}_1, \mathbf{p}_2)}$ depend strongly on $F_A(\tau)$~[Eqs.~\eqref{eq:alphaA2} and \eqref{eq:alphap1p2tau}]. The sign and magnitude of the small shifts from $\alpha_{\Delta \tau}^{(A^2)}$ follow from the vector potential integral in Fig.~\ref{fig:integral}, as this phase difference reflects the variation in $F_A(\tau)$ at ionization and rescattering times (circles and squares in Fig.~\ref{fig:integral}, respectively).

Imposing the conditions for constructive interference on $\alpha_{\Delta \tau}^{(\mathbf{p}_1, \mathbf{p}_2)}$ leads to straight-line fringes for linearly polarized fields given by
\begin{equation}
  p_{2\parallel} = \frac{-p_{1\parallel}[F_A(t'+\Delta\tau)-F_A(t')] + 2n\pi}{F_A(t+\Delta\tau)-F_A(t)},
   \label{eq:temporalfringes}
\end{equation}
whose slope is highly dependent on the difference between $F_A(\tau)$ at the rescattering and tunnelling times associated with the interfering events. If the events are in the same half-cycle i.e., they share the same rescattering time, the fringes occur at $p_{2\parallel}=2n\pi$. With $n=0$, the fringes pass through the origin. Upon incoherent symmetrization, this can lead to chequerboard-type patterns depending on the width and spacing of the fringes [see Fig.~\ref{fig:temporal0}(c)]. This also occurs for fields where the integral of the vector potential is equal at the rescattering time for both events - for example in Figs.~\ref{fig:integral}(a) and (d) as indicated by the burgundy and purple squares. In theory, for a given field and phase, one can infer the relative magnitude of the gradients for each pair of interfering events by inspection of $F_A(\tau)$. However, since temporal shifts are washed out in the fully coherent distributions [Fig.~\ref{fig:coherent}], further discussion is neglected here. 
The $\alpha_{\delta t}^{\mathbf{p_2}}$ phase shift also contains an $F_A$-dependent term, which will lead to fringes parallel to the $p_{2\parallel}$ axis.

\subsection{Exchange-Temporal Interference}

\begin{figure}[!htbp]
\centering
\includegraphics[width=\columnwidth]{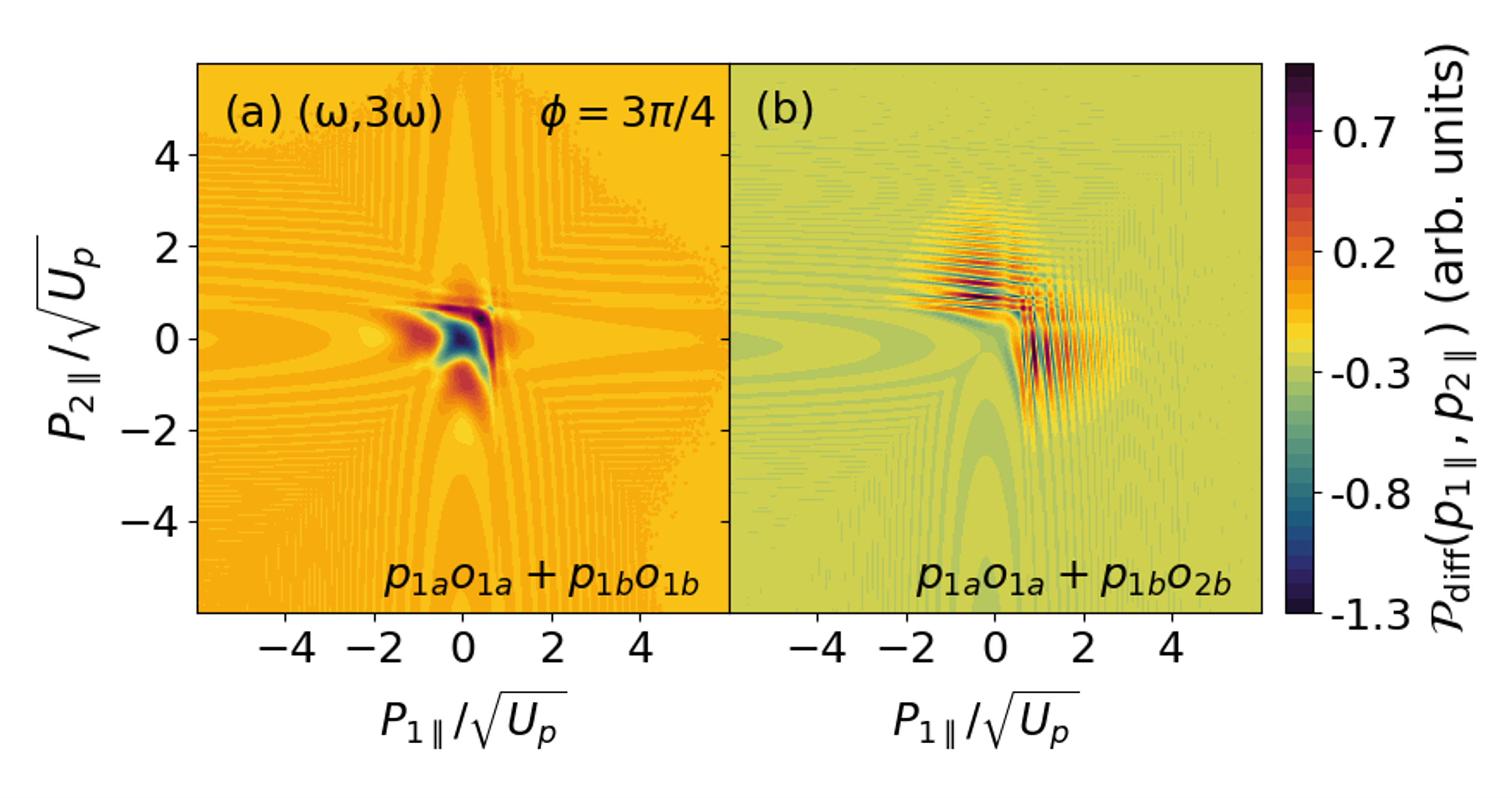}
    \caption{Difference between RESI distributions in which both exchange and event interference occur and the fully incoherent distribution calculated without prefactors for all possible events with $m \neq n$ and the $(\omega, 2\omega, \phi=3\pi/4)$ driving field. Other driving-field parameters are the same as in the corresponding panel of Figs.~\ref{fig:fields} and \ref{fig:coherent}. Each panel has been normalized to its maximum probability density.}
    \label{fig:combined135}
\end{figure}

\begin{figure}[!htbp]
\centering
\includegraphics[width=\columnwidth]{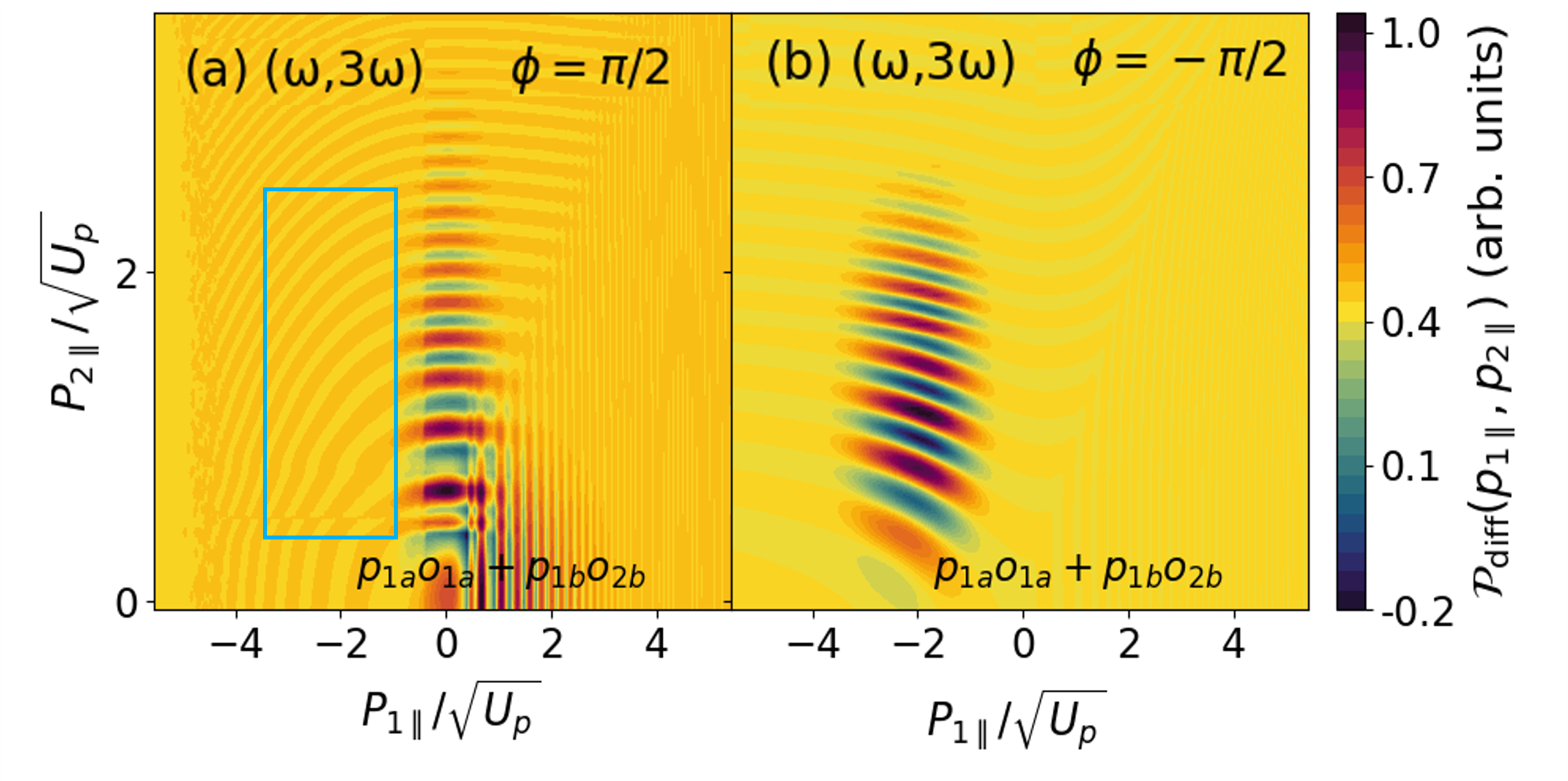}
    \caption{Difference between RESI distributions in which both exchange and event interference occur and the fully incoherent distribution calculated without prefactors for the $P_{1a}O_{1a}$ and $P_{1b}O_{2b}$ interfering events with   
    the $(\omega, 3\omega)$ driving field and $\phi=\pi/2$ (left) and $-\pi/2$ (right). Other driving-field parameters are the same as in the corresponding panels of Fig.~\ref{fig:fields} and \ref{fig:coherent}. Each panel has been normalized to its maximum probability density. The panels have been zoomed into the positive $p_{2\parallel}$ axes.}
    \label{fig:appendixcombined}
\end{figure}

The $\alpha_{\Delta\tau}^{(\mathbf{p}_1 \leftrightarrow \mathbf{p}_2)}$ exchange-temporal building block results in linear diagonal fringes with slopes that depend on the values of $F_A(\tau)$ at the rescattering and tunnelling times of the interfering events [Eq.~\eqref{eq:combinedfringes}]. For a given field and phase, the relative magnitude of the fringe gradients arising from a pair of interfering events can be inferred from Fig.~\ref{fig:integral} by considering the difference in values at times denoted by the squares and triangles. 

This is highly dependent on the shape of the electric field. For example, with $(\omega, 2\omega, \phi=\pi/2)$ the slopes of fringes arising from $P_{1a}O_{1a} + P_{1b}O_{1b}$ and $P_{1a}O_{1a} + P_{1b}O_{2b}$ are identical [see Fig.~\ref{fig:combined90}], and $O_{1/2b}$ both have the same electric field amplitude.  In contrast, with $(\omega, 2\omega, \phi=3\pi/4)$, the electric field amplitude of $O_{2b}$ is larger than $O_{1b}$ resulting in fringes from $P_{1a}O_{1a} + P_{1b}O_{2b}$ being slightly steeper and more closely spaced than $P_{1a}O_{1a} + P_{1b}O_{1b}$ as shown in Fig.~\ref{fig:combined135}.

With the $(\omega, 3\omega)$ field and $\phi=\pi/2$ and $\phi=-\pi/2$, fringes arising from $P_{1a}O_{1a}+P_{1b}O_{2b}$ and $P_{1a}O_{2a}+P_{1b}O_{1b}$ have inverse gradients, since the electric field amplitudes of these peaks are the inverse of each other. This means that when the combined interference is isolated from these combinations of events, we expect to see fringes with the same slope but in opposite directions -see Fig.~\ref{fig:appendixcombined}. For these phases, different momentum regions are illuminated; however, the blue box in panel (a) indicates the region in which the slopes in both plots are equal and opposite. This effect is quite subtle since the fringes are almost horizontal, but one may, in theory, alter the field to achieve more dramatic slopes. 

\end{document}